\documentclass[a4paper,fleqn]{cas-sc}
\usepackage[numbers,sort&compress]{natbib}

\usepackage{subcaption}
\usepackage{algorithm}
\usepackage{algpseudocode}
\usepackage{setspace}
\usepackage{makecell}

\usepackage{threeparttable}
\usepackage{lineno} 

\usepackage{adjustbox}
\usepackage{graphicx}
\usepackage{epstopdf}

\algrenewcommand\algorithmicindent{0.5em}

\usepackage{xcolor}
\newcommand{\rev}[1]{\textcolor{black}{#1}}

\def\tsc#1{\csdef{#1}{\textsc{\lowercase{#1}}\xspace}}
\tsc{WGM}
\tsc{QE}
\tsc{EP}
\tsc{PMS}
\tsc{BEC}
\tsc{DE}
\begin{document}
\let\WriteBookmarks\relax
\def\floatpagepagefraction{1}
\def\textpagefraction{.001}

% \linenumbers

\shorttitle{Gamma Process Degradation Modeling and Performance-Driven Opportunistic Maintenance}

\shortauthors{Haohao SHI}

\title [mode = title]{Gamma Process Degradation Modeling and Performance-Driven Opportunistic Maintenance Optimization for LED Lighting Systems
}
\author[1]{Haohao Shi}
\cormark[1]
\ead{haohao.shi@hdr.qut.edu.au}
\credit{Writing - original draft, Visualization, Software, Methodology, Conceptualization.}
\affiliation[1]{organization={ Faculty of Engineering, Queensland University of Technology},
    city={Brisbane},
    % citysep={}, % Uncomment if no comma needed between city and postcode
    postcode={4000}, 
    state={Queensland},
    country={Australia}}

\author[1]{Huy Truong-Ba}
\credit{Writing - review \& editing, Supervision, Funding acquisition, Methodology, Conceptualization}

\author[1]{Michael E. Cholette}
\credit{Writing - review \& editing, Supervision, Funding acquisition, Methodology, Conceptualization}

%Fourth author
%fifth author
\author[2]{Brenden Harris}
\credit{Writing - review \& editing, Funding acquisition, Data support}
\author[2]{Juan Montes}
\credit{Writing - review \& editing, Data support}
%sixth author
\author[1]{Tommy H.T. Chan}
\credit{Writing - review \& editing, Supervision, Funding acquisition}

\affiliation[2]{organization={Fredon Queensland},
    city={Brisbane},
    % citysep={}, % Uncomment if no comma needed between city and postcode
    postcode={4119}, 
    state={Queensland},
    country={Australia}}

\cortext[cor1]{Corresponding author: Haohao Shi}

\begin{highlights}
\item A performance-driven, simulation-in-the-loop framework optimizes opportunistic maintenance for LED lighting systems.
\item Bayesian inference calibrates a non-homogeneous Gamma process from LM-80 data.
\item A competing-failure formulation integrates package degradation and driver failures into a unified luminaire state.
\item A deficiency ratio quantifies long-term violations of illuminance requirements.
\item A surrogate model replaces repeated Radiance evaluations for scalable Monte Carlo policy search.
\end{highlights}

\begin{abstract}
Large-scale LED lighting systems consist of multiple LED luminaires, each subject to gradual package degradation and abrupt driver failures. Their acceptability is governed by long-term spatio-temporal illuminance compliance on the working plane, which is shaped collectively by all luminaire states, rather than by individual luminaire reliability alone. Current practice, however, verifies compliance only at commissioning through static indices, which cannot capture intermittent or persistent violations accumulated over the operating horizon. This paper proposes a performance-driven, simulation-in-the-loop framework for opportunistic maintenance optimization of LED lighting systems. Package degradation is modeled by a non-homogeneous Gamma process aligned with the IESNA TM-21 lumen maintenance projection method. Driver failures are described by a Weibull lifetime model. The two mechanisms are then integrated into a unified luminaire state via a competing-failure formulation. Model parameters are calibrated from IESNA LM-80 test data via Bayesian inference, with uncertainty propagated from stress levels to service conditions. A programmatic workflow integrates degradation modeling, ray-tracing simulation, and performance evaluation, mapping stochastic luminaire trajectories to the working plane illuminance field. To characterize long-term system performance, a deficiency ratio is defined to quantify the fraction of operational time during which average illuminance or uniformity requirements are violated. To enable scalable Monte Carlo evaluation, a surrogate model with high predictive fidelity replaces repeated Radiance evaluations. The opportunistic maintenance policy is then optimized in a multi-objective setting, balancing performance deficiency and maintenance effort. A realistic office-zone case study demonstrates the framework and reveals Pareto trade-offs for maintenance decision support.
\end{abstract}

% Keywords
% Each keyword is separated by \sep

\maketitle

\section{Introduction}

Light-emitting diode (LED) lighting systems are widely used in buildings because they offer high efficacy \citep{wen_-situ_2024,khalid_advances_2025}, long service life \citep{tan_chapter_2023, ibrahim_bayesian_2021}, and flexible control \citep{cui_distributed_2026}. LED lighting systems inherently place system performance at the center of operation and maintenance management because it directly affects occupant well-being and cognitive performance \citep{elhami_three-objective_2025}. In practice, system performance is not defined by the condition of any single luminaire, but by whether the illuminance field on the working plane (WP) satisfies prescribed performance requirements. Typically, these requirements are expressed in terms of maintained illuminance and illuminance uniformity, as specified in lighting standards such as Australian/New Zealand Standard (AS/NZS) 1680.1 \citep{joint_technical_committee_lg-001_as_2006}, Illuminating Engineering Society of North America (IESNA) Lighting Handbook \citep{judith_iesna_2000}, and British Standard European Norm (BS EN) 12464 \citep{the_british_standards_institution_light_2021}.

These requirements are evaluated over all discrete calculation points on the WP illuminance field \citep{joint_technical_committee_lg-001_as_2006, judith_iesna_2000, the_british_standards_institution_light_2021}, making them inherently both spatial and temporal. Spatially, each LED luminaire contributes to all calculation points simultaneously, and the illuminance field is further shaped by room geometry, surface reflectances, lighting system layout, and photometric distributions \citep{bean_lighting_2014}. Temporally, the illuminance field evolves with the real-time states of all LED luminaires. Consequently, \rev{although the degradation of individual luminaires may be statistically independent,} their collective evolution leads to strongly coupled, system-level performance changes. Together, these spatio-temporal characteristics and performance requirements make system-level acceptability fundamentally a performance-oriented question, rather than a simple aggregation of unit reliabilities.

However, in current engineering practice, these performance requirements are typically evaluated only at the commissioning stage. Compliance is verified for the as-designed, initial condition by evaluating static indices such as average illuminance and uniformity over calculation points on the WP illuminance field \citep{joint_technical_committee_lg-001_as_2006, judith_iesna_2000, the_british_standards_institution_light_2021}. Over long operating horizons, degradation and failures of luminaires accumulate and cause intermittent or persistent violations, which static indices alone cannot capture. Despite the practical importance of long-term compliance, large-scale LED lighting systems still lack a widely adopted system-level criterion for characterizing performance non-compliance over time and a metric for quantifying the duration of requirement violations over long-term operation.

Evaluating long-term system performance further requires time-indexed state trajectories of LED luminaires. Each luminaire is governed by two competing failure mechanisms: gradual degradation of LED packages and abrupt outages caused by driver failures \citep{fan_combined_2021}. For packages, the deterministic IESNA TM-21 lumen maintenance projection method (TM-21) \citep{iesna_ansi_2021} is widely used in practice, whereas stochastic degradation processes such as Gamma processes have been adopted to additionally capture path uncertainty \citep{ling_bayesian_2019, fan_gamma_2021, fan_prognostics_2021}. For drivers, Weibull-based lifetime models \citep{zhang_viable_2018} and physics-of-failure approaches \citep{sun_stochastic_2018, fan_combined_2021} have been explored. Despite these advances, several limitations remain. The two mechanisms are typically modeled in isolation, whereas an integrated luminaire state reflecting both gradual degradation and abrupt failures is needed for subsequent system-level analysis. Moreover, existing stochastic models are largely data-driven and do not align with the physically motivated exponential trend embodied in TM-21. In addition, because field observations are scarce over service lifetimes (often up to 60{,}000 hours \citep{iesna_ansi_2021}), accelerated degradation tests (ADT), such as the IESNA LM-80 lumen maintenance test (LM-80) \citep{iesna_approved_2008}, are typically required; however, the propagation of calibrated parameter uncertainty from stress conditions to service conditions is not always made explicit \citep{rocchetta_survey_2024}.

Mapping luminaire states to the WP illuminance field poses further challenges. Commercial lighting design tools such as DIALux Evo \citep{dialux_evo_2026}, Relux \citep{reluxdesktop_2026}, and AGi32 \citep{agi32_2026} are widely used for commissioning-time design, where a single fixed-state configuration is evaluated to verify compliance. However, these graphical user interface (GUI)-driven tools are not structured for programmatic invocation in Monte Carlo-driven long-term performance evaluation. Radiance \citep{shakespeare_rendering_2004}, a command-line ray-tracing engine, offers programmatic invocation that enables repeated performance evaluation driven by time-indexed stochastic state trajectories. Nevertheless, traditional design workflows are not intended for long-term performance evaluation. An integrated degradation--simulation--metric methodology is therefore required to enable Monte Carlo performance evaluation. Moreover, each Radiance evaluation remains computationally expensive, rendering large-scale Monte Carlo evaluation and policy search prohibitive. A surrogate-based acceleration therefore becomes essential at practical scales.

With these evaluation capabilities in place, maintenance policies can be optimized in a performance-driven manner. However, early research on lighting-related maintenance focused on retrofit-oriented optimization, where lamp replacement costs and their effects on energy savings and payback period were evaluated to identify optimal lighting schemes \citep{salata_maintenance_2015, malatji_multiple_2013, wang_multi-objective_2014}. Within this scope, control-theoretic frameworks have been developed to schedule lamp replacements for retrofitted lighting systems \citep{wang_control_2014, ye_optimal_2014, ye_optimal_2015}. The optimization scope was further expanded by jointly optimizing replacement timing and maintenance intensity as control inputs \citep{wang_optimal_2015, wang_multistate-based_2017}, and by incorporating luminous-flux degradation \citep{ikuzwe_energy-maintenance_2020} into the decision framework. However, these studies formulated maintenance as a finite-horizon scheduling problem within retrofit projects, rather than as the optimization of maintenance policy parameters governing long-term system operation, such as preventive maintenance (PM) intervals. Moreover, the optimization objectives focused on energy savings and economic return, rather than system-level lighting performance compliance with standard requirements.

In parallel, opportunistic maintenance has a mature methodological foundation for multi-unit systems, where maintenance actions are coordinated across units to exploit economic dependence \citep{gorenstein_predictive_2022, yang_mission_2024}, structural or functional relationships \citep{hu_risk_2014, wang_opportunistic_2023}, and other stochastic couplings, such as quality-dependent degradation effects \citep{shi_opportunistic_2023, wang_joint_2025}. \rev{Recent predictive maintenance studies have also moved toward system-centred re-optimization \citep{zhang_system-centred_2025}.} However, these studies typically evaluate system impact in terms of reliability, availability, or cost. In contrast, for lighting systems, system performance is governed by a spatial illuminance field on the WP. The system-level benefit of maintaining one luminaire is inherently coupled with the states and spatial layout of other luminaires. Adapting opportunistic maintenance to lighting systems therefore requires policy evaluation against spatial illuminance compliance rather than conventional reliability or cost metrics.

Motivated by these gaps, this paper develops a performance-driven simulation-in-the-loop framework for opportunistic maintenance (OM) optimization of large-scale LED lighting systems. The framework explicitly links luminaire-level degradation and failure processes to system-level compliance on the WP illuminance field, enabling maintenance decisions to be evaluated in terms of long-term system performance rather than luminaire reliability alone. By combining stochastic degradation/failure modeling, illuminance-field mapping, long-term system-level compliance assessment, and policy optimization within a unified formulation, the proposed framework enables opportunistic maintenance to be studied in a manner that is specific to LED lighting systems. The main contributions are summarized as follows:
\begin{itemize}
    \item \textit{Performance-driven OM framework:} a performance-driven opportunistic maintenance optimization framework for LED lighting systems is developed, in which a system-level deficiency ratio metric is introduced for multi-objective policy evaluation balancing system performance and maintenance effort.

    \item \textit{Physics-motivated degradation modeling and ADT-to-service-condition extrapolation:} a non-homogeneous Gamma process with a physics-motivated exponential mean trend is developed for LED package degradation; a Weibull-based lifetime model is introduced for LED driver failures, and the two are combined through a competing-failure formulation to define the overall luminaire state. An ADT-to-service-condition extrapolation and Bayesian inference workflow is employed to estimate degradation behavior across multiple stress levels and propagate uncertainty to service conditions.

    \item \textit{Programmatic performance-mapping methodology with surrogate-based scalable acceleration:} an integrated degradation--simulation--metric methodology is established to propagate stochastic luminaire state trajectories to the working plane illuminance field for performance metric assessment; a surrogate-based performance mapping is integrated into the methodology to replace repeated ray-tracing evaluations, thereby reducing computational overhead and enabling large-scale policy evaluation and optimization.
\end{itemize}

The remainder of this paper is organized as follows. Section~2 develops the degradation and failure models and presents the ADT-to-service-condition extrapolation and Bayesian inference workflow. Section~3 introduces the system-level performance mapping and proposes a deficiency ratio metric for assessing long-term system performance. Section~4 formulates the performance-driven maintenance optimization framework and details the simulation-based policy evaluation and surrogate-based acceleration. Section~5 presents a realistic case study with calibration, optimization results, and sensitivity analyses. Section~6 concludes the paper and outlines future directions. Table~\ref{tab:notation_problem} summarizes the key notation and core abbreviations used in this manuscript.

\begin{table}[width=.9\linewidth,cols=2,pos=h]
    \caption{Key notation and core abbreviations used in the manuscript.}
    \label{tab:notation_problem}
    \centering
        \begin{tabular*}{\tblwidth}{@{} LL@{} }
            \toprule
            Symbol/Abbreviation & Description \\
            \midrule
            WP & Working plane \\
            ADT & Accelerated degradation tests \\
            DR & Deficiency ratio \\
            OM & Opportunistic maintenance \\
            PM & Preventive maintenance \\
            CM & Corrective maintenance \\
            $J$ & Number of luminaires in the lighting system \\
            $N$ & Number of working plane calculation points \\
            $s=1,\dots,S$ & Monte Carlo replication index \\
            $k=1,\dots,K$ & Event index \\
            $T_{\text{over}}$ & Operational horizon \\
            $T_{\text{PM}}$ & Preventive maintenance (PM) interval (decision variable) \\
            $H_{\text{OM}}$ & Opportunistic maintenance (OM) threshold (decision variable) \\
            $T_{\text{OM}}$ & Equivalent OM age trigger, $T_{\text{PM}}(1-H_{\text{OM}})$ \\
            $\mathbf{L}(t)$ & Luminaire degradation-state vector at time $t$ \\
            $\mathbf{E}(t)$ & Illuminance vector on the working plane at time $t$ \\
            $E_{\text{avg}}(t)$ & Average illuminance at time $t$ \\
            $U(t)$ & Illuminance uniformity at time $t$ \\
            $S_{E}, S_{U}$ & Performance requirements (standards) \\
            $T_{\text{defi}}(k)$ & Deficiency duration within $(t_{k-1},t_k)$ \\
            $R_{\text{DR}}$ & Deficiency ratio objective over $[0,T_{\text{over}}]$ \\
            $R_{\text{DR},s}$ & Deficiency ratio for simulation run $s$ over $[0,T_{\text{over}}]$ \\
            $R_{\text{SDR}}$ & Mean deficiency ratio over $S$ simulation runs \\
            $N_{\text{tv}}$ & Total-number-of-site-visits objective \\
            $N_{\text{tv},s}$ & Total number of site visits in simulation run $s$ \\
            $N_{\text{STV}}$ & Mean total number of site visits over $S$ simulation runs \\
            $N_{\text{tr}}$ & Total-number-of-replacements objective \\
            $N_{\text{tr},s}$ & Total number of replacements in simulation run $s$ \\
            $N_{\text{STR}}$ & Mean total number of replacements over $S$ simulation runs \\
    \bottomrule
    \end{tabular*}
\end{table}

\section{Degradation modeling and Bayesian calibration of LED luminaires}
This section develops luminaire-level models for the two dominant failure mechanisms of an LED luminaire, namely gradual LED package degradation (manifested as light-output loss) and abrupt LED driver failures. Package degradation is modeled by a non-homogeneous Gamma process whose mean evolution follows a physically motivated exponential form. Driver lifetime is modeled by a Weibull distribution. The two mechanisms are assumed to be statistically independent at the luminaire level and are integrated through a competing-failure formulation to define the overall luminaire state for subsequent system-level performance analysis. Since field data are scarce due to the long service life of LED packages, ADT is employed; a stress--acceleration relationship is introduced to link model parameters across temperatures and to extrapolate them to service conditions. Finally, Bayesian inference is employed to estimate model parameters jointly across multiple stress levels and to quantify parameter uncertainty for downstream Monte Carlo simulation and maintenance optimization.

\subsection{Degradation and failure models for LED packages and drivers} \label{sec:Component_level}
In an LED luminaire, the LED package (hereafter, package) is responsible for light generation, whereas the LED driver (hereafter, driver) provides electrical power conversion and regulation. Package degradation is gradual and typically observed as a loss of normalized light output over time, while driver failures are often abrupt and result in a complete loss of functionality. The corresponding stochastic models are presented next, followed by their integration into a unified luminaire state.

\paragraph{LED package degradation: a non-homogeneous Gamma process.}
Package degradation is quantified via the normalized light output. Following the widely adopted $L_{70}$ criterion from the IESNA, a degradation failure is defined as the condition in which the light output falls below 70\% of its initial value \citep{iesna_ansi_2021}.

Let $\mathbf{X}(t) = \left[ X_{1}(t), X_{2}(t), \dots, X_{J}(t) \right]$ denote the package degradation states at time $t$, where
\begin{equation}
    X_{j}(t) = 1 - P_{j,\text{out}}(t), \quad j=1,2,\ldots,J,
    \label{eq:LO}
\end{equation}
and $0 \leq P_{j,\text{out}}(t) \leq 1$ represents the normalized light output of the $j$-th package (with $P_{j,\text{out}}(0)=1$). Hence, $L_{70}$ corresponds to $P_{j,\text{out}}(t) < 0.7$, equivalently $X_j(t) > 0.3$.

For each package $j$, $\{X_j(t), t\ge 0\}$ is modeled as a non-homogeneous Gamma process. This choice is motivated by the physical characteristics of LED lumen degradation. Specifically, lumen degradation is driven by cumulative and irreversible damage mechanisms \citep{zhao_progress_2022, letson_reviewreliability_2023}, leading to a monotonic degradation process. The Gamma process enforces non-negative increments and is therefore more directly aligned with the monotonic nature of the latent degradation path. This modeling choice is also consistent with prior studies on LED degradation \citep{sun_stochastic_2018, ling_bayesian_2019, fan_gamma_2021, fan_prognostics_2021}.

More formally, for any time instants $0 \le t_1 < t_2 < \cdots < t_K$, the increments
$X_j(t_2) - X_j(t_1), \, X_j(t_3) - X_j(t_2), \ldots, X_j(t_K) - X_j(t_{K-1})$
are independent and non-negative, with $X_j(0)=0$ almost surely. Moreover, for $t>s$,
\begin{equation}
    X_j(t)-X_j(s) \sim \mathrm{Ga}\bigl(\alpha(t)-\alpha(s), \beta\bigr),
\end{equation}
where $\alpha(\cdot)$ is a non-decreasing shape function and $\beta$ is the rate parameter. Accordingly, the marginal distribution of $X_j(t)$ is Gamma with shape $\alpha(t)$ and rate $\beta$, and its probability density function (PDF) is
\begin{equation}
    f_{X_j(t)}(x)= \frac{\beta^{\alpha(t)}}{\Gamma(\alpha(t)) } x^{\alpha(t) - 1} \exp\left( -\beta x \right), \quad x >0,
    \label{eq:XPDF}
\end{equation}
where $\Gamma(\cdot)$ denotes the Gamma function.

Compared with the standard exponential model used in TM-21 \citep{iesna_ansi_2021}, the Gamma process allows uncertain degradation paths to be represented. As discussed above, TM-21 assumes an exponential average light-output decay under constant stress. This is physically motivated: thermal factors are a primary cause of package degradation, and thermal accumulation inside the package can exhibit an exponential trend dependent on temperature \citep{tan_chapter_2_2023}. Therefore, a time-dependent exponential shape function is introduced to build a non-homogeneous Gamma process.
\begin{equation}
    \alpha(t) = A \exp(bt),
\end{equation}
where $A$ and $b$ are constants and are assumed invariant across stress levels under the linear acceleration assumption introduced later. This choice yields an exponential mean degradation trajectory:
\begin{equation}
    \mathbb{E}[X_j(t)] = \frac{\alpha(t)}{\beta},
\end{equation}
and for $t>s$, the expected increment is
\begin{equation}
    \mathbb{E}[X_j(t) - X_j(s)] = \frac{\alpha(t)-\alpha(s)}{\beta}.
\end{equation}

\paragraph{LED driver failures: a Weibull lifetime model.}
Driver malfunctions are characterized by a sudden and complete loss of functionality. This behavior is appropriately modeled using a traditional two-state approach: operational versus failed, where the failure time of drivers is assumed to follow a Weibull distribution \citep{zhang_viable_2018}. The PDF is given by:
\begin{equation}
    f_{T_{d}}(t) = \frac{\eta}{\lambda} \left( \frac{t}{\lambda} \right)^{\eta-1} \exp \left( - \left( \frac{t}{\lambda} \right)^{\eta} \right), t \geq 0,
    \label{eq:WBL}
\end{equation}
where $\lambda$ is the scale parameter and $\eta$ is the shape parameter.

\paragraph{Competing failure integration for luminaire state.}
To evaluate the overall state of an LED luminaire (comprising both the package and the driver), let $\mathbf{L}(t) = [L_{1}(t), L_{2}(t), \ldots, L_J(t)]$ denote the luminaire degradation states in the lighting system. The degradation state of the $j$-th luminaire at time $t$ is defined as 
\begin{equation}
    \begin{aligned}
        L_j(t)  & = 1 - \left( 1-X_j(t) \right) \cdot \left( 1 - S_j(t) \right) \\
                & = S_j(t) \cdot \left( 1 - X_j(t) \right) + X_j(t),
        \label{eq:FL}
    \end{aligned}    
\end{equation}
where $S_j(t)$ is the driver failure indicator:
\begin{equation}
    S_j(t) =
    \begin{cases}
        1 & \text{if the driver has failed before } t \\
        0 & \text{otherwise}
    \end{cases}.
\end{equation}

In particular, this formulation reflects that driver failure is treated as an immediate luminaire failure condition: once the driver fails, the luminaire is considered fully failed ($L_j(t)=1$); otherwise, the state is governed by the gradual package degradation $X_j(t)$. Equivalently,
\begin{align}
    L_j(t) =
    \begin{cases}
        1  & \text{if the driver has failed before } t \\
        X_j(t) & \text{otherwise}
    \end{cases}.
    \label{eq:LRB}
\end{align}

This competing-failure formulation captures both gradual package degradation and abrupt driver failures, enabling consistent luminaire state trajectories to be generated for subsequent system-level performance evaluation and maintenance optimization.

\subsection{Linear accelerated model and parameter extrapolation}
Because typical LED packages age slowly and have lifetimes exceeding 60{,}000 hours \citep{iesna_ansi_2021}, field failure data are scarce within practical observation periods. Therefore, ADT is widely used to obtain sufficient degradation data within a practical time frame. The LM-80 method \citep{iesna_approved_2008} is the industry-standard constant-stress ADT for LED packages and is routinely performed prior to commercial release.

Linear acceleration theory assumes that the underlying failure mechanism remains invariant across stress levels. Under this assumption, the Gamma process shape function is taken as invariant, while the rate parameter varies with stress (e.g., temperature) \citep{tobias_applied_2012, elsayed_reliability_2020,  zhou_two-stage_2025}. Let $\beta(T)$ denote the rate parameter at case temperature $T$. The Arrhenius model is adopted to describe the temperature dependence \citep{tobias_applied_2012, elsayed_reliability_2020,jiang_inference_2019,rocchetta_uncertainty_2024}: 
\begin{equation}
    \beta(T) = C \exp\left(\frac{E_a}{k_B T} \right),
\end{equation}
where $C$ is a pre-exponential factor, $E_a$ is the activation energy (eV), $k_B$ is the Boltzmann constant ($8.62\times 10^{-5}$ eV/K), and $T$ is the absolute temperature (K). Taking logarithms yields
\begin{equation}
    \ln \beta(T) = \ln C + \frac{E_a}{k_BT}    
\end{equation}

Following completion of the ADT, an acceleration model is fitted to the high-stress data and then extrapolated to service conditions. The acceleration factor (AF) is defined as
\begin{equation}    
    AF  =  \frac{t({T_{\text{stress}}})} {t({T_{\text{normal}}})} 
        =  \frac{\beta ({T_{\text{stress}}})} {\beta (T_{\text{normal}})},
\end{equation}
where $t({T_{\text{stress}}})$ and $t({T_{\text{normal}}})$ are failure times under stress and normal temperatures, $T_{\text{stress}}$ and $T_{\text{normal}}$, respectively. 

In addition, AF-based extrapolation yields only a nominal (point) estimate of the service-condition rate parameter and does not quantify parameter uncertainty. To address this limitation, Bayesian parameter inference is adopted in the next subsection to capture epistemic uncertainty under service conditions while retaining the linear acceleration relationship across stress levels.

\subsection{Bayesian parameter inference under multiple stress conditions} \label{sec:BPI}
A Bayesian inference framework is adopted to estimate the parameters of the non-homogeneous Gamma process model and to quantify uncertainties associated with the parameter estimates and the degradation process. Observations collected under different case temperatures are flattened and combined into a unified dataset of the form:
\begin{equation}
    \mathscr{D} = \{(x_{k-1},\, x_k, \, t_{k-1},\, t_k,\, T_k) \mid k=1,2,\ldots,M\},
\end{equation}
where $x_k$ is the observed degradation value at time $t_k$ under (constant) stress temperature $T_k$. Under the non-homogeneous Gamma process model, 
\begin{equation}
    X_k - X_{k-1} \sim \mathrm{Ga} \bigl(\alpha(t_k)-\alpha(t_{k-1}),\beta(T_k)\bigr)    
\end{equation}
with the exponential shape function $\alpha(t) =\exp (\ln A +bt)$ and the temperature-dependent rate function $\beta(T) = \exp(\ln C + E_a/(k_B T))$.

In Bayesian inference, the unknown model parameters are treated as random variables characterized by prior distributions, which encode prior engineering knowledge or physical constraints. These priors are updated with the observed data $\mathscr{D}$ to yield the posterior distribution. Let $\boldsymbol{\theta} = (\ln A,b, \ln C, E_a )$ denote the transformed parameters. Given the observations, the \textit{likelihood} function is expressed as
\begin{equation}
    f(\mathscr{D} | \boldsymbol{\theta}) = \prod_{k=1}^{M} f_{\Gamma} (x_k -x_{k-1} | \alpha(t_k; \boldsymbol{\theta})-\alpha(t_{k-1}; \boldsymbol{\theta}),\beta(T_k; \boldsymbol{\theta}))
\end{equation}
where $f_{\Gamma} (\cdot)$ denotes the Gamma PDF. By Bayes' theorem, the posterior distribution is given by:
\begin{equation}
    p(\boldsymbol{\theta} | \mathscr{D}) = \frac{f(\mathscr{D}|\boldsymbol{\theta}) p(\boldsymbol{\theta})}{\int f(\mathscr{D}|\boldsymbol{\theta}) p(\boldsymbol{\theta}) d\boldsymbol{\theta}}
\end{equation}
where the denominator represents the evidence.

To propagate both epistemic and aleatory uncertainties, a two-stage Monte Carlo scheme is employed. First, $n_{\mathrm{path}}$ parameter sets ${\boldsymbol{\theta}_{l} = (\ln A_{l},b_{l}, \ln C_{l}, {E_a}_{l})}_{l=1}^{n_{\mathrm{path}}}$ are drawn from the posterior distribution using Markov chain Monte Carlo (MCMC). Then, for each draw $\boldsymbol{\theta}_{l}$, $n_{\mathrm{ps}}$ independent Gamma process trajectories are generated:
\begin{equation}
    G(t)_{(l,m)} - G(s)_{(l,m)}  \sim \mathrm{Ga}\left( \alpha_l(t) - \alpha_l(s), \beta_l(T) \right), \quad m = 1,\dots, n_{\mathrm{ps}}.
    \label{eq:ddraw}
\end{equation}

Consequently, a total of $n_{\mathrm{path}} \times n_{\mathrm{ps}}$ trajectories are generated to represent virtual luminaires accounting for both parameter uncertainty and process variability. Maintenance policies are applied to these luminaires, and the optimization objectives are estimated by averaging over trajectories. This approach ensures that the variability of degradation outcomes is adequately characterized for each plausible parameter, thereby reducing estimator noise and avoiding undue influence of extreme single-path realizations. 

The inferred luminaire state trajectories $\mathbf{L}(t)$ provide the time-varying status of individual luminaires and serve as inputs to the system-level lighting analysis in the next section, where dynamic illuminance distributions and performance indices are evaluated over the WP.

\section{System-level performance mapping and deficiency ratio metric for LED lighting systems} \label{sec:SPM}

Conventional lighting performance assessments are predominantly static, focusing on whether commissioning-time metrics meet relevant standards. Such assessments do not capture performance fluctuations over the service life of the system caused by luminaire degradation and failures, which may lead to risks in occupant safety, visual comfort, and productivity. In this section, a physics-based simulation model is used to establish a system-level mapping from luminaire states to the WP illuminance field, and two standard performance indices are adopted to quantify static performance. These static indices are then recast into performance-deficiency measures through the performance deficiency duration (PDD) and the deficiency ratio (DR), enabling evaluation of the long-term dynamic performance of the system for subsequent maintenance optimization.

\subsection{System-level performance mapping and static performance indices}
\begin{figure}
    \centering
    \begin{subfigure}[t]{0.3\textwidth}
        \centering
        \includegraphics[height=3cm, keepaspectratio]{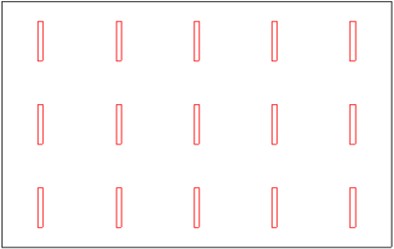}
        \caption{Luminaire layout of the lighting system}
        \label{fig:DLS}
    \end{subfigure}
    \hfill
    \begin{subfigure}[t]{0.3\textwidth}
        \centering
        \includegraphics[height=3cm, keepaspectratio]{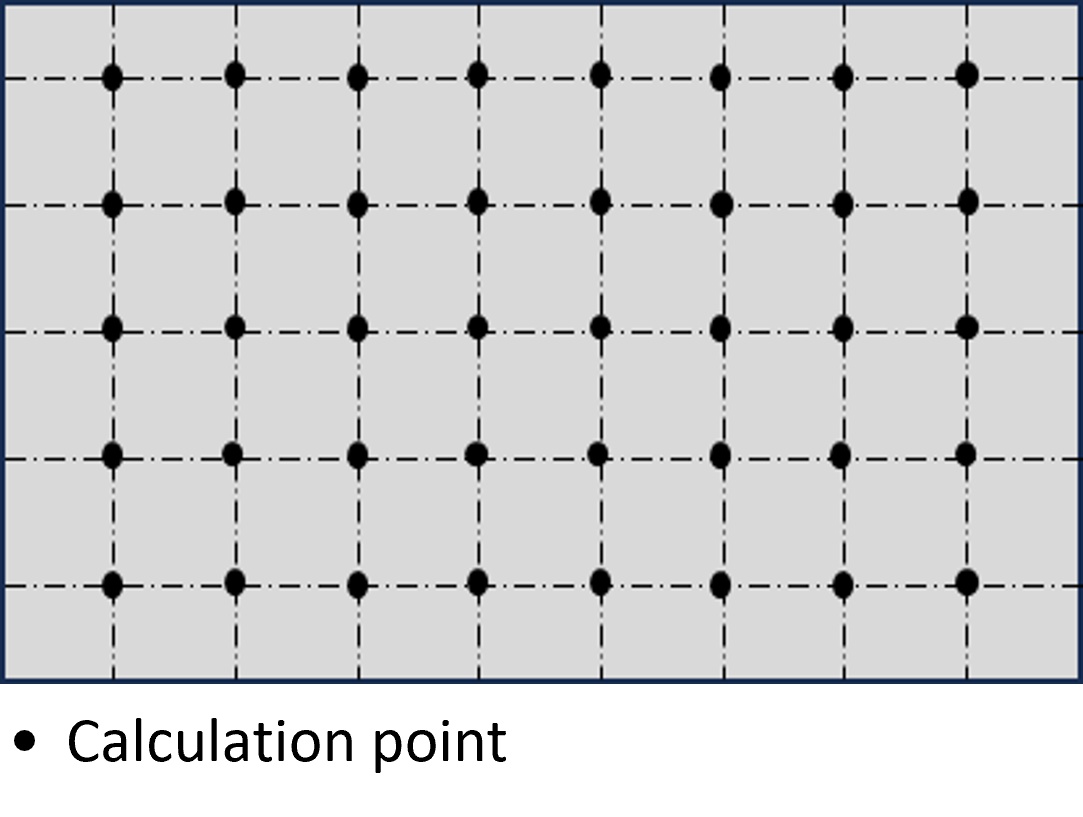}
        \caption{Working plane calculation grid}
        \label{fig:TMG}
    \end{subfigure}
    \hfill
    \begin{subfigure}[t]{0.3\textwidth}
        \centering
        \includegraphics[height=3cm, keepaspectratio]{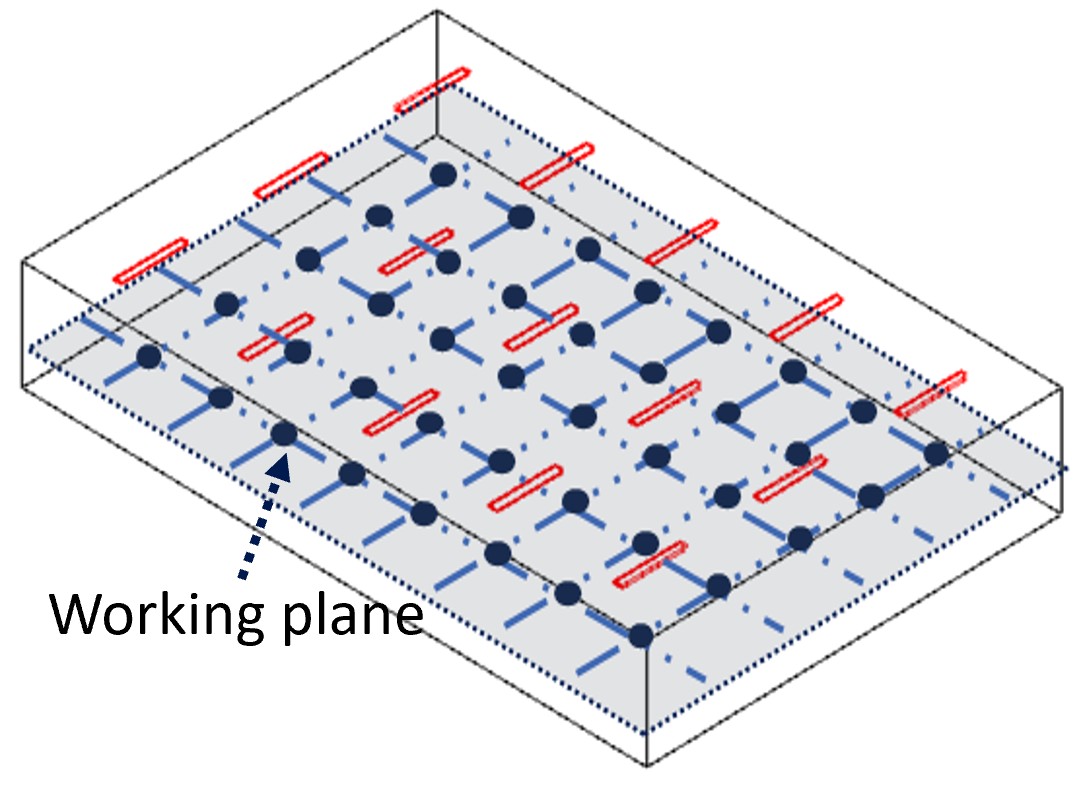}
        \caption{Schematic of the working plane}
        \label{fig:DWP}
    \end{subfigure}
    \caption{Schematic of the LED lighting system layout, working plane calculation grid, and working plane used for performance evaluation.}
    \label{fig:ALEDL}
\end{figure}

The illuminance distribution on a horizontal WP provides a practical basis for assessing the system-level performance of an LED lighting system in indoor environments. The WP is defined as a virtual horizontal plane, typically located at a height of approximately 0.7--0.85~m above the floor to represent desk height for tasks such as writing and data processing. As illustrated in \autoref{fig:ALEDL}, the gray plane represents the WP, red rectangles denote LED luminaires, and black markers denote calculation points used for quantitative performance assessment.

A physics-based ray-tracing simulator, Radiance, is employed to compute the WP illuminance field from the building geometry, material properties, luminaire photometry, and luminaire states. Radiance can process 3D models generated by Building Information Modeling (BIM) tools and extract geometric and material properties for lighting analysis. Its simulation engine combines Monte Carlo ray tracing with deterministic techniques \citep{shakespeare_rendering_2004, bean_lighting_2014}, and its command-line interface facilitates integration with Python for batch evaluation.

\begin{figure}
    \centering
    \begin{subfigure}[b]{0.6\textwidth}
        \centering
        \includegraphics[width=\textwidth]{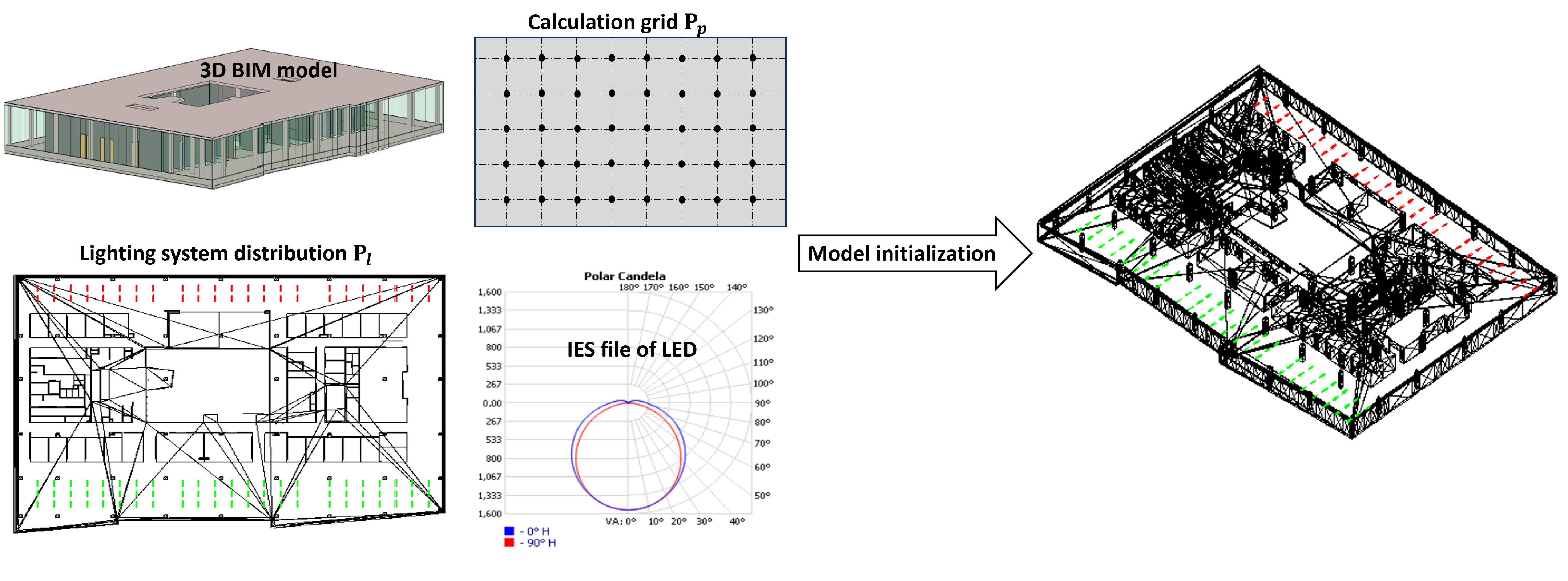} 
        \caption{Model initialization}
        \label{fig:MI}
    \end{subfigure}
    \begin{subfigure}[b]{0.6\textwidth}
        \centering
        \includegraphics[width=\textwidth]{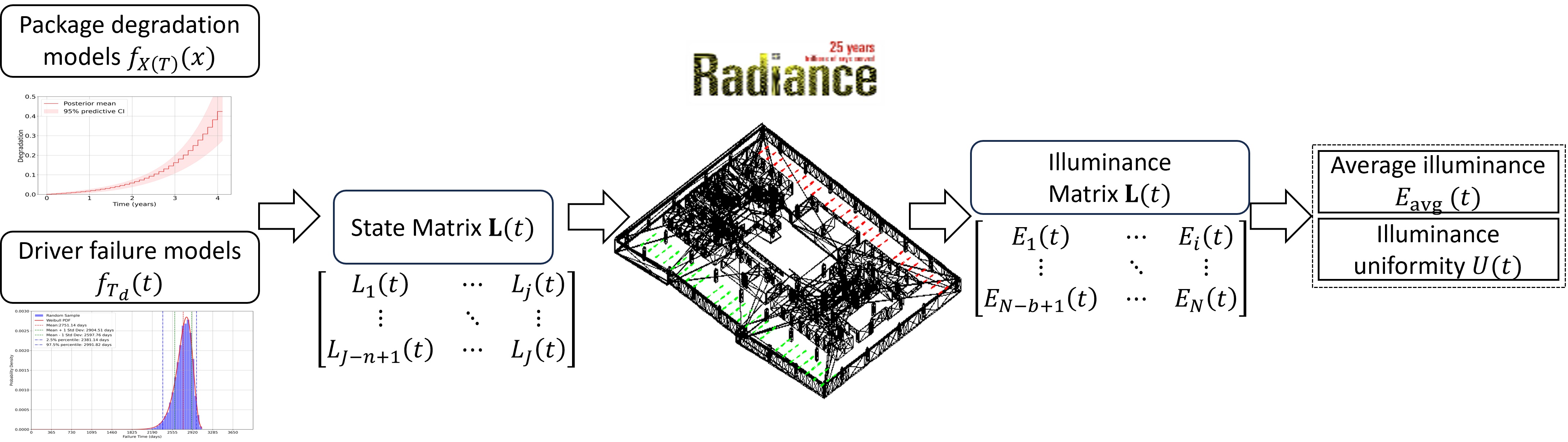} 
        \caption{Illuminance evaluation and performance analysis}
        \label{fig:LQA}
    \end{subfigure}
\caption{Radiance-based workflow for mapping luminaire degradation states to working plane illuminance and performance indices.}
\label{fig:FLA}
\end{figure}

Accurate lighting source modeling is essential for reliable illuminance prediction. In this work, luminaire photometry is specified using IESNA LM-63 photometric (.IES) files \citep{iesna_iesnalm-63-02_2002} and combined with a 3D BIM-based geometry/material model, luminaire layout, and a WP calculation grid to establish a physics-based performance mapping in Radiance. Preparing these 3D inputs (geometry and material assignment, photometric specification, luminaire placement, and grid definition) is an engineering-intensive step; however, once the model is established, it can be repeatedly evaluated at discrete time points as the luminaire degradation trajectories evolve over the operational life. These inputs are used in Radiance to establish the performance-mapping model, as depicted in \autoref{fig:MI}.

As summarized in \autoref{fig:LQA}, the Radiance-based performance mapping is executed through a programmatic workflow: stochastic luminaire state trajectories generated in Python are used to parameterize the model at discrete time points, and the resulting WP illuminance outputs are automatically parsed for performance evaluation. This automation is essential for the large-scale Monte Carlo evaluation and optimization presented in later sections.

Let the luminaire degradation state vector be defined by \rev{Equations~(\ref{eq:LO})--(\ref{eq:LRB})} as
\begin{equation}
    \mathbf{L}(t) = \left[ L_1(t), L_2(t), \dots, L_J(t)\right],
    \label{eq:SMLS}
\end{equation}
where $j=1,2,\dots,J$ denotes the $j$-th luminaire. Given $\mathbf{L}(t)$, the performance mapping model outputs the corresponding WP illuminance vector
\begin{equation}
    \mathbf{E}(t) = \left[ E_1(t), E_2(t), \dots, E_N(t) \right], 
    \label{eq:IMWP}
\end{equation}
where $E_i(t)$ is the illuminance at the $i$-th calculation point, measured in lux (lx), i.e., luminous flux per unit area, and $i=1,2,\dots,N$.

To quantify static performance at time $t$, two standard illuminance-based indices are adopted \citep{joint_technical_committee_lg-001_as_2006, judith_iesna_2000, the_british_standards_institution_light_2021}. The average illuminance is
\begin{equation}
    E_{\text{avg}}(t)= \frac{1}{N} \sum_{i=1}^{N} E_{i}(t),
    \label{eq:avg}
\end{equation}
and the illuminance uniformity (hereafter, uniformity) is defined as
\begin{equation}
    U(t) = \frac{E_{\text{min}}(t)}{E_{\text{avg}}(t)},
    \label{eq:NUI}
\end{equation}
where $E_{\text{min}}(t)=\min_{1\le i\le N} E_i(t)$.

Let $S_E$ and $S_U$ denote the standard requirements (thresholds) for average illuminance and uniformity, respectively. Static compliance at time $t$ is assessed by
\begin{equation}
    E_{\text{avg}}(t)\ge S_E,\qquad U(t)\ge S_U.
\label{eq:static_req}
\end{equation}

\begin{figure}
    \centering
    \begin{subfigure}[t]{0.3\textwidth}
        \centering
	\includegraphics[width=\textwidth]{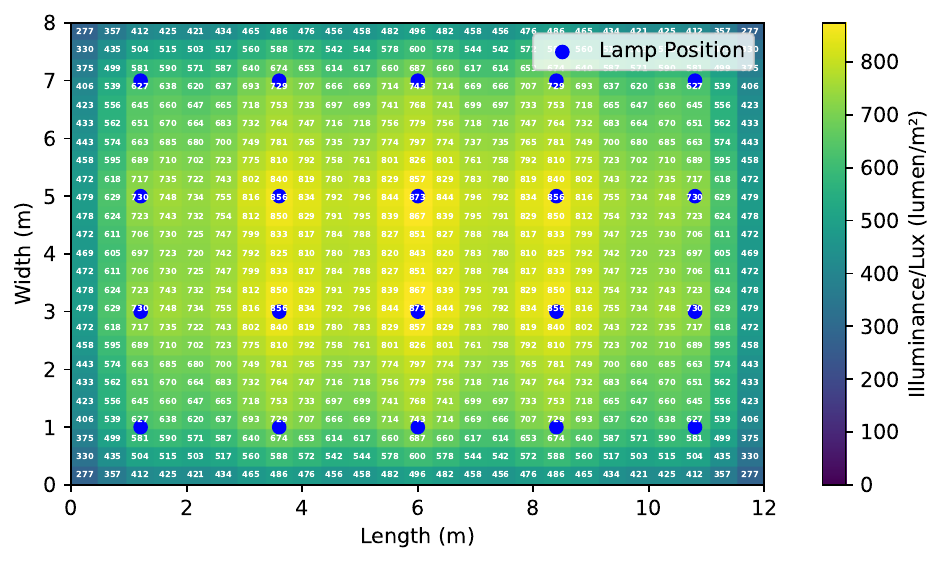} 
	\caption{The illuminance map for a brand-new lighting system}
	\label{fig:IM1} 
    \end{subfigure}
    \hfill
    \begin{subfigure}[t]{0.3\textwidth}
        \centering
        \includegraphics[width=\textwidth]{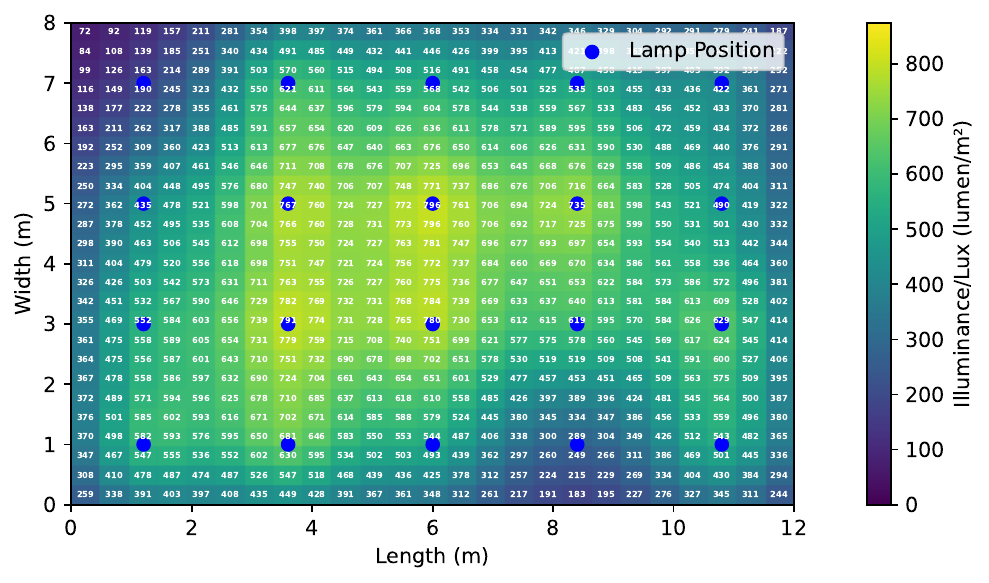}
        \caption{The illuminance map at day 1259}
        \label{fig:IM2}
    \end{subfigure}
    \hfill
    \begin{subfigure}[t]{0.3\textwidth}
        \centering
        \includegraphics[width=\textwidth]{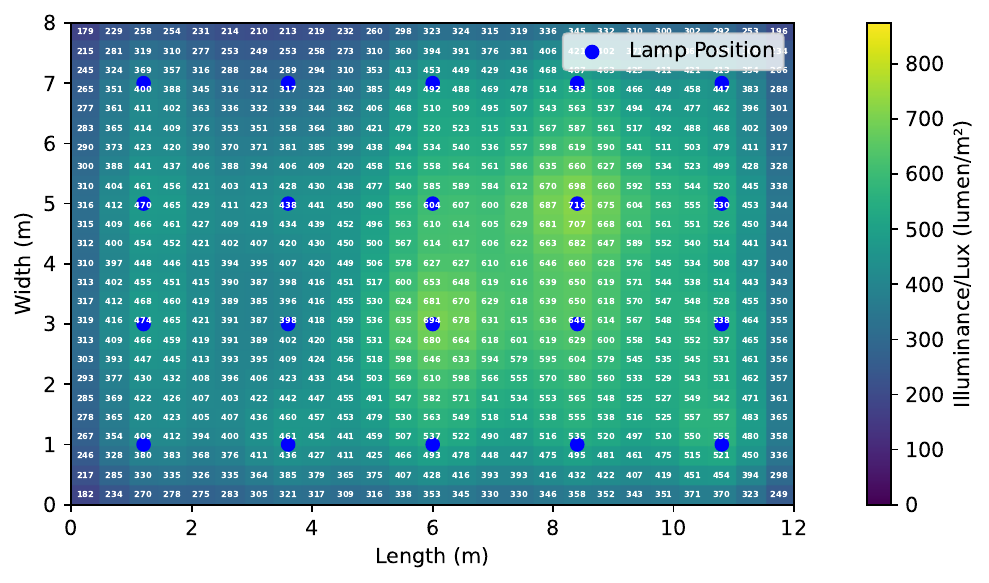}
        \caption{The illuminance map at day 1882}
        \label{fig:IM3}
    \end{subfigure}
    \caption{Evolution of the working plane illuminance distribution under luminaire degradation and failures.}
    \label{fig:IWIF}
\end{figure}

Illuminance maps provide a visual representation of the spatio-temporal performance of the lighting system. In \autoref{fig:IWIF}, warmer colors indicate higher illuminance levels, and the progressive color and spatial-pattern changes over time reflect degradation- and failure-induced performance drift. This motivates the need for a metric for evaluating long-term dynamic performance beyond commissioning-time indices.

\subsection{Long-term system performance evaluation and deficiency ratio metric}\label{sec:deficiency_ratio}

To evaluate the long-term performance of the system, the temporal evolution of luminaire states over an operational horizon is represented by the luminaire state trajectory matrix
\begin{equation}
    \mathbf{L}_{s} = \left[ \mathbf{L}(t_{1}), \mathbf{L}(t_{2}), \dots, \mathbf{L}(t_{K}) \right] ^T,
    \label{eq:GL}
\end{equation}
where $\{t_k\}_{k=1}^{K}$ is a discrete time grid, and $\mathbf{L}(t_{k})$ is the system state vector at time $t_k$. Based on the performance mapping model, the corresponding illuminance sequence matrix is
\begin{equation}
    \mathbf{E}_{s} = \left[ \mathbf{E}(t_{1}), \mathbf{E}(t_{2}), \dots, \mathbf{E}(t_{K}) \right]^T.
    \label{eq:IM}
\end{equation}

Using \rev{Equations~(\ref{eq:avg})--(\ref{eq:static_req})}, the static indices at each time point are computed as $\{E_{\text{avg}}(t_k), U(t_k)\}_{k=1}^{K}$. For each interval $(t_{k-1}, t_k)$, define $\Delta t_k = t_k - t_{k-1}$ and introduce binary deficiency indicators for average illuminance and uniformity:
\begin{equation}
    I_E(k)=\mathbb{I}\!\left(E_{\text{avg}}(t_k) < S_E\right),\qquad
    I_U(k)=\mathbb{I}\!\left(U(t_k) < S_U\right),
\end{equation}
where $\mathbb{I}(\cdot)$ is the indicator function. The interval-level deficiency durations associated with the two indices over $(t_{k-1}, t_k)$ are defined by
\begin{equation}
    T_{\text{E\_defi}}(k)=I_E(k)\,\Delta t_k,\qquad
    T_{\text{U\_defi}}(k)=I_U(k)\,\Delta t_k.
\end{equation}

The interval-level PDD is then defined as the maximum of the two deficiency durations:
\begin{equation}
    T_{\text{defi}}(k) = \max(T_{\text{E\_defi}}(k), T_{\text{U\_defi}}(k)),
    \label{eq:TDEFI}
\end{equation}

The use of the maximum operator reflects a requirement-based interpretation of system performance consistent with current lighting standards. Under this interpretation, violation of any individual criterion (e.g., average illuminance or uniformity) within an interval is sufficient to render the system non-compliant over that interval. This formulation provides a simplified and interpretable duration-based extension of standard commissioning indices, although it does not differentiate the magnitude of deviation from the required thresholds or the relative severity across different criteria. More refined formulations, such as deviation-weighted or severity-aware penalties, would require additional modeling assumptions and are left to future work.

\begin{figure}
	\centering 
	\includegraphics[width=0.5\textwidth]{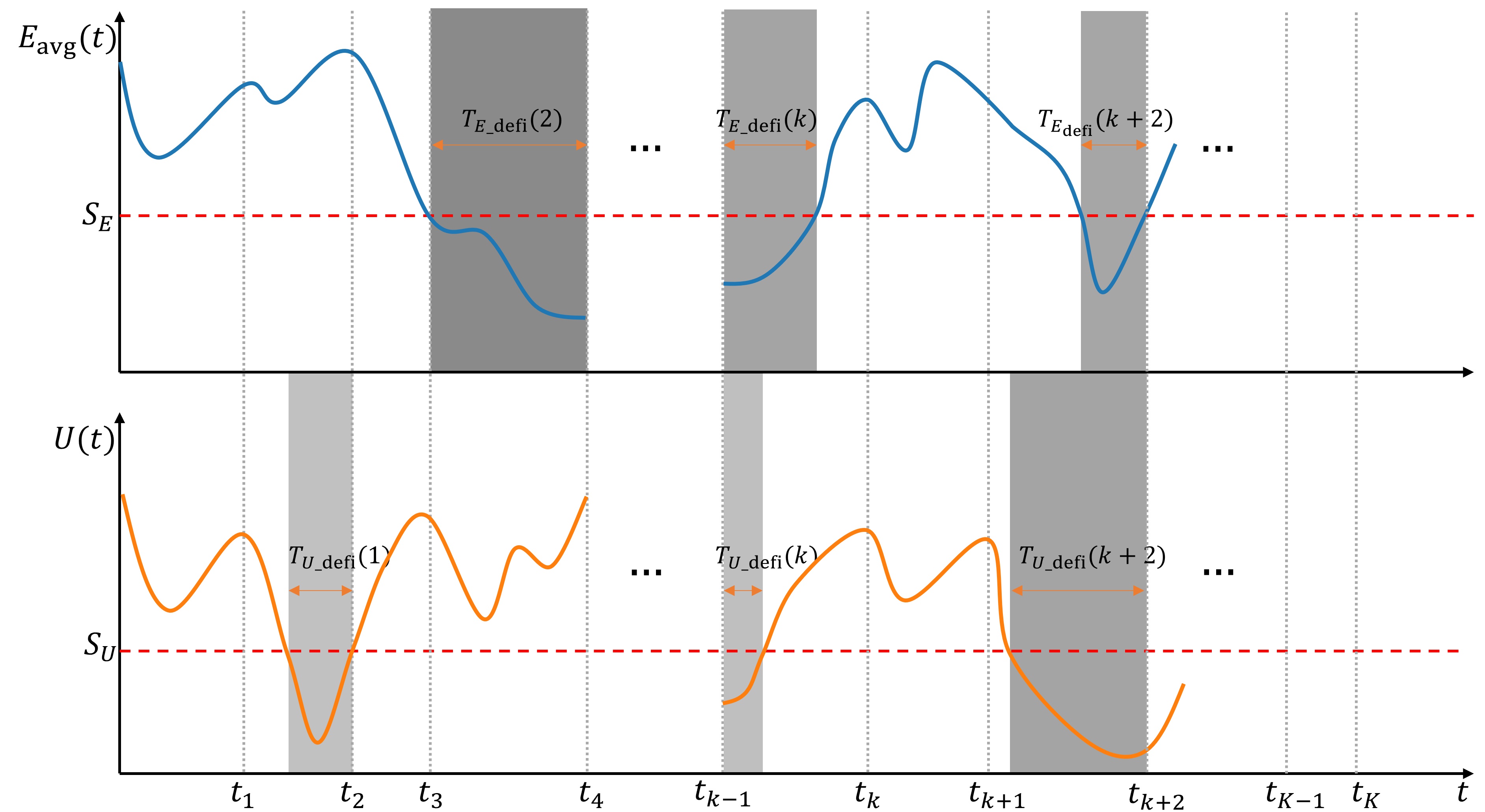} 
	\caption{Static performance indices, standard requirements, and deficiency periods over time}
	\label{fig:LQI} 
\end{figure}

Under this discrete-time approximation, the maximum operator treats the interval as performance-deficient if at least one of the two requirements is violated. An illustration of index trajectories, requirement thresholds, and resulting deficiency periods is shown in \autoref{fig:LQI}.

Finally, the deficiency ratio (DR) is defined as the fraction of operational time during which the system violates at least one performance requirement:
\begin{equation}
    R_{\text{DR}} = \frac{\sum_{k=1}^{K} T_{\text{defi}}(k)}{T_{\text{over}}}.
    \label{eq:DR}
\end{equation}
where $T_{\text{over}}$ is the total operating horizon.

The DR recasts static illuminance-based indices into a duration-based metric for evaluating long-term system performance, which is subsequently used as a system-level objective for maintenance optimization.

Large-scale Monte Carlo evaluation under maintenance policies requires repeated mapping from $\mathbf{L}(t_k)$ to $\mathbf{E}(t_k)$. To enable efficient optimization, the Radiance-based mapping is accelerated using a surrogate model (see Section~\ref{sec:SAIE}).

\section{Performance-driven maintenance optimization for LED lighting systems}

This section develops a simulation-in-the-loop framework to optimize opportunistic maintenance policies for LED lighting systems subject to stochastic luminaire degradation and failures. A unit-based time-window policy is formulated to exploit maintenance dependency across luminaires. Maintenance strategies are evaluated through the discrete-event maintenance simulation module (\autoref{alg:Discrete_event} \rev{in Appendix~\ref{app:alg_details}}) and the performance evaluation module (\autoref{alg:PEM} \rev{in Appendix~\ref{app:alg_details}}). Monte Carlo simulation is then used to estimate the trade-offs among the DR, the total number of site visits, and the total number of replacements. To make large-scale evaluation computationally feasible, illuminance evaluation within the loop is accelerated using a surrogate model that replaces repeated Radiance calls.

\subsection{Opportunistic maintenance policy formulation}\label{sec:OMP}
In an LED lighting system, performance degradation and failures are primarily driven by two mechanisms: gradual LED package degradation (light-output loss) and abrupt LED driver failures (complete outage). Because packages and drivers are typically treated as non-repairable field units, maintenance is modeled as luminaire replacement.

Corrective maintenance (CM) is initiated when a luminaire is reported as failed or unacceptable in service, commonly due to a complete outage caused by driver failure or noticeable dimming caused by package degradation (with the degradation failure definition following the $L_{70}$ criterion in Section~\ref{sec:Component_level}). Such reactive interventions can be costly because they require unscheduled site visits and access coordination, introduce safety and logistical constraints, and may allow degraded lighting performance to persist until intervention.

PM is therefore scheduled at fixed intervals to reduce disruptive CM events. Nevertheless, fixed-interval PM and purely reactive CM can be inefficient: PM may replace luminaires prematurely, whereas CM may occur only after unacceptable performance has already persisted, increasing the frequency of unscheduled visits and cumulative performance shortfalls. To exploit dependence among maintenance activities, an opportunistic maintenance (OM) strategy is adopted: when a maintenance visit occurs (triggered by either CM or PM), additional luminaires that are close to their next scheduled PM are replaced concurrently. By consolidating actions into fewer visits, OM reduces visit frequency and improves maintenance resource utilization.

Accordingly, CM is applied for driver failures (complete loss of function) and package degradation failures, while PM is performed periodically every $T_{\text{PM}}$. Both CM and PM events create an opportunity to execute OM. Following the condition-index concept in \citet{shi_opportunistic_2023}, a maintenance opportunity window (MOW) is defined for luminaire $j$ as the ratio of the remaining time until its next scheduled PM to the PM interval:
\begin{equation}
    H_{\text{mow},j} = \frac{T_{\text{re},j}} {T_{\text{PM}} },
    \label{eq:hmov}
\end{equation}
where $T_{\text{re},j}$ is the remaining time until luminaire $j$ reaches its next scheduled PM time. The OM decision is determined by comparing $H_{\text{mow},j}$ with a predefined threshold $H_{\text{OM}}$:
\begin{equation}
    H_{j}(t) =
    \begin{cases}
        1,  & \text{if } H_{\text{mow},j} \leq H_{\text{OM}}, \\[1mm]
        0,  & \text{if } H_{\text{mow},j} > H_{\text{OM}}.
    \end{cases}
    \label{eq:OMS}
\end{equation}

\begin{figure} 
	\centering 
	\includegraphics[width=0.6\textwidth]{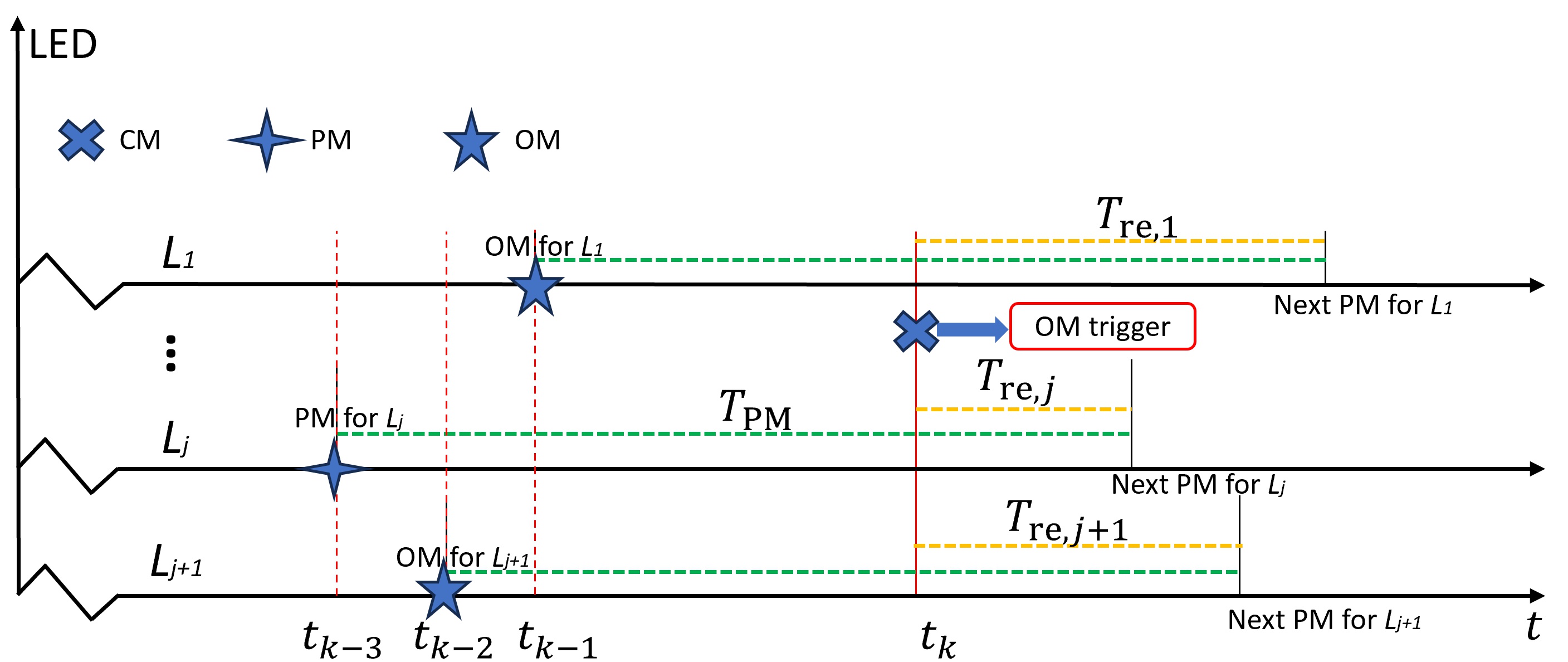} 
	\caption{Schematic illustration of the opportunistic maintenance policy.}
	\label{fig:SDOM} 
\end{figure}

As illustrated in \autoref{fig:SDOM}, a CM event occurring between $t_{k-1}$ and the next scheduled PM triggers the OM policy. For any luminaire $j$ satisfying $H_{\text{mow},j} \leq H_{\text{OM}}$, replacement is performed opportunistically during the same visit.

The total number of site visits over the operational horizon is defined as
\begin{equation}
    N_{\text{tv}} = N_{\text{cm,vis}} +  N_{\text{pm,vis}},
    \label{eq:TV}
\end{equation}
and the total number of replacements is
\begin{equation}
    N_{\text{tr}} = N_{\text{cm}} + N_{\text{pm}} + N_{\text{om}},
    \label{eq:TR}
\end{equation}
where $N_{\text{pm,vis}}$ and $N_{\text{cm,vis}}$ denote the numbers of PM and CM visits, and $N_{\text{cm}}$, $N_{\text{pm}}$, and $N_{\text{om}}$ denote the numbers of luminaires replaced via CM, PM, and OM, respectively.

\subsection{Multi-objective optimization formulation}\label{sec:MOOM}
The OM threshold and the PM interval jointly determine the aggressiveness of opportunistic replacement. A multi-objective optimization problem is therefore formulated over these two decision variables to balance long-term lighting performance against maintenance effort.

The maintenance policy aims to sustain high system performance while limiting maintenance workload. Three objectives are considered: the long-term DR $R_{\text{DR}}$ (Section~\ref{sec:deficiency_ratio}), the total number of site visits $N_{\text{tv}}$ (\autoref{eq:TV}), and the total number of luminaire replacements $N_{\text{tr}}$ (\autoref{eq:TR}). Two decision variables govern these objectives:
\begin{enumerate}
    \item OM threshold $H_{\text{OM}} \in [0,1]$, which determines whether additional luminaires are replaced opportunistically during a CM or PM visit. A larger $H_{\text{OM}}$ expands the set of luminaires eligible for opportunistic replacement, which can reduce future visit frequency and mitigate performance shortfalls, but may increase the number of replacements.
    \item PM interval $T_{\text{PM}} \in (0, T_{\text{over}}]$, which specifies the periodic PM schedule. Shorter intervals generally improve performance by limiting degradation accumulation, but increase visit frequency and replacements; longer intervals reduce planned visits but may increase corrective interventions and performance deficiency.
\end{enumerate}

Accordingly, the performance-driven maintenance optimization problem is formulated as
\begin{equation}
    \begin{aligned}
        & \min_{H_{\text{OM}},\, T_{\text{PM}}} \left( R_{\text{DR}},\, N_{\text{tv}},\, N_{\text{tr}} \right), \\
        & \text{subject to} \quad 0 \leq H_{\text{OM}} \leq 1, \quad 0 < T_{\text{PM}} \leq T_{\text{over}},
    \end{aligned}
    \label{eq:MOOM}
\end{equation}
where $T_{\text{over}}$ is the operational horizon and $R_{\text{DR}}$ quantifies the proportion of time during which system performance falls below standard requirements.

To make the maintenance rule interpretable, the two decision variables can be expressed as an equivalent age-based trigger. From \rev{Equations~(\ref{eq:hmov})--(\ref{eq:OMS})}, OM is executed when the remaining time to the next PM is within a fraction $H_{\text{OM}}$ of the PM interval, which is equivalent to an OM age threshold
\begin{equation}
    T_{\text{OM}} = T_{\text{PM}}\left(1-H_{\text{OM}}\right).
    \label{eq:TOM}
\end{equation}
Here, $T_{\text{PM}}$ sets the maximum allowable age before the next scheduled PM action, and $T_{\text{OM}}$ specifies how early a luminaire becomes eligible for opportunistic replacement within each PM cycle. A larger $H_{\text{OM}}$ yields a smaller $T_{\text{OM}}$, corresponding to a more aggressive opportunistic strategy.

In principle, $R_{\text{DR}}$, $N_{\text{tv}}$, and $N_{\text{tr}}$ could be mapped to monetary costs and aggregated into a single objective. However, the required cost coefficients (e.g., labor logistics, access constraints, contractor pricing, and service-level penalties) are highly site- and operator-specific and are unavailable in this study. Instead, a multi-objective formulation is adopted: $N_{\text{tv}}$ serves as a proxy for fixed/setup effort associated with site access and visit organization, while $N_{\text{tr}}$ reflects variable workload related to replacement actions. The set of Pareto-optimal (non-dominated) solutions \citep{keller_multi-objective_2019} is then identified to characterize the trade-off among performance deficiency, the total number of site visits, and the total number of replacements.

\subsection{Simulation-based policy evaluation via discrete-event Monte Carlo}

Closed-form evaluation of long-term lighting performance risk under competing failure modes and discrete-event OM rules is generally intractable, particularly when system acceptability depends on a spatial illuminance field obtained from ray tracing. Monte Carlo simulation therefore becomes the practical engine to propagate both parameter uncertainty (from Bayesian calibration) and process uncertainty (stochastic degradation/failures) to system-level performance metrics. This enables statistically robust comparison of maintenance policies and quantification of trade-offs beyond mean behavior, including variability and tail risk relevant to asset management.

A Monte Carlo simulation framework is developed to evaluate candidate maintenance policies under stochastic luminaire degradation and failures. For each policy setting (a $(T_{\text{PM}}, H_{\text{OM}})$ pair), $S$ independent simulation runs are performed. Each run consists of two coupled modules: (i) a discrete-event maintenance simulator that generates the event timeline and the resulting luminaire state trajectory, and (ii) a performance evaluation module that maps the trajectory to the long-term deficiency metric defined in Section~\ref{sec:deficiency_ratio}.

\paragraph{Discrete-event maintenance simulation.}
For a given simulation run $s$, the system evolution is represented by a sequence of events. Each luminaire $j$ is associated with an event vector that stores candidate times of maintenance-related and performance-recording events. At event index $k$, the event vector of luminaire $j$ is defined as
\begin{equation}
    \mathbf{t}_{s,k,j}  = \left[ t_{s,k,j}^{\text{pm}},\; t_{s,k,j}^{\text{cm,df}},\; t_{s,k,j}^{\text{cm,pf}},\; t_{s,k,j}^{\text{event-end}},\; t_{s,k,j}^{\text{record}} \right]^T,
    \label{eq:EV}
\end{equation}
where
\begin{itemize}
    \item $t_{s,k,j}^{\text{pm}}$ is the next scheduled PM time,
    \item $t_{s,k,j}^{\text{cm,df}}$ is the predicted CM time due to driver failure,
    \item $t_{s,k,j}^{\text{cm,pf}}$ is the predicted CM time due to package degradation failure,
    \item $t_{s,k,j}^{\text{event-end}}$ is the completion time of an ongoing maintenance action on luminaire $j$ (if any); otherwise it is set to $T_{\text{over}}$,
    \item $t_{s,k,j}^{\text{record}}$ is the next time at which system performance will be recorded for evaluating deficiency duration.
\end{itemize}

Stacking all luminaire event vectors yields the system event matrix at event index $k$:
\begin{equation}
    \mathbf{T}_{s,k}^{\text{evt}} =\bigl[ \mathbf{t}_{s,k,1},\; \mathbf{t}_{s,k,2},\; \dots,\; \mathbf{t}_{s,k,J} \bigr]^T.
    \label{eq:EMM}
\end{equation}

The next system event time is defined as the minimum over all entries of $\mathbf{T}_{s,k}^{\text{evt}}$:
\begin{equation}
    t_{s,k}^{\text{ts-evt}} = \min \mathbf{T}_{s,k}^{\text{evt}} 
    = \min_{\substack{j=1,2,\ldots,J }} \; \mathbf{t}_{s,k,j},
    \label{eq:tnsme}
\end{equation}
where the minimization is taken over the five candidate event times for every luminaire.

\begin{figure} 
	\centering 
	\includegraphics[width=0.7\textwidth]{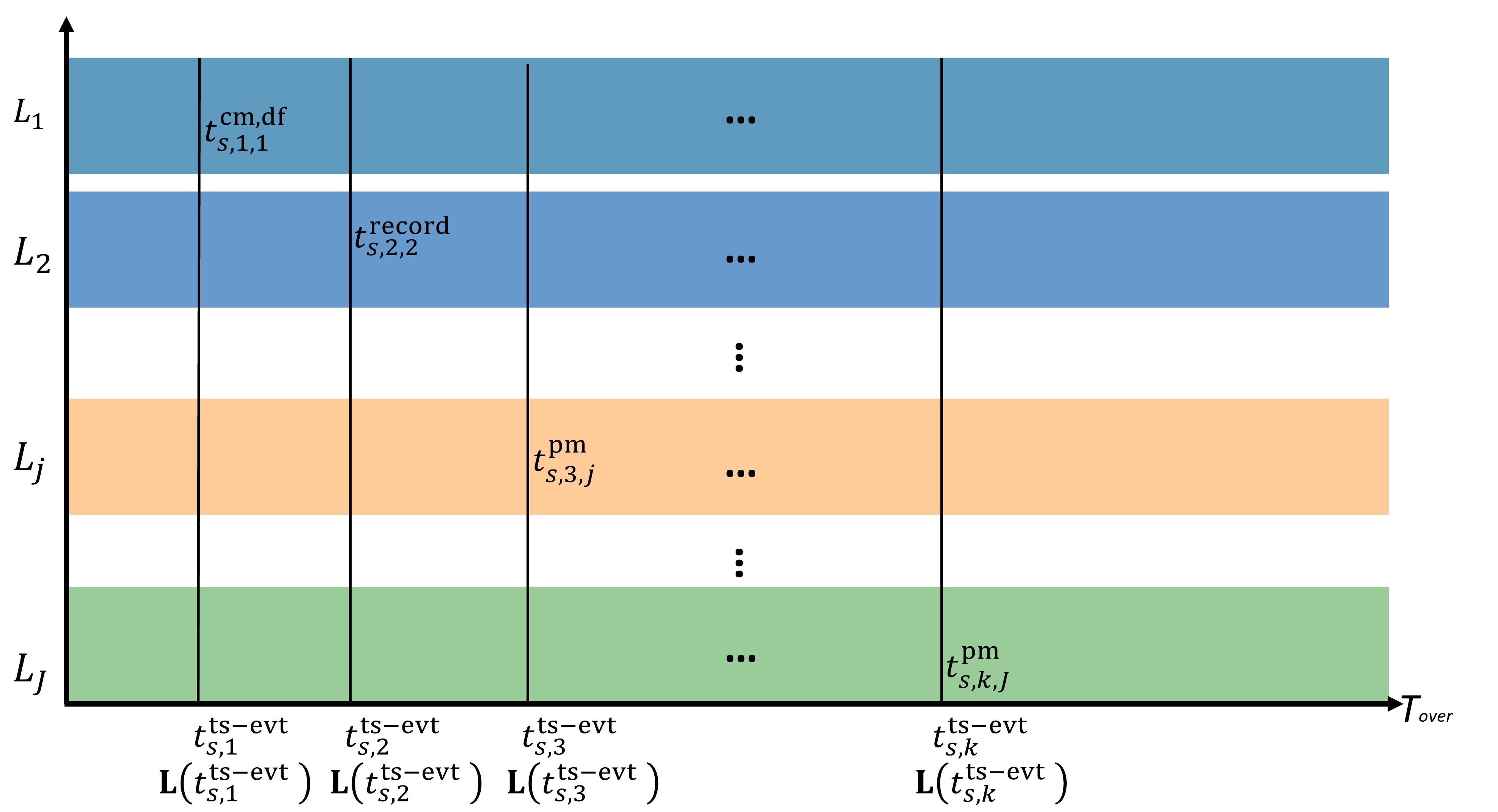} 
	\caption{Schematic diagram of the discrete-event simulation model.}
	\label{fig:SDEDM} 
\end{figure}

Given $t_{s,k}^{\text{ts-evt}}$, the corresponding event type (PM, CM due to driver failure, CM due to package degradation failure, maintenance completion, or performance recording) is identified, and the relevant state variables and event times are updated. For example, in \autoref{fig:SDEDM}, the first event is a CM triggered by driver failure on luminaire $L_1$; after replacement, the maintenance completion time is updated as
$t_{s,k,1}^{\text{event-end}} = t_{s,k}^{\text{ts-evt}} + \text{service time}$.
Once maintenance is completed, the event vector of the renewed luminaire is reinitialized and the system event matrix \autoref{eq:EMM} is updated for the next event index.

In addition to maintenance-triggered events, performance-recording events $t_{s,k,j}^{\text{record}}$ are inserted to ensure sufficient temporal resolution for computing interval-level PDD values, especially when maintenance events are sparse. The full discrete-event workflow is summarized in \autoref{alg:Discrete_event} \rev{in Appendix~\ref{app:alg_details}}.

The resulting strictly increasing event-time sequence for simulation $s$ is
\begin{equation}
    \mathbf{T}^{\text{ts-evt}}_{s} = \bigl[  t_{s,1}^{\text{ts-evt}},\; t_{s,2}^{\text{ts-evt}},\; \dots,\; t_{s,K}^{\text{ts-evt}} \bigr].
    \label{eq:GEVT}
\end{equation}

At each event time $t_{s,k}^{\text{ts-evt}}$, the system luminaire state vector $\mathbf{L}(t_{s,k}^{\text{ts-evt}})$ is logged. Stacking these vectors yields the state trajectory
\begin{equation}
    \mathbf{L}_{s} = \bigl[ \mathbf{L}(t_{s,1}^{\text{ts-evt}}),\; \mathbf{L}(t_{s,2}^{\text{ts-evt}}),\; \dots,\; \mathbf{L}(t_{s,K}^{\text{ts-evt}}) \bigr]^{T}.
    \label{eq:GLS}
\end{equation}

\paragraph{Performance evaluation and objective estimation.}
Given $\mathbf{L}_{s}$, the performance-mapping model produces the corresponding illuminance sequence matrix $\mathbf{E}_{s}$ (see \autoref{eq:IM}), from which the deficiency duration $T_{\text{defi},s}(k)$ is computed as described in Section~\ref{sec:deficiency_ratio}. The resulting deficiency ratio for run $s$ is
\begin{equation}
    R_{\text{DR},s} = \frac{\sum_{k=1}^{K} T_{\text{defi}, s}(k)}{T_{\text{over}}}.
    \label{eq:NODR}
\end{equation}
The pseudocode of the performance evaluation module is provided in~\autoref{alg:PEM} \rev{in Appendix~\ref{app:alg_details}}.

For each run $s$, the numbers of CM and PM visits ($N_{\text{cm,vis},s}$ and $N_{\text{pm,vis},s}$) and the counts of luminaire replacements via CM, PM, and OM ($N_{\text{cm},s}$, $N_{\text{pm},s}$, and $N_{\text{om},s}$) are recorded. The total numbers of site visits and replacements are computed as
\begin{align}
    N_{\text{tv},s} &= N_{\text{cm,vis},s} + N_{\text{pm,vis},s},
    \label{eq:TV_sim} \\
    N_{\text{tr},s} &= N_{\text{cm},s} + N_{\text{pm},s} + N_{\text{om},s}.
    \label{eq:TR_sim}
\end{align}

Averaging over $S$ independent runs yields the estimated objectives for the policy:
\begin{align}
    N_{\text{STV}} &= \frac{1}{S} \sum_{s = 1}^{S} N_{\text{tv}, s},
    \label{eq:TTV}\\
    N_{\text{STR}} &= \frac{1}{S} \sum_{s = 1}^{S} N_{\text{tr}, s},
    \label{eq:TTR}\\
    R_{\text{SDR}} &= \frac{1}{S} \sum_{s=1}^{S} R_{\text{DR},s}.
    \label{eq:MODR}
\end{align}

For each policy setting, the objective triplet $(R_{\text{SDR}}, N_{\text{STV}}, N_{\text{STR}})$ is obtained. To reduce the influence of Monte Carlo variability, a sequential statistical screening step based on one-sided Welch's $t$-tests is then applied before reporting the retained candidate policies.

Specifically, for each candidate policy $i$ and each compared policy $j$, one-sided Welch's $t$-tests are performed for each objective with hypotheses
\begin{equation}
   H_{0}:\ \mu_{j} \leq \mu_{i}
   \quad\text{versus}\quad
   H_{1}:\ \mu_{j} > \mu_{i},
\end{equation}
where $\mu$ denotes the Monte Carlo estimated mean of the corresponding objective, and smaller values are preferred. Rejecting $H_0$ indicates that policy $i$ performs significantly better than policy $j$ on that objective.

A significance level of $\alpha_{\text{sv}}$ is adopted. For a given policy $i$, if there exists another compared policy $j$ such that $H_0$ cannot be rejected for all three objectives, then policy $i$ is removed by the screening rule. Equivalently, policy $i$ is retained only if, for every compared policy $j$, there exists at least one objective on which policy $i$ is significantly better than policy $j$. The full procedure is summarized in \autoref{alg:stat_screen} \rev{in Appendix~\ref{app:alg_details}}.

\subsection{Surrogate-based acceleration of performance evaluation} \label{sec:SAIE}
Direct illuminance evaluation via Radiance at every simulated event index (i.e., each $t_{s,k}^{\text{ts-evt}}$ in the event-time vector $\mathbf{T}^{\text{ts-evt}}_{s}$ defined in \autoref{eq:GEVT}) is computationally prohibitive; a surrogate-based performance mapping is therefore introduced to enable efficient simulation-in-the-loop optimization.

During system operation, each Monte Carlo replication produces hundreds to thousands of luminaire state vectors $\mathbf{L}(t_k)$ that must be mapped to WP illuminance for evaluating the static performance indices $E_{\text{avg}}(t_k)$ and $U(t_k)$. Repeated Radiance calls at all event times lead to an excessive computational burden. For example, for a working zone with $J=76$ LED luminaires and $N=265$ WP calculation points, a single Radiance evaluation takes on the order of one minute. When many policy settings and thousands of Monte Carlo replications are required, the overall runtime becomes prohibitive.

\paragraph{Physics-informed linear surrogate.}
For a fixed room geometry and material configuration, the illuminance at a given point is approximately an affine function of luminaire photometric output scaling, consistent with the superposition property of light transport. With luminaire degradation state $L_j(t)\in[0,1]$, the corresponding normalized output scaling vector is defined as
\begin{equation}
    \mathbf{Q}(t) = \mathbf{1}_J - \mathbf{L}(t),
\end{equation}
where $\mathbf{1}_J$ is an all-ones vector. The WP illuminance vector $\mathbf{E}(t)\in\mathbb{R}^{N}$ is then modeled as
\begin{equation}
    \mathbf{E}(t) = \mathbf{b}_E + \mathbf{C}_E\,\mathbf{Q}(t),
    \label{eq:surrogate_single}
\end{equation}
where $\mathbf{b}_E\in\mathbb{R}^{N}$ is an intercept term and $\mathbf{C}_E\in\mathbb{R}^{N\times J}$ is a coefficient matrix capturing the contribution of each luminaire to each calculation point under the fixed geometry/material setting.

Stacking the $K$ event times in simulation run $s$ yields the state trajectory matrix $\mathbf{L}_s\in\mathbb{R}^{K\times J}$ and the corresponding output-scaling matrix $\mathbf{Q}_s=\mathbf{1}_{K}\mathbf{1}_{J}^{\top}-\mathbf{L}_s$. The illuminance sequence matrix $\mathbf{E}_s\in\mathbb{R}^{K\times N}$ is
\begin{equation}
    \mathbf{E}_s = \mathbf{1}_{K}\,\mathbf{b}_E^{\top} + \mathbf{Q}_s\,\mathbf{C}_E^{\top},
    \label{eq:surrogate_stack}
\end{equation}
where $\mathbf{1}_{K}$ is an all-ones vector of length $K$.
This surrogate replaces Radiance in the performance evaluation module (\autoref{alg:PEM} \rev{in Appendix~\ref{app:alg_details}}) to compute $\mathbf{E}(t_k)$ efficiently, after which $E_{\text{avg}}(t_k)$ and $U(t_k)$ are obtained via \rev{Equations~(\ref{eq:avg})--(\ref{eq:NUI})}.

\begin{figure}
    \centering
    \begin{subfigure}[b]{0.2\textwidth}
        \centering
        \includegraphics[width=\textwidth]{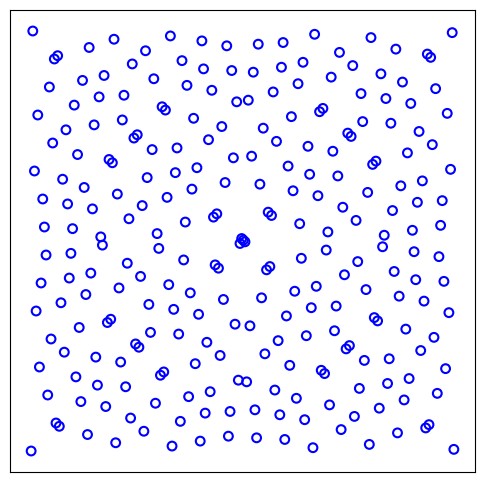}
        \caption{Sobol samples (2D projection)}
        \label{fig:DSD}
    \end{subfigure}
    \begin{subfigure}[b]{0.2\textwidth}
        \centering
        \includegraphics[width=\textwidth]{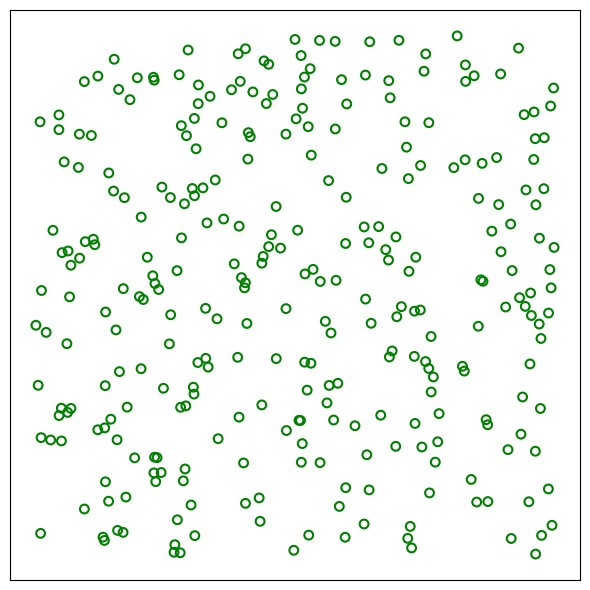}
        \caption{Pseudo-random samples (2D projection)}
        \label{fig:DGRD}
    \end{subfigure}
    \caption{Distribution comparison of Sobol and pseudo-random samples in a 2D projection \citep{sobol_construction_2011}}
    \label{fig:DCSRD}
\end{figure}

\paragraph{Training data generation and model fitting.}
The training dataset consists of paired samples $\{(\mathbf{L}^{(n)},\mathbf{E}^{(n)})\}_{n=1}^{N_{\text{data}}}$, where each illuminance vector $\mathbf{E}^{(n)}$ is produced by a one-time Radiance simulation under the corresponding luminaire state $\mathbf{L}^{(n)}$. Because $\mathbf{L}(t)\in[0,1]^J$ (\autoref{eq:LRB}), a scrambled Sobol sequence is employed to generate quasi-random state samples that cover the high-dimensional state space efficiently. \autoref{fig:DCSRD} compares the distribution of Sobol-generated samples with conventional pseudo-random samples in a 2D projection.

\begin{figure} 
	\centering 
	\includegraphics[width=0.7\textwidth]{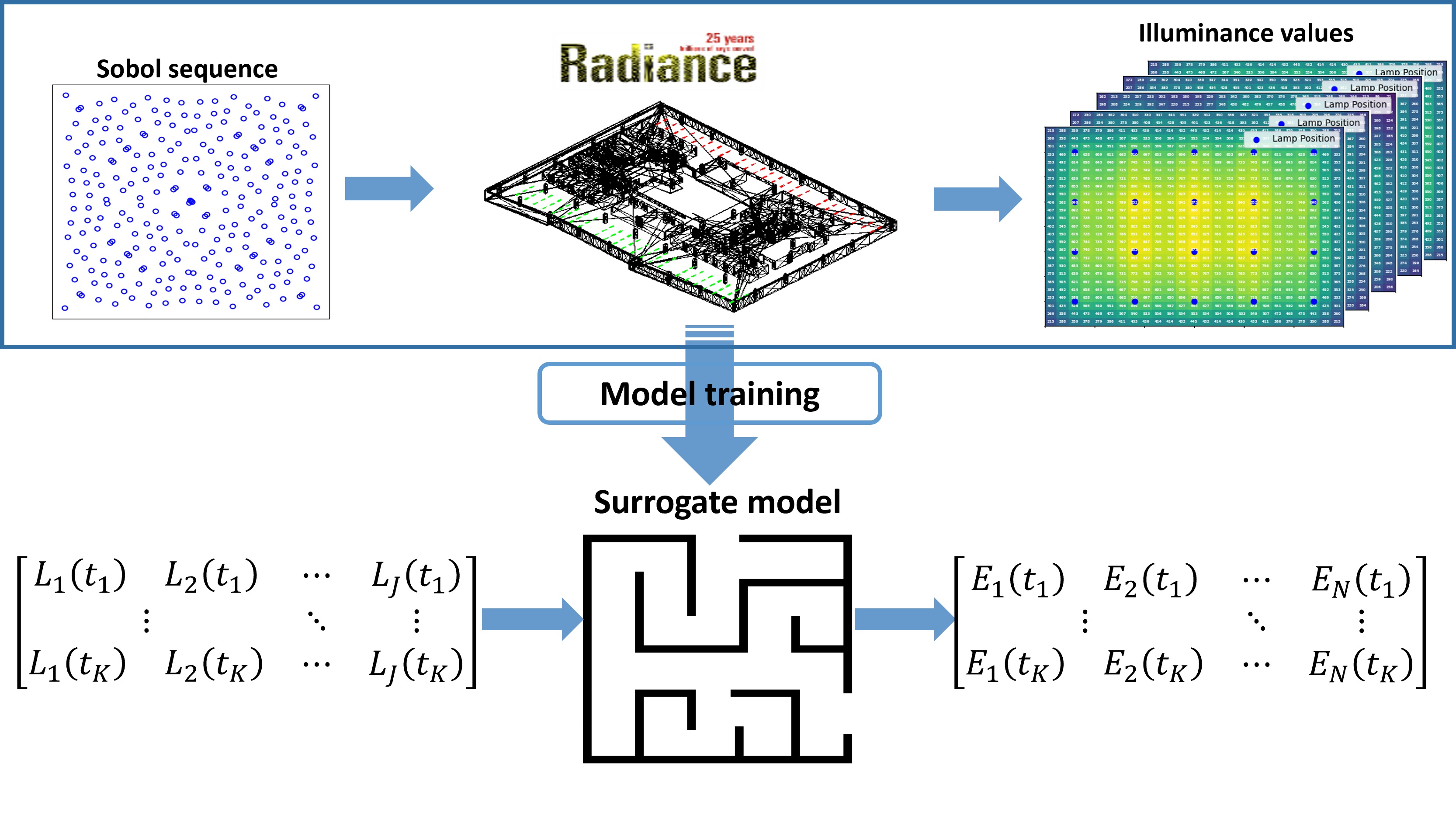} 
	\caption{Surrogate-based acceleration of the Radiance performance evaluation workflow.}
    \label{fig:ILAM} 
\end{figure}

The Sobol-generated luminaire states are input to Radiance to obtain illuminance outputs, and the resulting input--output pairs are used to fit the linear surrogate in \autoref{eq:surrogate_single} (e.g., by least squares). Once trained, the surrogate provides a computationally efficient replacement for Radiance in large-scale simulations, as illustrated in \autoref{fig:ILAM}.

\begin{figure} 
	\centering 
	\includegraphics[width=0.3\textwidth]{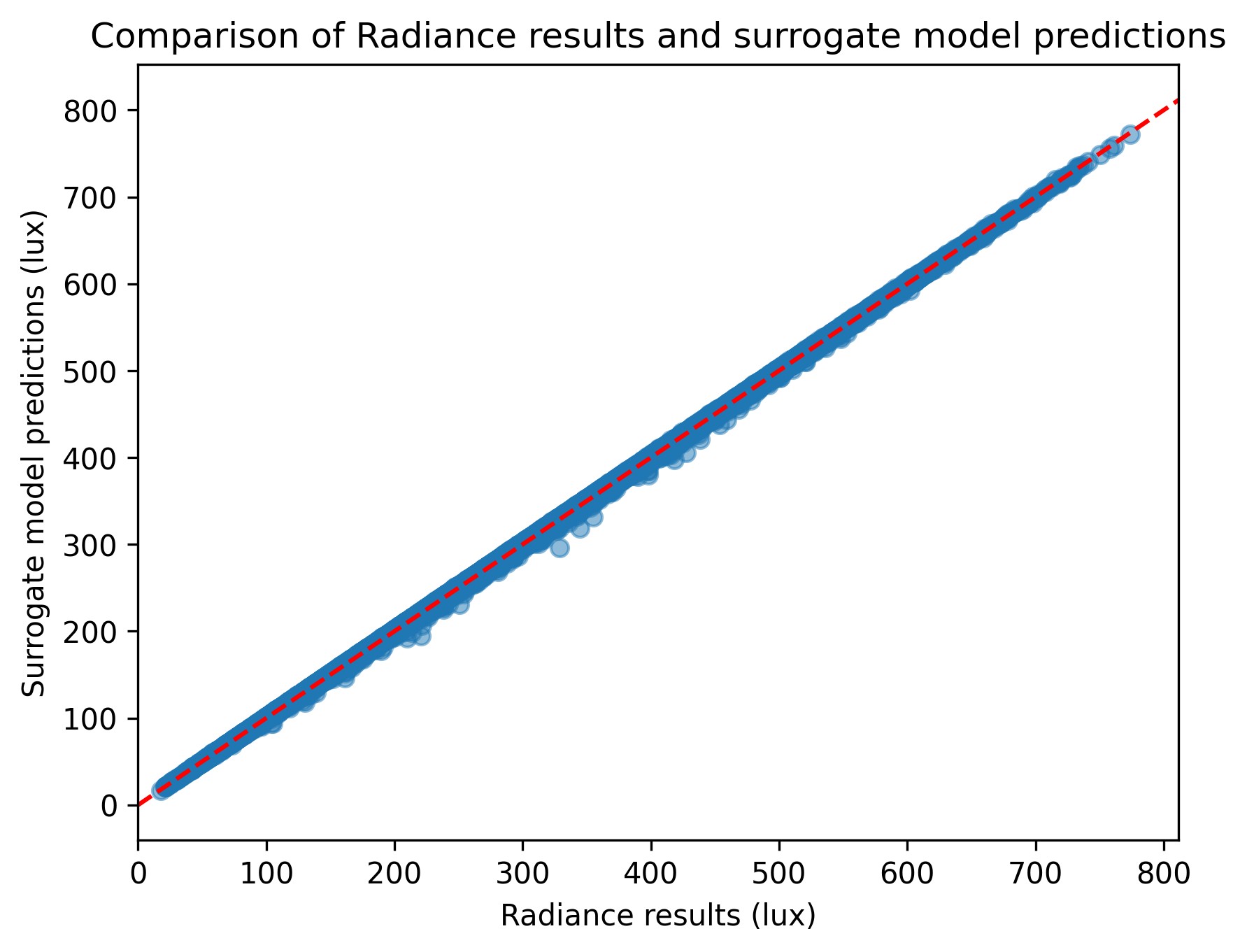} 
    \caption{Validation of the surrogate model against Radiance outputs on the test dataset.}
	\label{fig:CRSSP} 
\end{figure}

In this study, $N_{\text{data}}=2016$ samples are generated for the case with $J=76$ luminaires and $N=265$ calculation points, producing an input matrix of size $[2016,\,76]$ and an output matrix of size $[2016,\,265]$. The dataset is split into 80\% for training and 20\% for validation. As shown in \autoref{fig:CRSSP}, the red dashed line represents the ideal case in which the predicted values match the Radiance outputs. The predicted illuminance values closely match the Radiance outputs, yielding $R^2=0.9999$ (coefficient of determination), with a root mean square error (RMSE) of 0.990 and a mean absolute error (MAE) of 0.702. These results demonstrate the high accuracy of the surrogate model. In the subsequent maintenance simulation and optimization, the surrogate model is used to predict the illuminance over all points on the WP illuminance field for performance evaluation. Given the small prediction error at each point, the resulting surrogate approximation error is expected to have a negligible impact on the performance indices and, consequently, on maintenance optimization; a supplementary error propagation analysis is provided in Appendix~\ref{sec:surrogate_error}.

\paragraph{Computational benefit.}
As summarized in \autoref{tab:CECDPEM}, the surrogate incurs a one-time offline cost for training data generation and model fitting, but reduces the online cost of mapping luminaire states to illuminance from the order of minutes per snapshot (Radiance) to the order of milliseconds. This acceleration is essential for the large-scale Monte Carlo evaluation and multi-policy search required by the simulation-in-the-loop maintenance optimization framework.
\begin{table}[width=.9\linewidth,cols=3,pos=h]
    \caption{Order-of-magnitude wall-clock estimates for performance evaluation (representative setting)}
    \label{tab:CECDPEM}
    \begin{tabular*}{\tblwidth}{@{} LLL@{} }
        \toprule
        Task & Radiance (direct) & Surrogate-based \\ 
        \midrule
        Offline training data generation & -- & 32 hours \\
        Offline model fitting            & -- & 1 minute \\
        Online evaluation: 10{,}000 simulations$^\dagger$ & $\sim$700 days & $\sim$20 seconds \\ 
        Online policy search: 100 policy settings$^\ddagger$ & $\sim$ decades & $\sim$33 minutes \\
        \bottomrule
    \end{tabular*}

\noindent\footnotesize{
$^\dagger$ Estimates are based on a representative case with $J=76$ luminaires and $N=265$ calculation points, where one Radiance evaluation for one luminaire state snapshot takes $\mathcal{O}(1)$ minute. Each simulation run evaluates performance at $K$ event times, and the reported time corresponds to evaluating $10{,}000$ simulations for one policy setting. 
$^\ddagger$ Policy-search time multiplies the per-policy evaluation by 100 policy settings; values are order-of-magnitude estimates intended to highlight scalability rather than exact hardware-dependent runtimes.}
\normalsize
\end{table}

The surrogate model is constructed for a fixed room and lighting configuration. The mapping from luminaire states to working plane illuminance depends on room geometry, surface reflectance, luminaire photometry, luminaire layout, and the working plane calculation grid. When these configuration-related inputs change, the corresponding mapping should be re-established and the surrogate model should be re-trained accordingly. Developing a more generalizable surrogate would require explicitly parameterizing these inputs, which would substantially increase the input dimensionality and require much larger training datasets to ensure sufficient coverage of the expanded configuration space. A generalized surrogate that explicitly parameterizes these configuration-related inputs is left for future work.

\section{Case study: model calibration and maintenance optimization results}
The proposed performance-driven maintenance optimization framework is instantiated on a representative office lighting system. A real building zone is modeled in Radiance, degradation and failure models are calibrated using LM-80 accelerated tests, and large-scale Monte Carlo evaluation is conducted to identify Pareto-optimal opportunistic maintenance policies.

\subsection{Case description and simulation setup}
\begin{figure}
    \centering
    \begin{subfigure}[b]{0.3\textwidth}
	\centering
	\includegraphics[width=\textwidth]{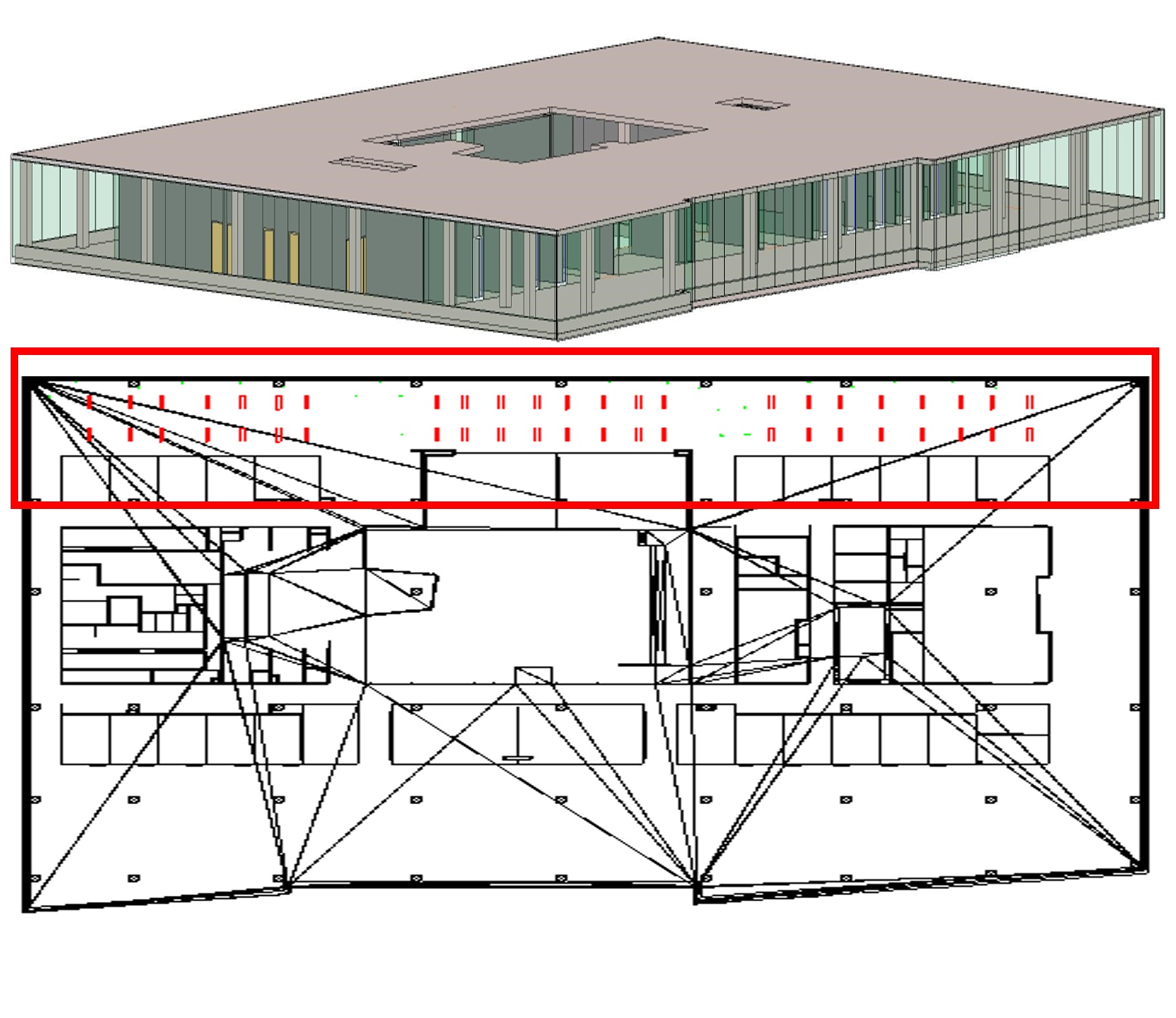}
    \caption{Application office level, with Zone 1 highlighted}
        \textit{Note:} The area in the red box is Zone 1
	\label{fig:RLDFCS}
    \end{subfigure}
    \begin{subfigure}[b]{0.3\textwidth}
        \centering
        \includegraphics[width=\textwidth]{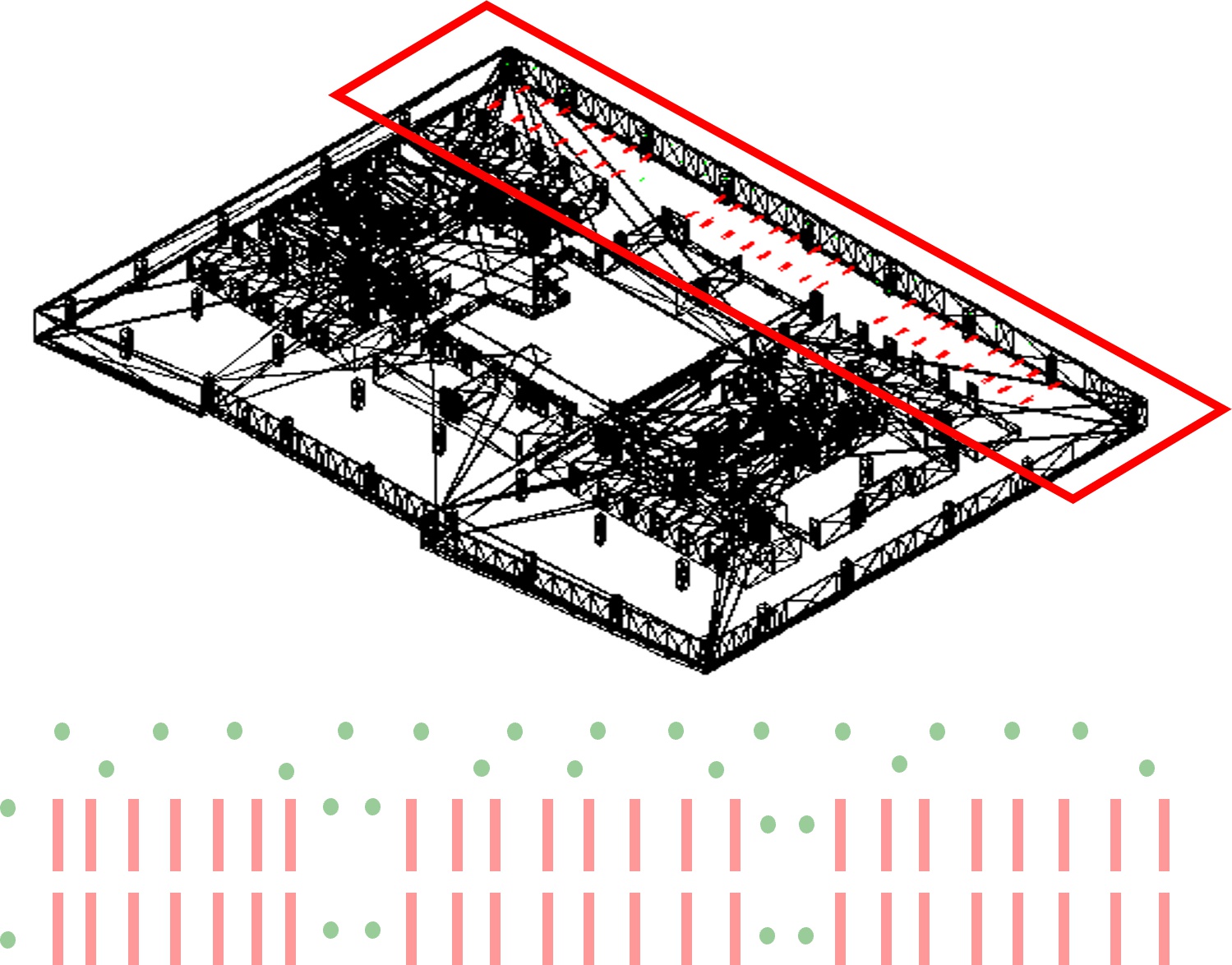}
        \caption{Luminaire layout of Zone 1}
            \textit{Note:} Red rectangles denote B7 and green circles denote D13 luminaires.
        \label{fig:DLLS}
    \end{subfigure}
    \begin{subfigure}[b]{0.3\textwidth}
        \centering
        \includegraphics[width=\textwidth]{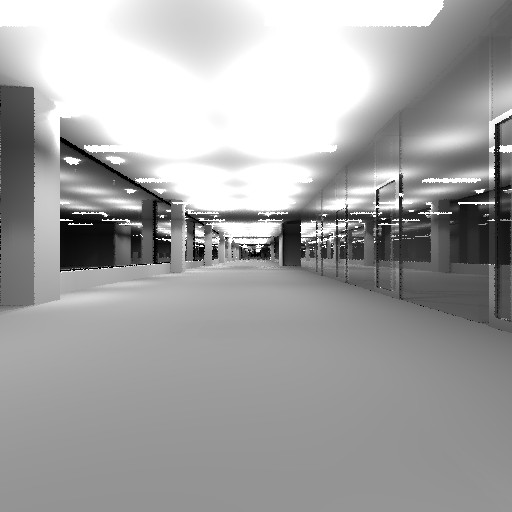}
        \caption{Radiance-rendered view of Zone 1}
        \label{fig:VZ}
    \end{subfigure}
    \begin{subfigure}[b]{0.8\textwidth}
        \centering
        \includegraphics[width=\textwidth]{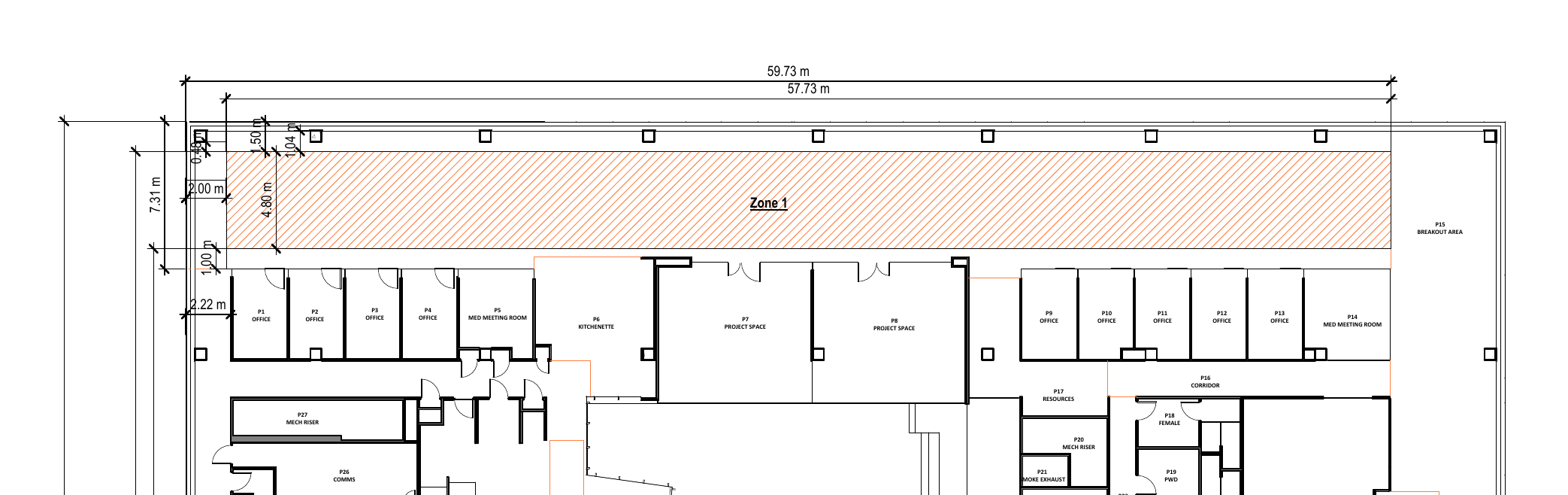}
        \caption{Geometric dimensions of Zone 1}
        \label{fig:DZ}
    \end{subfigure}
    \caption{Application case for Zone 1, including the BIM model, luminaire layout, Radiance rendering, and geometric dimensions.}
    \label{fig:DACL}
\end{figure}

To demonstrate the proposed framework, a working zone (Zone~1) in an office building at Queensland University of Technology (QUT) is considered. Zone~1 is designated for writing, reading, and data processing tasks, as shown in \autoref{fig:DACL}. The zone measures 65.38~m in length and 6.80~m in width. The WP is defined at a height of 0.80~m above the finished floor and spans 57.73~m $\times$ 4.80~m, consistent with AS/NZS~1680.1:2006 requirements \citep{joint_technical_committee_lg-001_as_2006}. The lighting system comprises $J=76$ LED luminaires of two types: 46 B7 and 30 D13 luminaires (\autoref{fig:DLLS}). Calculation points on the WP are placed on a 1~m grid, satisfying the maximum spacing requirement in \citep{the_british_standards_institution_light_2021}. Surface reflectances for the ceiling, walls, and floor are set to 0.7, 0.5, and 0.2, respectively, following AS/NZS~1680.1:2006 \citep{joint_technical_committee_lg-001_as_2006}. The maintained illuminance requirement is set to $S_{E}=500$~lux and the uniformity requirement to $S_{U}=0.6$, also consistent with AS/NZS~1680.1:2006 \citep{joint_technical_committee_lg-001_as_2006}.

Package degradation parameters are estimated using the Bayesian procedure in Section~\ref{sec:BPI} and LM-80 test data. Since only B7 package test data are available, the resulting posterior is applied to both B7 and D13 packages to demonstrate the end-to-end framework; this simplifying assumption will be relaxed in future work when D13-specific test data become available. In the absence of standardized lifetime test data and maintenance records for LED drivers, baseline driver lifetimes are specified using Weibull distributions calibrated to match the mean time to failure (MTTF) of the package degradation model (Setting~S1). Sensitivity analyses are then performed by perturbing the Weibull parameters to represent earlier or later driver failure regimes (Settings~S2 and S3).

On-site CM replacement is assumed to take 3 days for package degradation failures and 2 days for driver failures. Each policy setting is evaluated using $S=10{,}000$ Monte Carlo replications. The lighting system is assumed to operate 12~hours per day over an operational horizon of $T_{\text{over}}=50$~years, corresponding to a typical building lifetime. PM intervals are varied from 365~days to 18{,}250~days (50~years) in steps of 365~days, and the OM thresholds are swept from 0.05 to 1.00 in steps of 0.05. Statistical filtering of candidate policies is performed at the significance level $\alpha_{\text{sv}}=0.05$. Key simulation settings are summarized in \autoref{tab:PFSM}.
\begin{table}[width=1\linewidth,cols=2,pos=h]
    \caption{Parameters for the maintenance optimization simulation study.}
    \label{tab:PFSM}
    \resizebox{\columnwidth}{!}{%
        \begin{tabular*}{\tblwidth}{@{} LL @{} }
            \toprule
            \textbf{Parameter} & \textbf{Value} \\
            \midrule
            Zone~1 length, width, height (m) & 65.38, 6.80, 3.55 \\
            Ceiling, wall, floor reflectance & 0.7, 0.5, 0.2 \\
            WP length, width, height (m) & 57.73, 4.80, 0.80 \\
            WP grid spacing (m) & 1 \\
            Number of B7, D13 luminaires & 46, 30 \\
            Maintained illuminance requirement $S_{E}$ (lux) & 500 \\
            Uniformity requirement $S_{U}$ & 0.6 \\
            CM service time (days): Package degradation ($d_{\text{CM-pf}}$), driver failure ($d_{\text{CM-df}}$) & 3, 2 \\
            Monte Carlo replications per policy ($S$) & 10{,}000 \\
            Statistical filtering level $\alpha_{\text{sv}}$ & 0.05 \\
            Operational horizon $T_{\text{over}}$ (years) & 50 \\
            PM interval range $T_{\text{PM}}$ (days) & {[}365, 18{,}250{]} \\
            OM threshold range $H_{\text{OM}}$ & {[}0.05, 1.00{]} \\
            \bottomrule
        \end{tabular*}%
    }
\end{table}

All simulations are parallelized on QUT's high-performance computing system using \textbf{\textit{joblib}}. For the full policy sweep, wall-clock runtimes are on the order of a few hours per PM interval batch under multi-core execution.

\subsection{Bayesian calibration using LM-80 data}
After defining the zone geometry, lighting requirements, and simulation settings, the package degradation model is calibrated from LM-80 data using Bayesian inference.

LM-80 tests were conducted at three temperature levels, $55^\circ\mathrm{C}$, $85^\circ\mathrm{C}$, and $105^\circ\mathrm{C}$, with 25 LED packages tested at each temperature. Each unit was inspected every 1{,}000~hours over a total duration of 10{,}000~hours. Inspection times were then converted into years for model fitting.

Diffuse Normal priors are assigned to the transformed parameters, $\ln A \sim \mathcal{N}(0, 10^2)$ and $\ln C \sim \mathcal{N}(0, 10^2)$, and half-Normal priors are assigned to the nonnegative parameters to enforce physically consistent monotone degradation behavior, $b \sim \mathcal{N}^+(0, 10^6)$ and $E_a \sim \mathcal{N}^+(0, 10^6)$. Posterior inference is performed using \textbf{\textit{CmdStanPy}} with four Markov chains, each run for 2{,}000 warm-up iterations and 2{,}000 sampling iterations.

\begin{figure}
    \centering
        \begin{subfigure}[b]{0.3\textwidth}
    	\centering
    	\includegraphics[width=\textwidth]{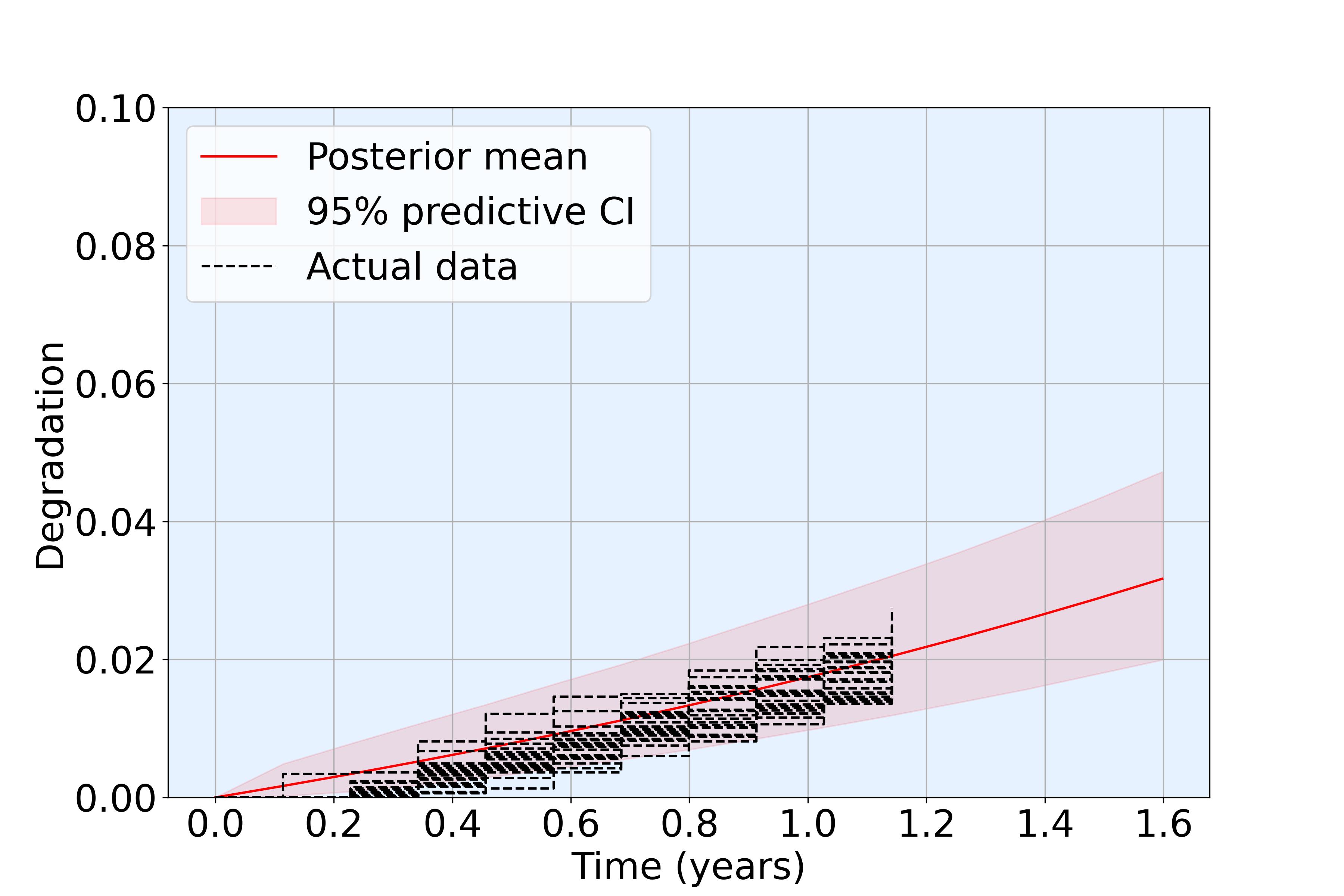}
        \caption{$55^\circ \mathrm{C}$ (held out; fitted on $85^\circ\mathrm{C}$ and $105^\circ\mathrm{C}$)}
    	\label{fig:PDT55}
        \end{subfigure}
    \hfill
    \begin{subfigure}[b]{0.3\textwidth}
        \centering
        \includegraphics[width=\textwidth]{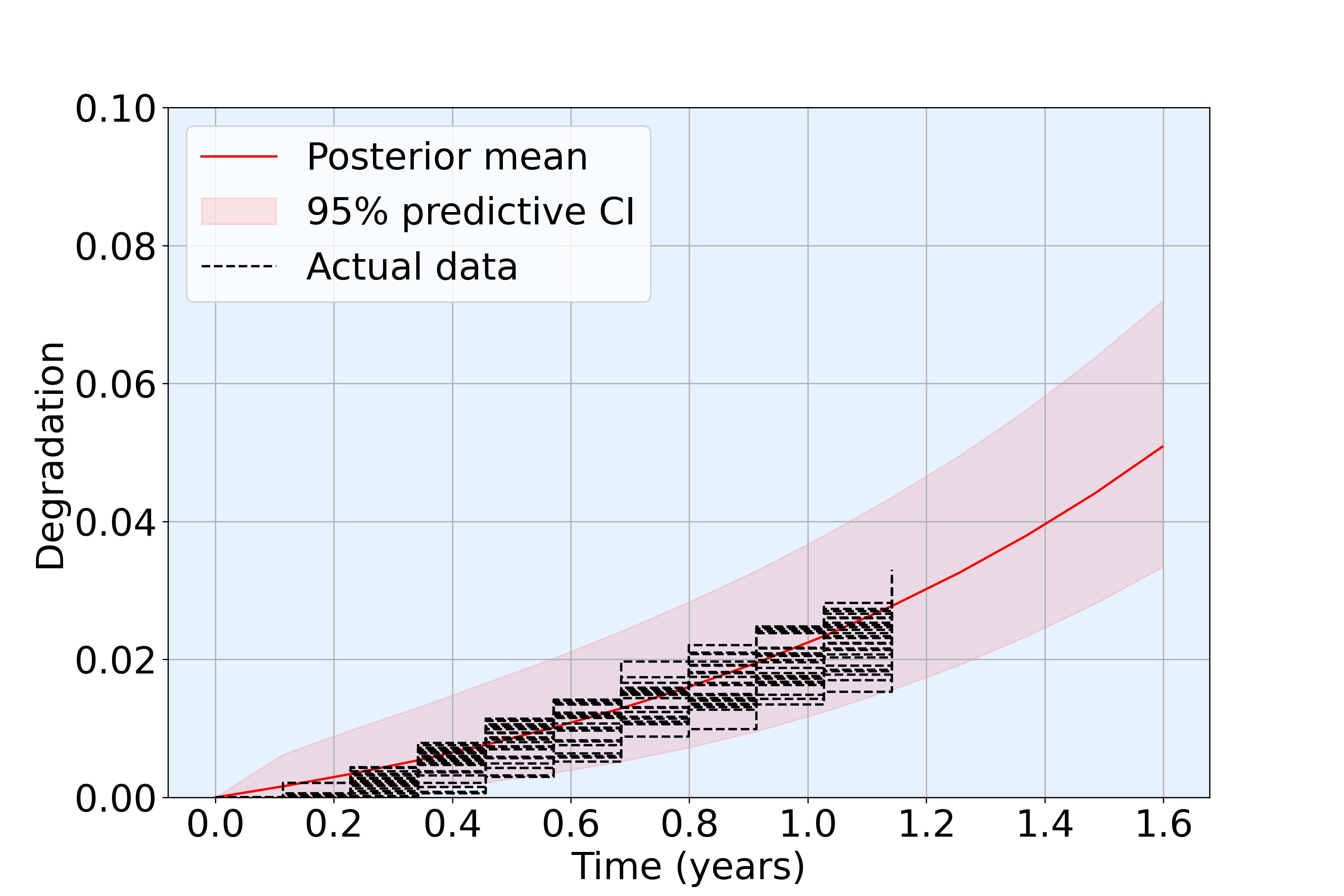}
        \caption{$85^\circ \mathrm{C}$ (held out; fitted on $55^\circ\mathrm{C}$ and $105^\circ\mathrm{C}$)}
        \label{fig:EPDT85}
    \end{subfigure}
    \hfill
    \begin{subfigure}[b]{0.3\textwidth}
        \centering
        \includegraphics[width=\textwidth]{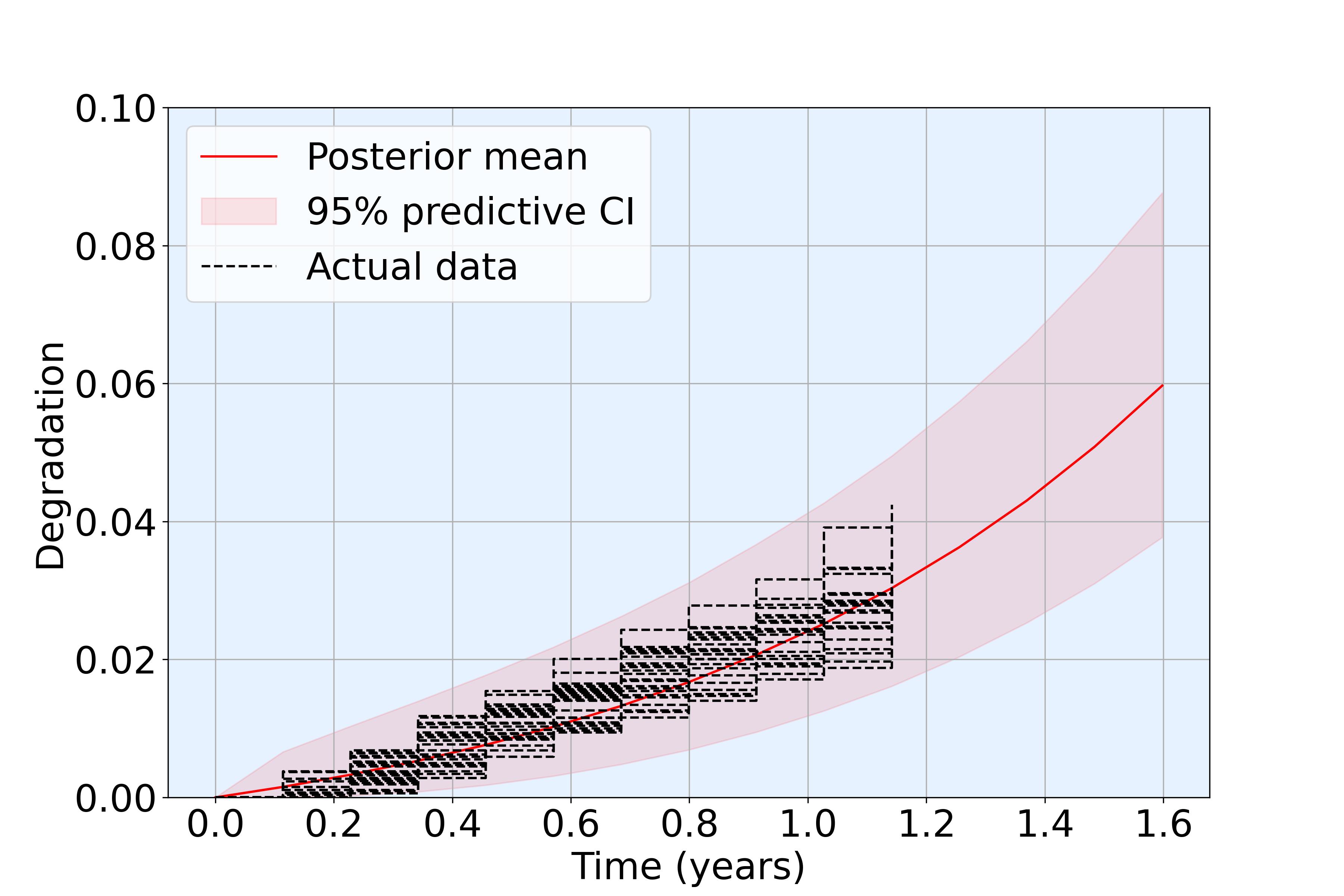}
        \caption{$105^\circ \mathrm{C}$ (held out; fitted on $55^\circ\mathrm{C}$ and $85^\circ\mathrm{C}$)}
        \label{fig:PDT105}
    \end{subfigure}
\caption{Hold-out validation of the calibrated model across stress levels.}
\label{fig:PEPDT}
\end{figure}

Hold-out validation is used to assess the extrapolation capability of the calibrated model across stress levels. This procedure also provides empirical support for the acceleration assumption. In this validation, each temperature is held out in turn from calibration and predicted by extrapolation from the posterior fitted on the remaining two temperatures. Specifically, three independent Bayesian calibrations are performed: (i) the $55^\circ\mathrm{C}$ trajectory is predicted by extrapolation from the posterior fitted on the $85^\circ\mathrm{C}$ and $105^\circ\mathrm{C}$ datasets; (ii) the $85^\circ\mathrm{C}$ trajectory is predicted by extrapolation from the posterior fitted on the $55^\circ\mathrm{C}$ and $105^\circ\mathrm{C}$ datasets; and (iii) the $105^\circ\mathrm{C}$ trajectory is predicted by extrapolation from the posterior fitted on the $55^\circ\mathrm{C}$ and $85^\circ\mathrm{C}$ datasets. As shown in \autoref{fig:PEPDT}, in all three cases the observed step trajectories at the held-out temperature are largely captured by the 95\% predictive interval (shaded band) and follow the posterior mean (solid red line). This consistent agreement across all three extrapolation directions supports both the non-homogeneous Gamma process model and the Bayesian calibration procedure.

\begin{figure}
    \centering
    \begin{subfigure}[b]{0.3\textwidth}
	\centering
	\includegraphics[width=\textwidth]{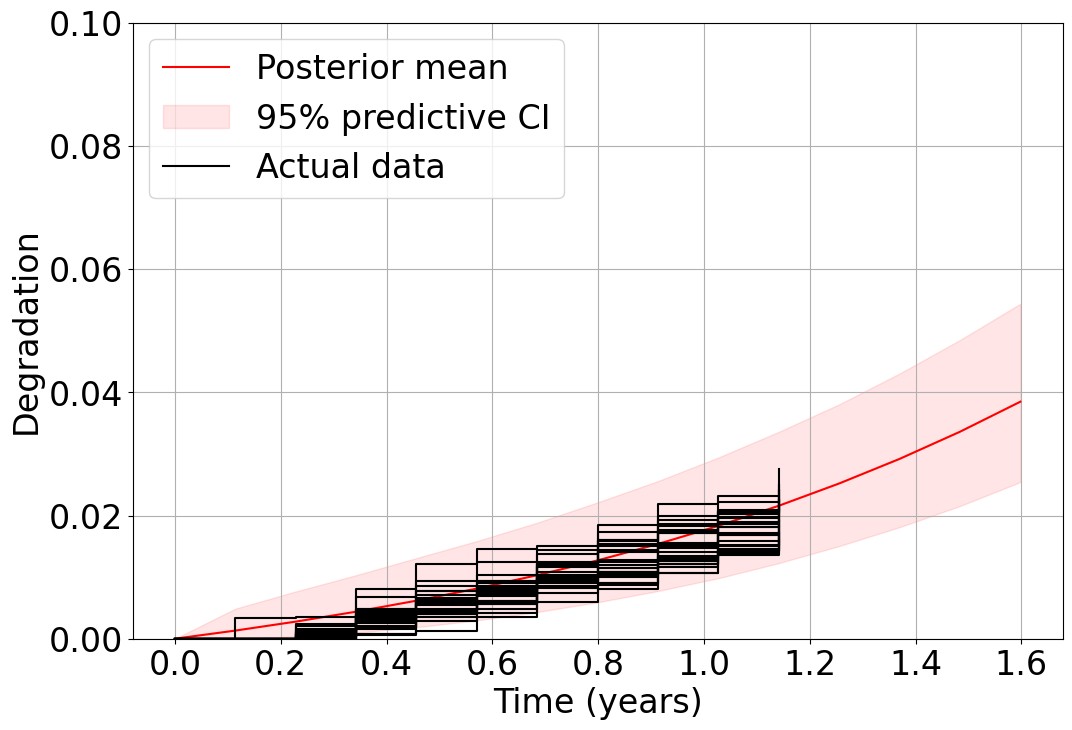}
	\caption{$55^\circ \mathrm{C}$}
	\label{fig:APDT55}
    \end{subfigure}
    \begin{subfigure}[b]{0.3\textwidth}
        \centering
        \includegraphics[width=\textwidth]{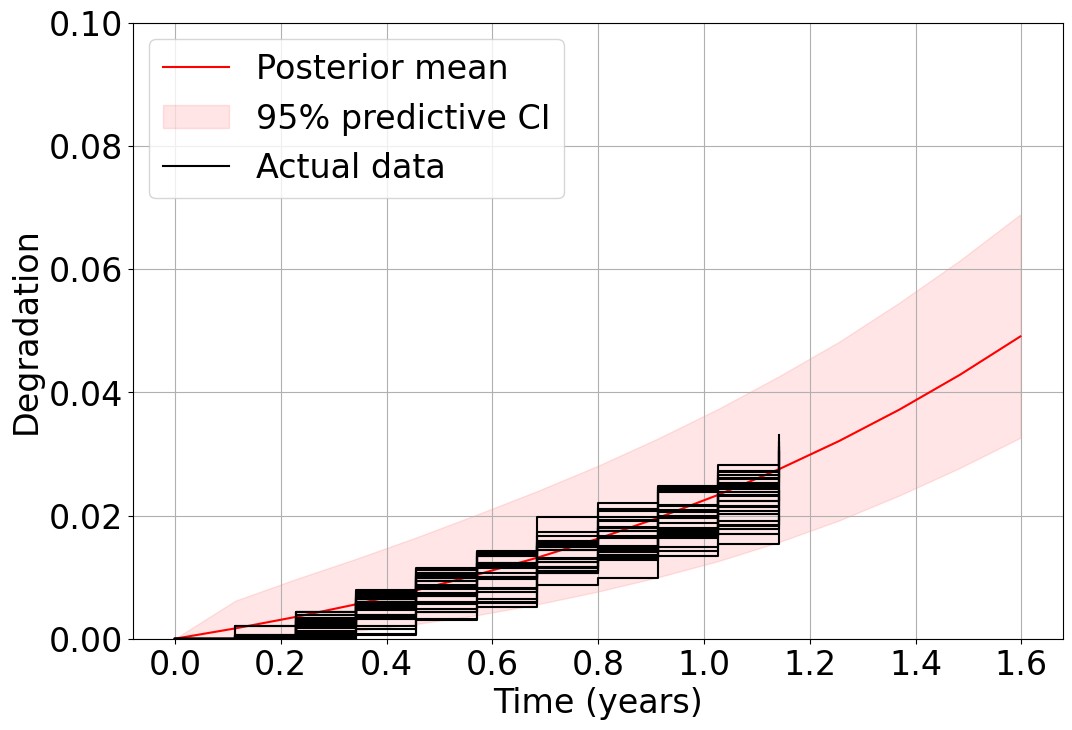}
        \caption{$85^\circ \mathrm{C}$}
        \label{fig:APDT85}
    \end{subfigure}
    \begin{subfigure}[b]{0.3\textwidth}
        \centering
        \includegraphics[width=\textwidth]{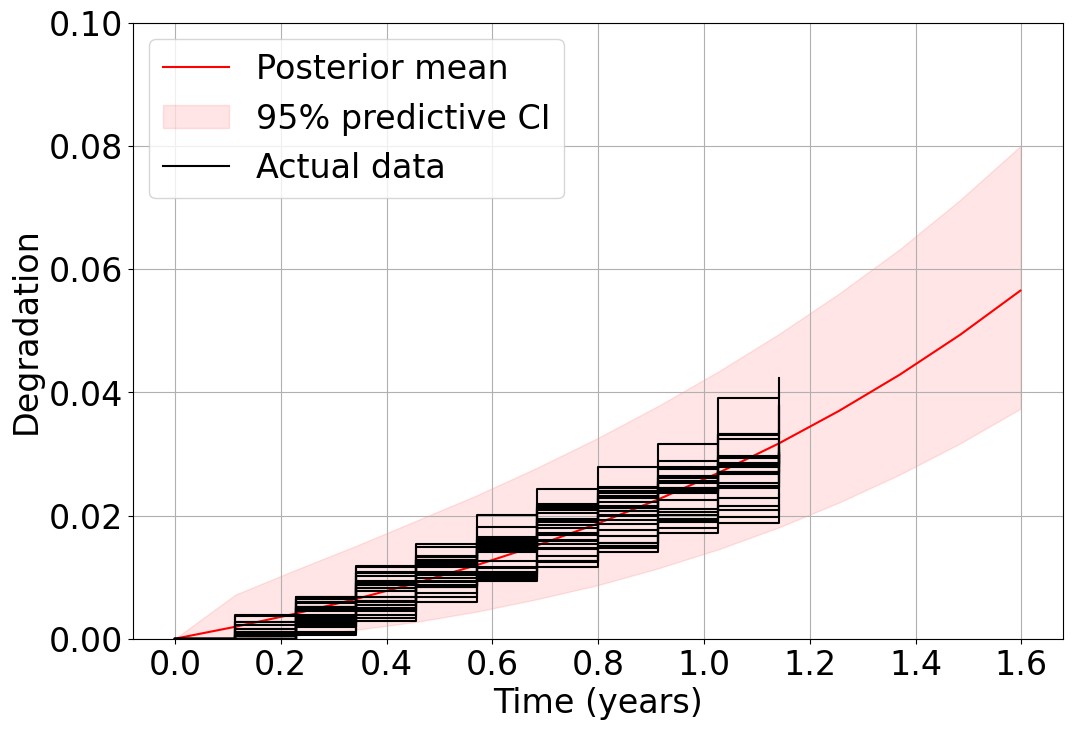}
        \caption{$105^\circ \mathrm{C}$}
        \label{fig:APDT105}
    \end{subfigure}
    \caption{Observed degradation data (black steps) and posterior-predictive trajectories (red mean and 95\% interval) at each temperature.}
    \label{fig:APDT}
\end{figure}

Finally, all three temperature datasets are combined to re-estimate the model parameters. \autoref{tab:EERCI} summarizes posterior means, 95\% credible intervals (CrIs), and effective sample size ($\mathrm{ESS}_{\text{bulk}}$ and $\mathrm{ESS}_{\text{tail}}$) diagnostics, with $\hat{R}$ statistics indicating good convergence ($\hat{R} < 1.01$). \autoref{fig:APDT} further compares the posterior-predictive trajectories against observations at each temperature, showing consistent agreement across the full accelerated-test range.
\begin{table}[width=.9\linewidth,cols=6,pos=h]
    \caption{Posterior means and 95\% CrIs for LM-80 calibration.}
    \label{tab:EERCI}
        \begin{tabular*}{\tblwidth}{@{} LLLLLL@{} }
            \toprule
            Parameters & Posterior mean & 95\% CrI & ESS$_{\text{bulk}}$ & ESS$_{\text{tail}}$ & $\hat{R}$ \\
            \midrule
            $\ln A$ & 2.2393 & [1.9472, 2.5366] & 2301.45 & 2369.31 & 1.0012 \\
            $b$     & 0.8841 & [0.7161, 1.0370] & 2330.92 & 2390.00 & 1.0013 \\
            $\ln C$ & 3.7446 & [2.7509, 4.7680] & 2592.95 & 2470.30 & 1.0020 \\
            $E_a$   & 0.0815 & [0.0505, 0.1112] & 2591.48 & 2367.46 & 1.0020 \\
            \bottomrule
        \end{tabular*} 
\end{table}

\subsection{Extrapolation to service conditions and baseline setting specification}
The posterior over the model parameters is propagated to service temperatures to obtain service-condition degradation dynamics, and driver lifetime distributions are specified to define the baseline setting.

Based on measured service conditions and practical relevance, $45^\circ\mathrm{C}$ is selected as the reference temperature for extrapolation to service conditions. Using the fitted stress--acceleration relationship, the extrapolated rate parameter is $\hat{\beta}(T=45^\circ\mathrm{C}) = 828.68$ with a 95\% CrI of $[700.77, 963.62]$.

\begin{figure}
    \centering
    \begin{subfigure}[b]{0.4\textwidth}
        \centering
        \includegraphics[width=\textwidth]{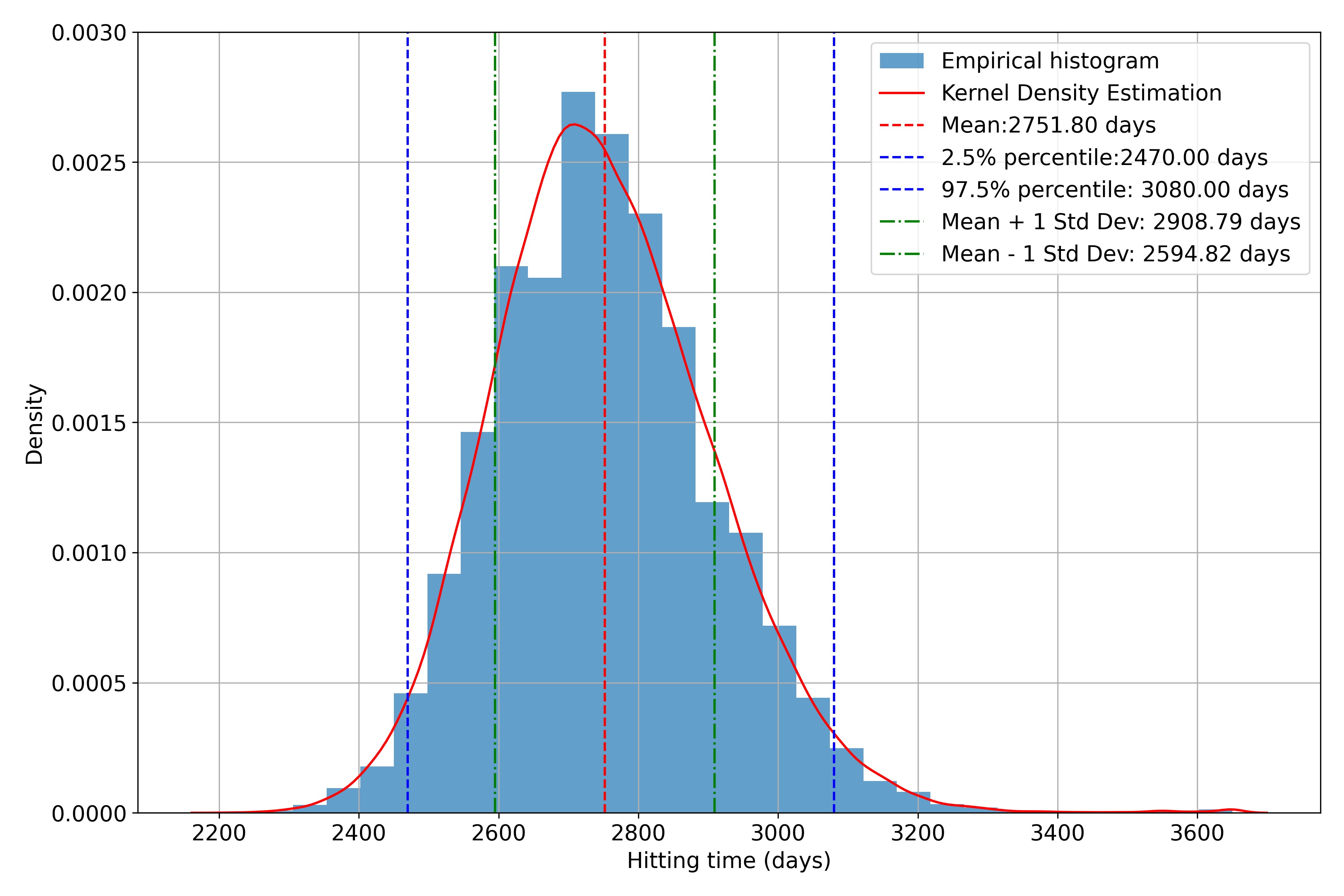}
        \caption{Gamma process lifetime distribution (S1)}
    \end{subfigure}
    \begin{subfigure}[b]{0.4\textwidth}
        \centering
        \includegraphics[width=\textwidth]{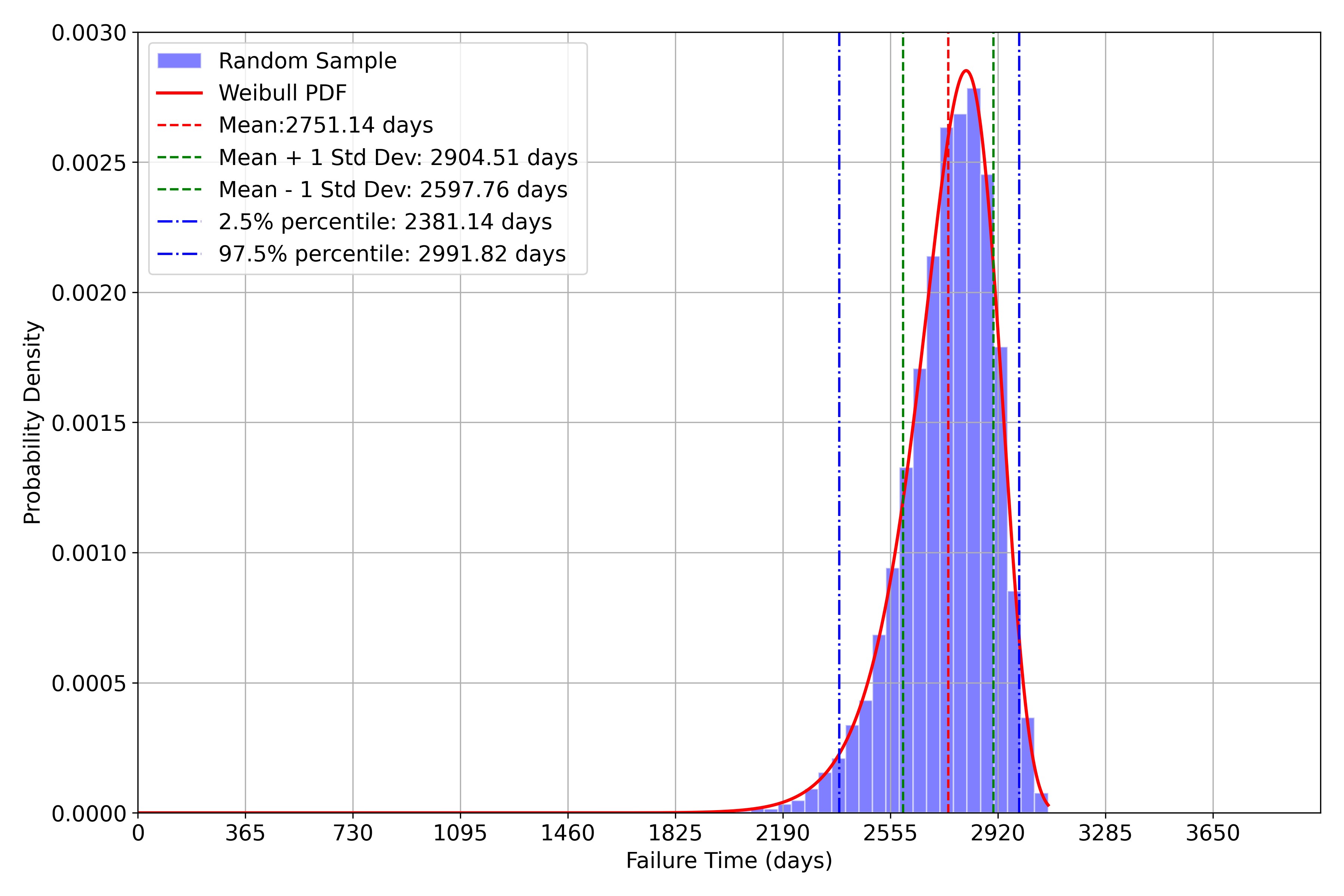}
        \caption{Weibull lifetime distribution (S1)}
    \end{subfigure}
    \caption{Lifetime distributions for LED packages and drivers under the baseline setting (S1).}
    \label{fig:LDPD}
\end{figure}

For LED drivers, a Weibull distribution is specified to match the package MTTF under the baseline setting (S1). Specifically, Weibull parameters $(21.82, 2818.09)$ are calibrated such that the driver MTTF aligns with the package MTTF. The resulting MTTFs for packages and drivers (7.54~years and 7.53~years, respectively) are illustrated in \autoref{fig:LDPD}. Alternative Weibull parameterizations are introduced later to assess sensitivity to assumptions.

\subsection{Maintenance optimization results}
With the calibrated degradation model and the baseline driver lifetime setting (S1), the proposed multi-objective optimization is evaluated over the grid of $(T_{\text{PM}}, H_{\text{OM}})$ policy settings. The full design includes 50 PM intervals and 20 OM thresholds (1,000 candidate policies). After statistical screening using one-sided Welch's $t$-tests, 36 policies are retained for the final Pareto analysis.

\begin{figure}
    \centering
    \begin{subfigure}[t]{0.48\textwidth}
	\centering 
	\includegraphics[width=\textwidth]{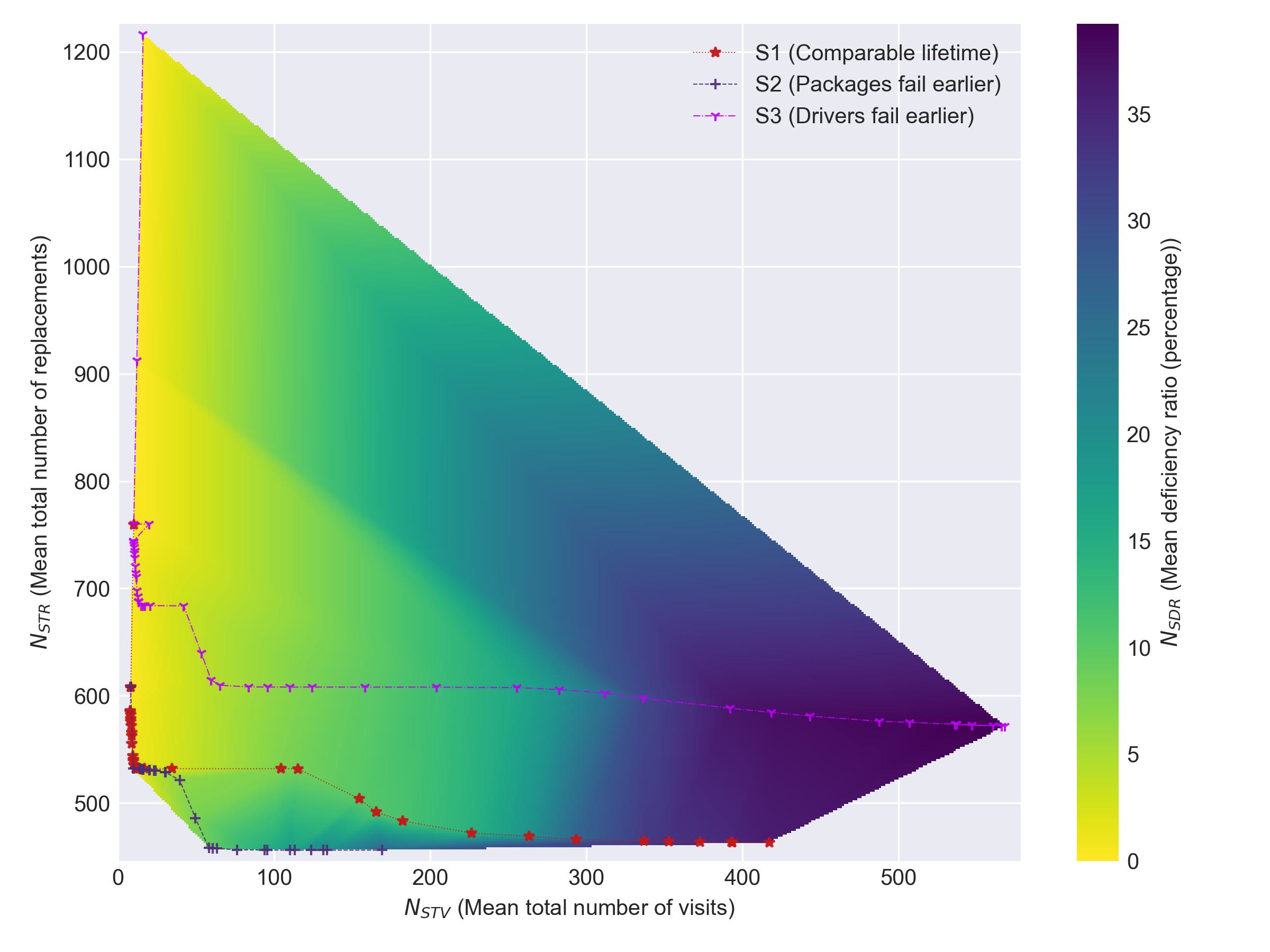} 
	\caption{Pareto front of all settings}
	\label{fig:PFS4} 
    \end{subfigure}
    \hfill
    \begin{subfigure}[t]{0.48\textwidth}
	\centering 
	\includegraphics[width=\textwidth]{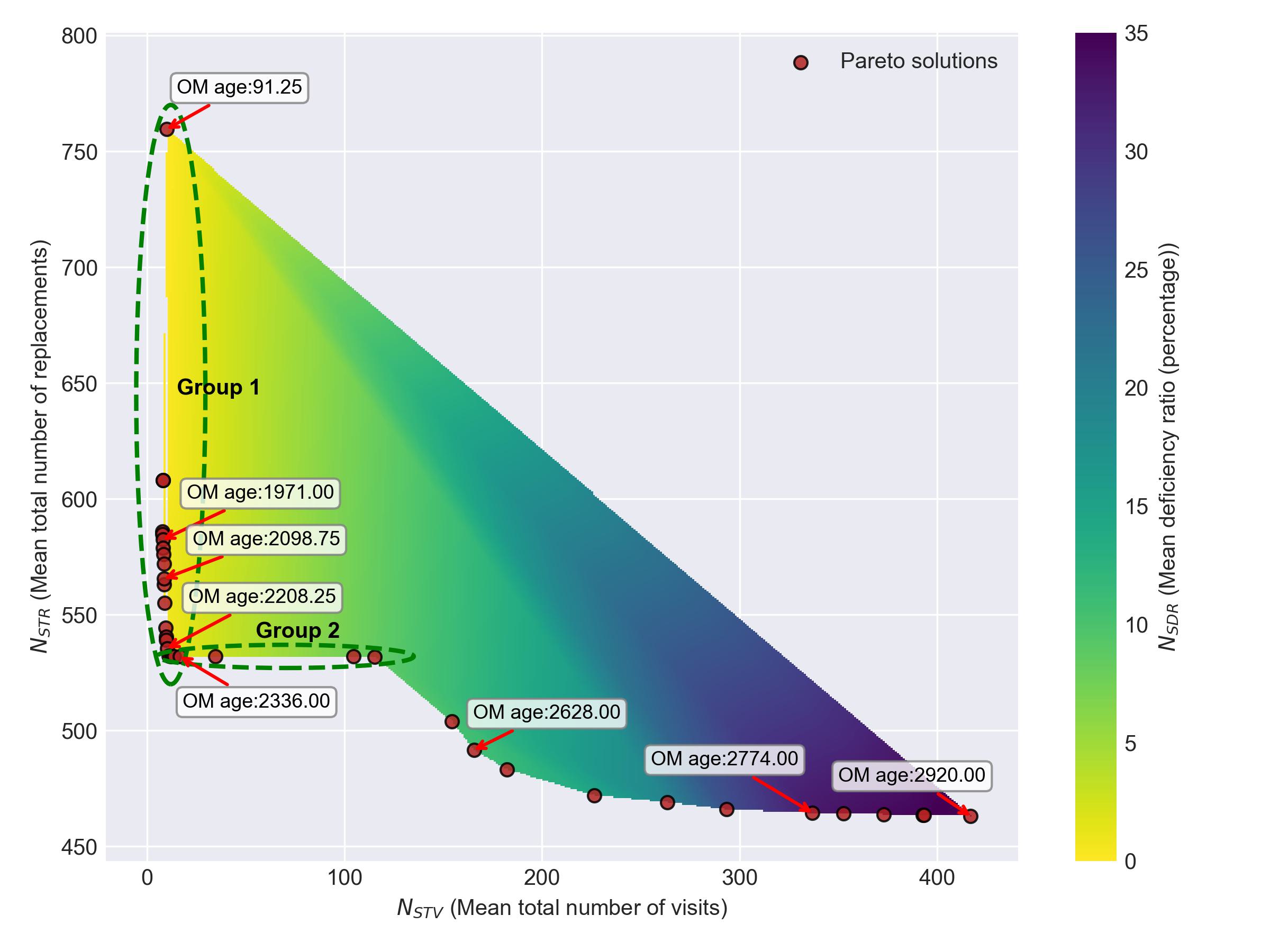} 
	\caption{Pareto front of baseline setting S1}
	\label{fig:PFS1} 
    \end{subfigure}

    \medskip
    \begin{subfigure}[t]{0.48\textwidth}
    \centering
	\includegraphics[width=\textwidth]{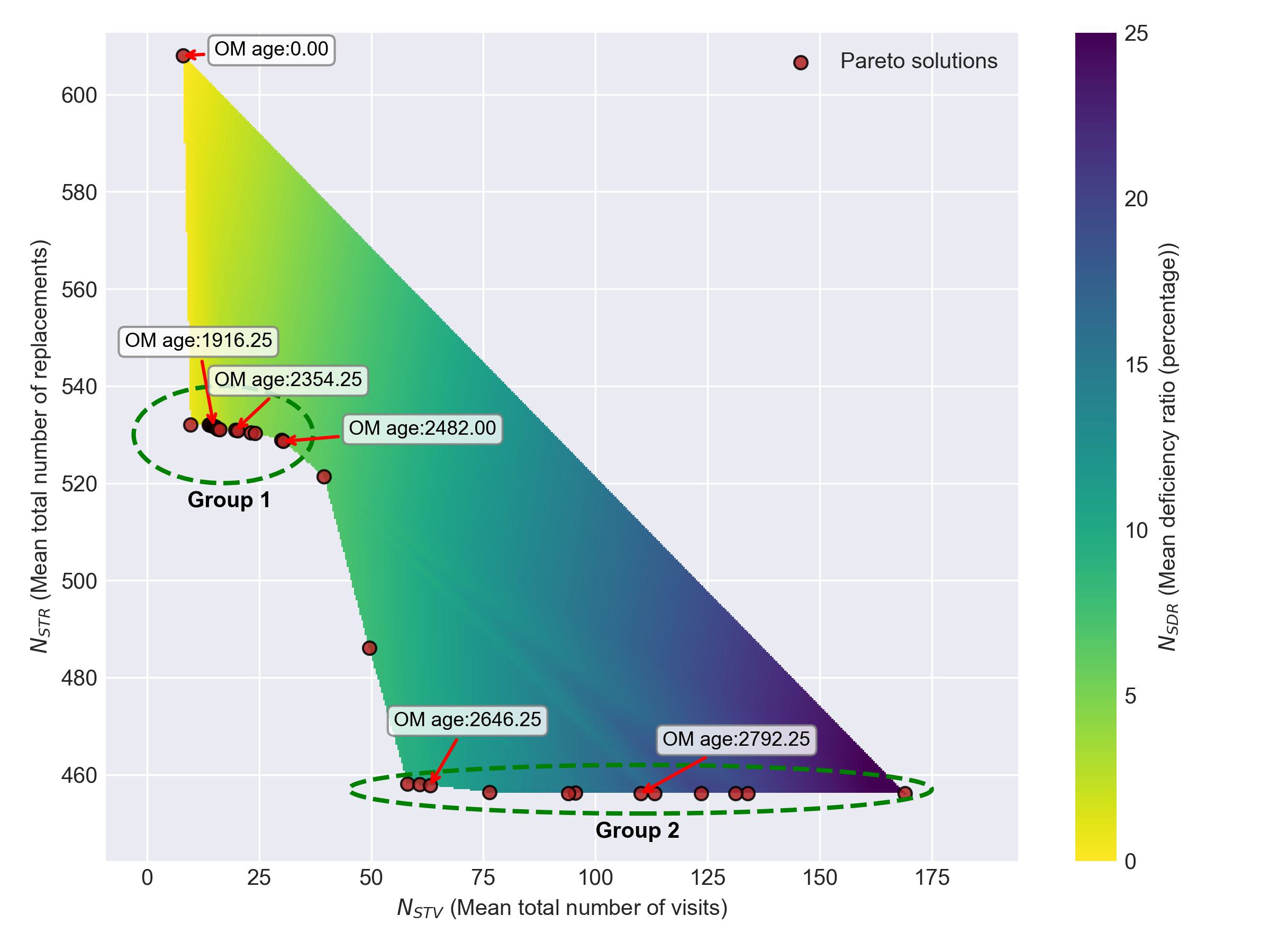} 
	\caption{Pareto front of S2}
	\label{fig:PFS2} 
    \end{subfigure}
    \hfill
    \begin{subfigure}[t]{0.48\textwidth}
	\centering 
	\includegraphics[width=\textwidth]{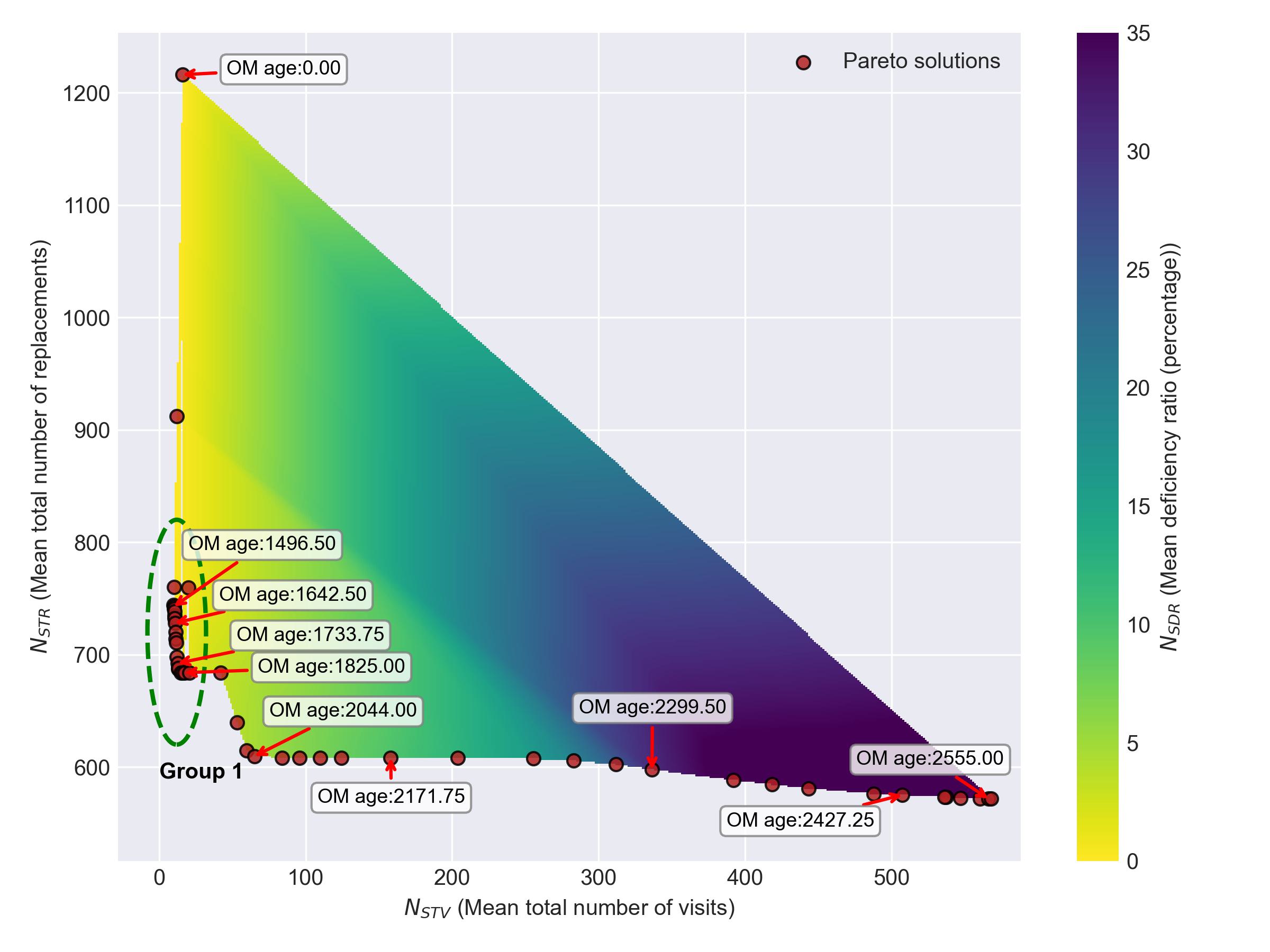} 
	\caption{Pareto front of S3}
	\label{fig:PFS3} 
    \end{subfigure}
    \caption{Pareto fronts for the baseline setting S1, S2 (packages fail earlier), and S3 (drivers fail earlier).}
    \label{fig:RS1}
\end{figure}

\autoref{fig:PFS4} depicts the resulting Pareto front in the three-dimensional objective space defined by the mean deficiency ratio ($\rev{R_{\text{SDR}}}$), the mean total number of site visits ($\rev{N_{\text{STV}}}$), and the mean total number of luminaire replacements ($\rev{N_{\text{STR}}}$). The background color illustrates variations in the mean deficiency ratio ($\rev{R_{\text{SDR}}}$), where blue regions correspond to higher values of $\rev{R_{\text{SDR}}}$, highlighting scenarios with poorer lighting performance. In contrast, yellow regions indicate lower values of $\rev{R_{\text{SDR}}}$, signifying scenarios with superior system performance. Across the Pareto front, reductions in the total number of luminaire replacements ($\rev{N_{\text{STR}}}$) are generally accompanied by increases in site visits ($\rev{N_{\text{STV}}}$), indicating that opportunistic maintenance becomes less aggressive and more temporally dispersed.

As shown in \autoref{fig:PFS1}, a useful interpretation emerges by relating Pareto solutions to the effective OM aggressiveness captured by the OM age $T_{\text{OM}}$ (Section~\ref{sec:MOOM}). The Pareto solutions can be classified into three main groups. In Group 1, as $T_{\text{OM}}$ increases, the mean total number of luminaire replacements ($\rev{N_{\text{STR}}}$) initially decreases substantially from approximately 759.5 to 532. During this transition, the mean total number of site visits ($\rev{N_{\text{STV}}}$) first decreases slightly from approximately 9.99 to 7.81 and then rises to approximately 13.68. This behavior occurs because, as $T_{\text{OM}}$ increases, fewer OM replacements are executed, leading to fewer replacements but poorer lighting performance. However, because $T_{\text{OM}}$ within this range remains below the 2.5th percentile of the lifetime distributions of the packages and drivers, most LEDs are still replaced through OM, and the increase in site visits ($\rev{N_{\text{STV}}}$) remains moderate.

When $T_{\text{OM}}$ reaches the MTTF of the two critical components, solutions transition to Group 2. The replacements initiated by CM visits partially offset the reduction in OM replacements, yielding a marginal reduction in $\rev{N_{\text{STR}}}$. As the OM policy adopts a more conservative stance, the system ``waits'' for enough CM visits to replace all LEDs. Once $T_{\text{OM}}$ exceeds the components' MTTF, the solutions migrate to Group 3 (the remaining points) and the OM policy becomes markedly more conservative. Consequently, both the mean total number of site visits ($\rev{N_{\text{STV}}}$) and the mean deficiency ratio ($\rev{R_{\text{SDR}}}$) begin to noticeably increase, indicating a clear trade-off: conservative policies reduce replacements but significantly compromise system performance.

Along the Pareto front, moving from the upper-left extreme toward the ``corner'' solutions yields a substantial reduction in luminaire replacements ($\rev{N_{\text{STR}}}$) with only modest increases in site visits ($\rev{N_{\text{STV}}}$). In practice, the corner solution represents an attractive compromise for facility managers, minimizing total maintenance interventions while maintaining acceptable system performance. Notably, the upper-left point corresponds to a PM policy. These results also suggest that a CM-driven OM policy might outperform a PM policy in practice, although the greater reliance on CM visits would place additional strain on spare parts inventory management.

In summary, the Pareto front reveals a cluster of solutions in which the chosen maintenance parameters ($T_{\text{PM}}$ and $H_{\text{OM}}$) achieve a balanced compromise among the conflicting objectives. 
As an offline optimization framework, the proposed approach provides a clear demarcation of viable maintenance strategies, offering decision-makers multiple options. 
From a decision-making perspective, the Pareto front can also be interpreted as representing solutions under different implicit cost-weighting preferences. 
For example, if the primary concern is to minimize disruptions and maintenance resources, a strategy with fewer site visits (and a slightly higher mean deficiency ratio) may be preferred. Conversely, if maintaining optimal system performance is paramount, then a strategy that involves more frequent site visits and replacements (thereby lowering $\rev{R_{\text{SDR}}}$) would be the logical choice.

\subsection{Cost comparison of benchmark and OM policies}
This subsection presents a cost-oriented comparison of benchmark and OM policies. To support this comparison, four policies are selected from the S1 Pareto front for detailed examination: Policies~1 and 2 serve as benchmark PM- and CM-dominated strategies, whereas Policies~3 and 4 are representative OM policies selected from the corner region of \autoref{fig:PFS1}. These policies are first compared in terms of their maintenance-action structure and are then evaluated under three representative unit-cost settings.
As shown in \autoref{tab:SPS}, Policy~1 is extracted from the upper-left extreme of \autoref{fig:PFS1}. This policy represents a near-pure PM solution and effectively operates as a group replacement policy, in which most renewals are carried out collectively through scheduled PM actions. It yields the lowest mean deficiency ratio, $\rev{R_{\text{SDR}}}$ ($0.0028\%$), but at the expense of the highest replacement burden, $\rev{N_{\text{STR}}}$ (759.5212). Policy~2 is extracted from the rightmost extreme of \autoref{fig:PFS1}. It represents a near-pure CM solution, in which renewals are driven almost entirely by CM events. This policy results in the poorest performance among all Pareto solutions ($36.0666\%$ mean deficiency ratio), despite a comparatively lower mean total number of replacements (463.1235). By contrast, Policies~3 and 4 are two representative OM policies selected from the corner region of \autoref{fig:PFS1}, where the trade-off among performance deficiency, site visits, and replacements is more balanced.

\begin{table}[htbp]

\caption{Selected Pareto solutions}
\label{tab:SPS}
    \begin{threeparttable}
    \resizebox{\linewidth}{!}{%
        \begin{tabular}{l c c c c c c c c c c c}
        \hline
        & Policy 
        & \makecell{$\rev{R_{\text{SDR}}}$ (Mean \\ deficiency \\ ratio (\%))} 
        & \makecell{$\rev{N_{\text{STV}}}$ (No. of \\ visits)} 
        & \makecell{$\rev{N_{\text{STR}}}$ (No. of \\replacements)} 
        & \makecell{OM \\ threshold (fraction)} 
        & \makecell{PM interval \\ (days)} 
        & OM age (days)
        & \makecell{Mean PM \\ replacements} 
        & \makecell{Mean CM \\ replacements}
        & OM (PM) 
        & OM (CM) \\ 
        \hline
        & 1  & 0.0028   & 9.9937    & 759.5212 & 0.95   & 1825  & 91.25 & 754.5204  & 0.0658   & 0       & 4.9350 \\ 
        & 2  & 36.0666  & 417.0330  & 463.1235 & 0.2    & 3650  & 2920  & 0         & 457.9810 & 0       & 5.1425  \\ 
        & 3  & 0.1283   & 8.0917    & 608      & 0.2    & 2190  & 1752  & 457.2635  & 2.0040    & 0.9075  & 147.8250 \\ 
        & 4  & 0.8719   & 12.0519   & 532.0176 & 0.8    & 11315 & 2263  & 0         & 12.2372  & 0       & 519.7804 \\ 
        \hline
        \end{tabular}%
    }
    \begin{tablenotes}
    \footnotesize
    \item[a] OM (PM): mean number of OM replacements triggered by PM visits;
    \item[b] OM (CM): mean number of OM replacements triggered by CM visits.
    \item[c] Policy~1 represents the near-pure PM benchmark extracted from the upper-left extreme of the S1 Pareto front;
    \item[d] Policy~2 represents the near-pure CM benchmark extracted from the right-most extreme of the S1 Pareto front.
    \end{tablenotes}
    \end{threeparttable}

\end{table}

To compare these four policies on a common cost basis, a normalized cost--benefit analysis is conducted using the policy statistics reported in \autoref{tab:SPS}. Since detailed site-specific monetary data and energy-cost models are not available in this study, the mean total number of site visits and the mean total number of replacements are used as cost proxies. Specifically, a normalized total cost is defined as $C = c_v \rev{N_{\text{STV}}} + c_r \rev{N_{\text{STR}}}$, where $\rev{N_{\text{STV}}}$ and $\rev{N_{\text{STR}}}$ denote the mean numbers of site visits and replacements, respectively, and $c_v$ and $c_r$ represent the unit visit cost and unit replacement cost. Three representative unit-cost settings are considered, namely $(c_r,c_v)=(1,1)$, $(1,100)$, and $(100,1)$, and the corresponding results are summarized in \autoref{tab:costbenefit}.

As shown in \autoref{tab:costbenefit}, the relative attractiveness of the representative policies changes with the assumed cost structure. Under equal unit costs, Policy~4 yields the lowest normalized total cost among the four policies while maintaining a low deficiency ratio. When site visits are relatively more expensive, Policy~3 yields the lowest normalized total cost among the four policies. By contrast, when replacement cost dominates, Policy~2 yields the lowest direct maintenance cost, but only at the expense of a substantially worse performance level, as reflected by its much higher deficiency ratio. Overall, the representative OM solutions maintain much better lighting performance levels than the near-pure CM benchmark while still reducing normalized cost relative to the near-pure PM benchmark, demonstrating a more balanced cost--performance trade-off.

\begin{table}[htbp]
\caption{Normalized cost--performance comparison for the selected policies}
\label{tab:costbenefit}
    \begin{threeparttable}
    \resizebox{\linewidth}{!}{%
    \begin{tabular}{c l c c c c c c}
    \hline
    Policy & Maintenance strategy &
    \makecell{$\rev{R_{\text{SDR}}}$\\(Mean deficiency\\ratio (\%))} &
    \makecell{OM\\threshold (fraction)} &
    \makecell{PM interval\\(days)} &
    \makecell{Normalized total cost\\$(c_r,c_v)=(1,1)$} &
    \makecell{Normalized total cost\\$(c_r,c_v)=(1,100)$} &
    \makecell{Normalized total cost\\$(c_r,c_v)=(100,1)$} \\
    \hline
    1 & Near-pure PM benchmark & 0.0028 & 0.95 & 1825    & 769.5149  & 1758.8912  & 75962.1137 \\
    2 & Near-pure CM benchmark & 36.0666 & 0.2  & 3650   & 880.1565  & 42166.4235 & $\mathbf{46729.3830}^{\dagger}$ \\
    3 & Representative OM solution & 0.1283 & 0.2  & 2190   & 616.0917  & $\mathbf{1417.1700}^{\dagger}$  & 60808.0917 \\
    4 & Representative OM solution & 0.8719 & 0.8  & 11315  & $\mathbf{544.0695}^{\dagger}$  & 1737.2076  & 53213.8119 \\
    \hline
    \end{tabular}%
    }
    \begin{tablenotes}
    \footnotesize
    \item[$^\dagger$] Minimum value in each column.
    \end{tablenotes}
    \end{threeparttable}

\end{table}

\subsection{Operational interpretation of representative policies}
\begin{figure}
    \centering
    \begin{subfigure}[t]{0.8\textwidth}
        \centering
	\includegraphics[width=\textwidth]{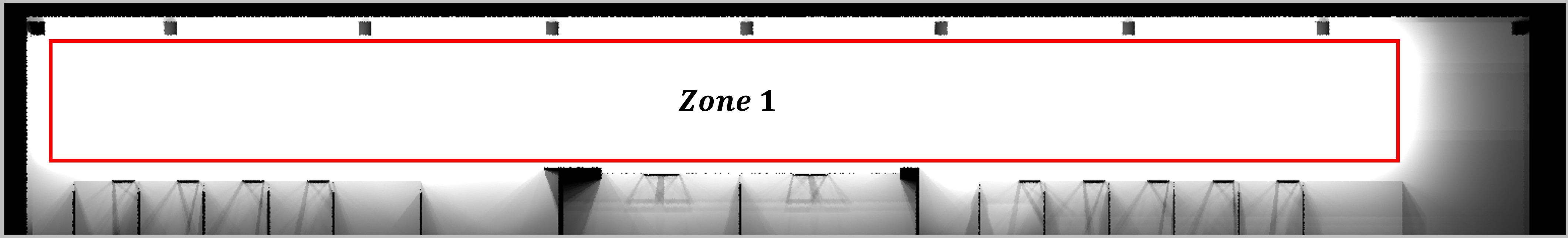} 
	\caption{Working plane of Zone 1 with a brand-new lighting system }
	\label{fig:FRM} 
    \end{subfigure}
    \begin{subfigure}[t]{0.8\textwidth}
        \centering
	\includegraphics[width=\textwidth]{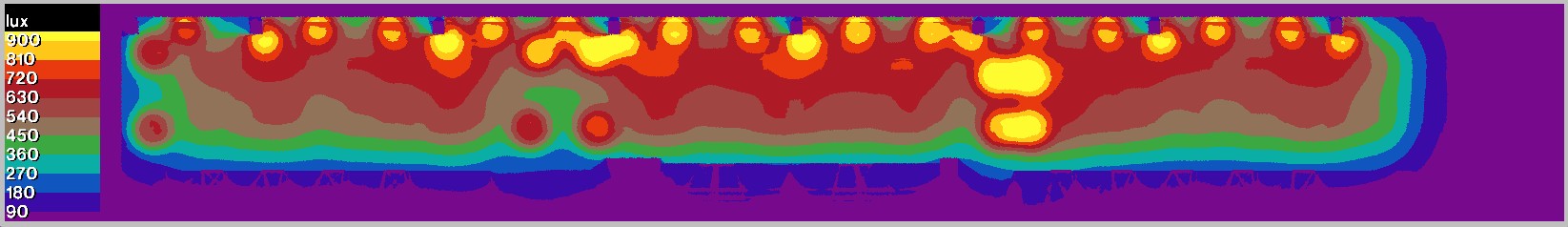} 
	\caption{Illuminance map for a brand-new lighting system}
	\label{fig:RM1} 
    \end{subfigure}
    \begin{subfigure}[t]{0.8\textwidth}
        \centering
        \includegraphics[width=\textwidth]{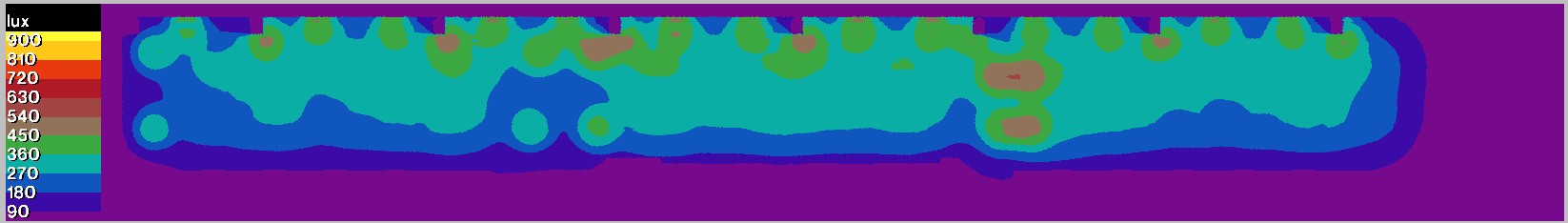}
        \caption{Illuminance map with degraded luminaires}
        \label{fig:RM2}
    \end{subfigure}
   \begin{subfigure}[t]{0.8\textwidth}
        \centering
        \includegraphics[width=\textwidth]{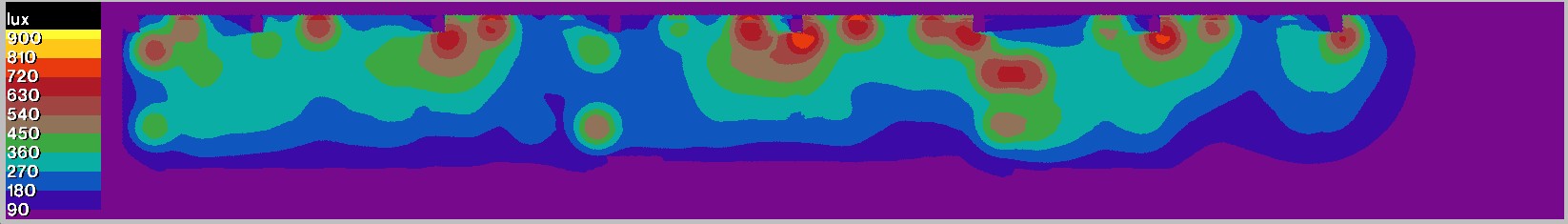}
        \caption{Illuminance map just after a maintenance event}
        \label{fig:RM3}
    \end{subfigure}
    \caption{Performance fluctuations of the lighting system under a representative maintenance scenario for Zone 1.}
    \label{fig:RPFMS}
\end{figure}

To illustrate the operational meaning of the performance objective, \autoref{fig:RPFMS} presents illuminance maps from one representative simulation trajectory. Warmer colors indicate higher illuminance values, while cooler colors indicate dimmer regions on the WP. As degradation accumulates and failures occur, the illuminance distribution becomes less uniform and the average level decreases (e.g., \autoref{fig:RM2}), increasing the mean deficiency ratio. After a maintenance event (\autoref{fig:RM3}), the map moves closer to the reference state (\autoref{fig:RM1}), and the growth of performance deficiency is arrested.

\begin{figure}
    \centering
    \begin{subfigure}[t]{0.45\textwidth}
        \centering
	\includegraphics[width=\textwidth]{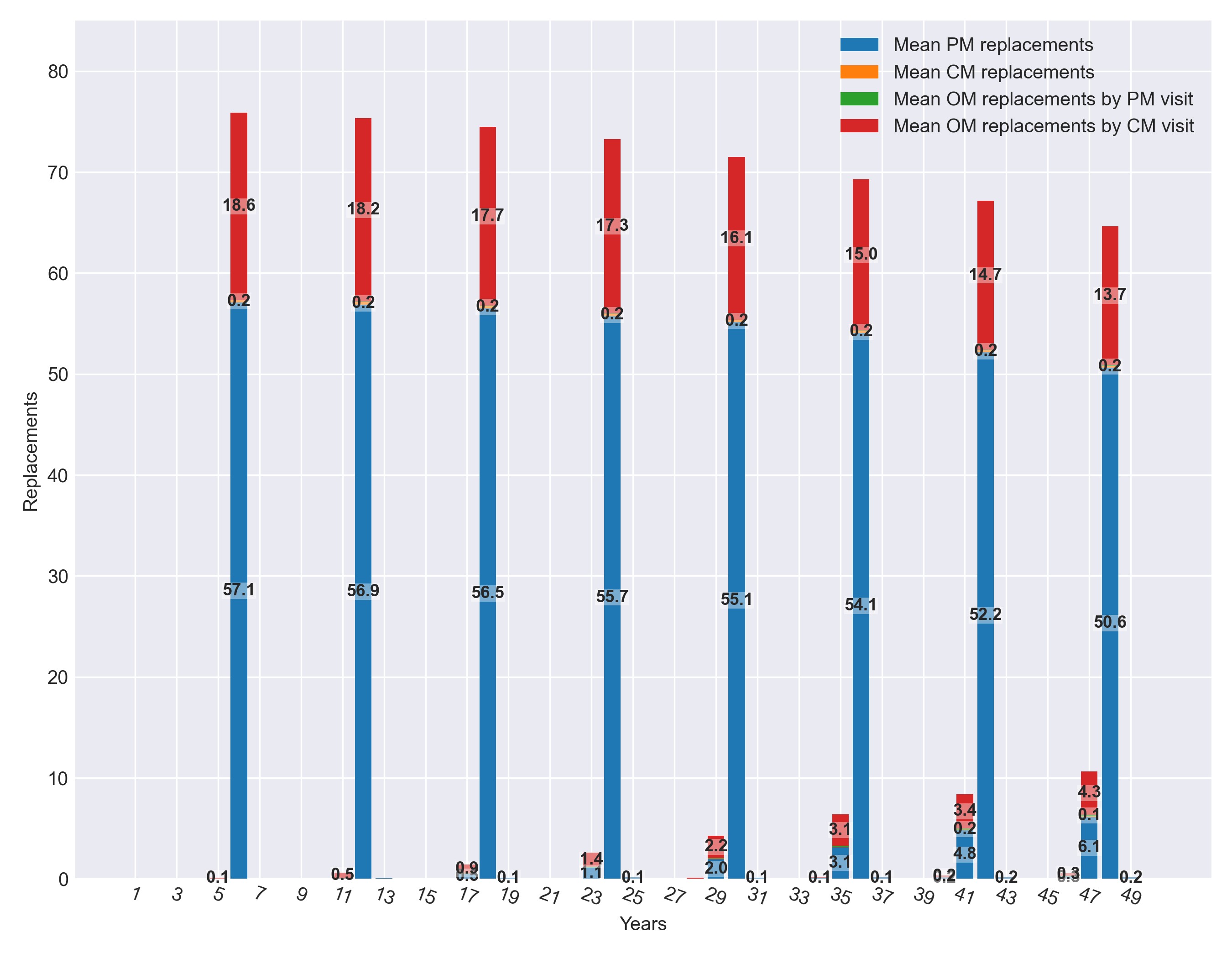} 
	\caption{Stacked bar chart for OM $=0.2$ and PM $=2190$ days ($T_{\text{OM}}=1752$ days)}
	\label{fig:SBC1} 
    \end{subfigure}
    \hfill
    \begin{subfigure}[t]{0.45\textwidth}
        \centering
	\includegraphics[width=\textwidth]{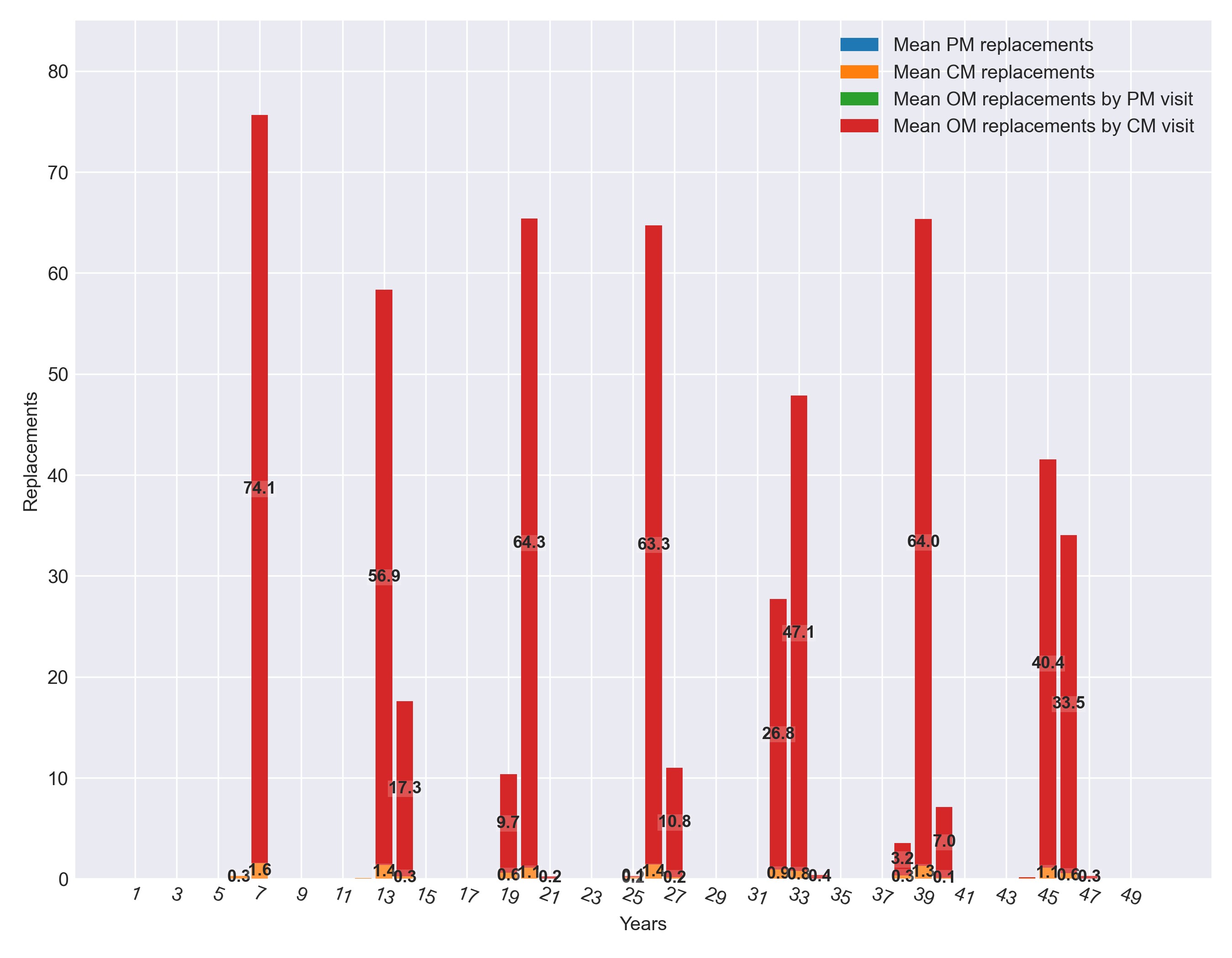} 
	\caption{Stacked bar chart for OM $=0.8$ and PM $=11315$ days ($T_{\text{OM}}=2263$ days)}
	\label{fig:SBC2} 
    \end{subfigure}
    \caption{Stacked bar charts for the selected policies}
    \label{fig:SBCSS}
\end{figure}

From the Pareto set, two representative policies are selected to demonstrate how different parameter choices alter the composition of maintenance actions. Their decision variables and objective values are summarized in \autoref{tab:SPS}, and the corresponding annual mean replacement breakdowns are shown in \autoref{fig:SBCSS}. For the first representative policy, the OM threshold is 0.2 and the PM interval is 2190 days -- below the 2.5th percentile of the package/driver lifetime distribution -- yielding a PM-driven OM policy. Although the OM age $T_{\text{OM}}$ (1752 days) is relatively short -- implying an aggressive OM policy -- the low variance in component lifetimes means that most replacements still occur at scheduled PM visits. As shown in \autoref{fig:SBC1}, a PM action every 6 years (2190 days) replaces about 70\% of the LEDs in the system, and any remaining units are opportunistically swapped out during CM visits in the same year. Over successive cycles, both the PM and OM-by-CM replacements shift to occur immediately before or after each 6-year PM cycle, reflecting the stochastic variability inherent in the degradation models. While this variability complicates precise maintenance planning, it provides facility managers with lead time information for spare parts provisioning and workforce allocation to better mitigate unexpected failures.

For the second representative policy, the PM interval is 11315 days -- above the 97.5th percentile of component lifetimes -- and the OM threshold is 0.8 ($T_{\text{OM}} = 2263$ days). This produces a CM-driven OM policy dominated by failure-triggered replacements. As \autoref{fig:SBC2} illustrates, the mean lifetimes are 7.5 years and $T_{\text{OM}}$ is 6.2 years, so nearly the entire LED lighting fleet is renewed within a two-year window every 6--7 years via CM and OM-by-CM replacements. The inherent uncertainty in stochastic degradation models causes these renewal events to occur slightly earlier or later around the 6- to 7-year mark, yielding actionable insights for asset managers to optimize maintenance lead times, inventory strategies, and overall resource planning.

\subsection{Sensitivity analysis: impacts of service temperature and driver reliability assumptions}
\begin{figure}
    \centering
    \begin{subfigure}[t]{0.45\textwidth}
        \centering
	\includegraphics[width=\textwidth]{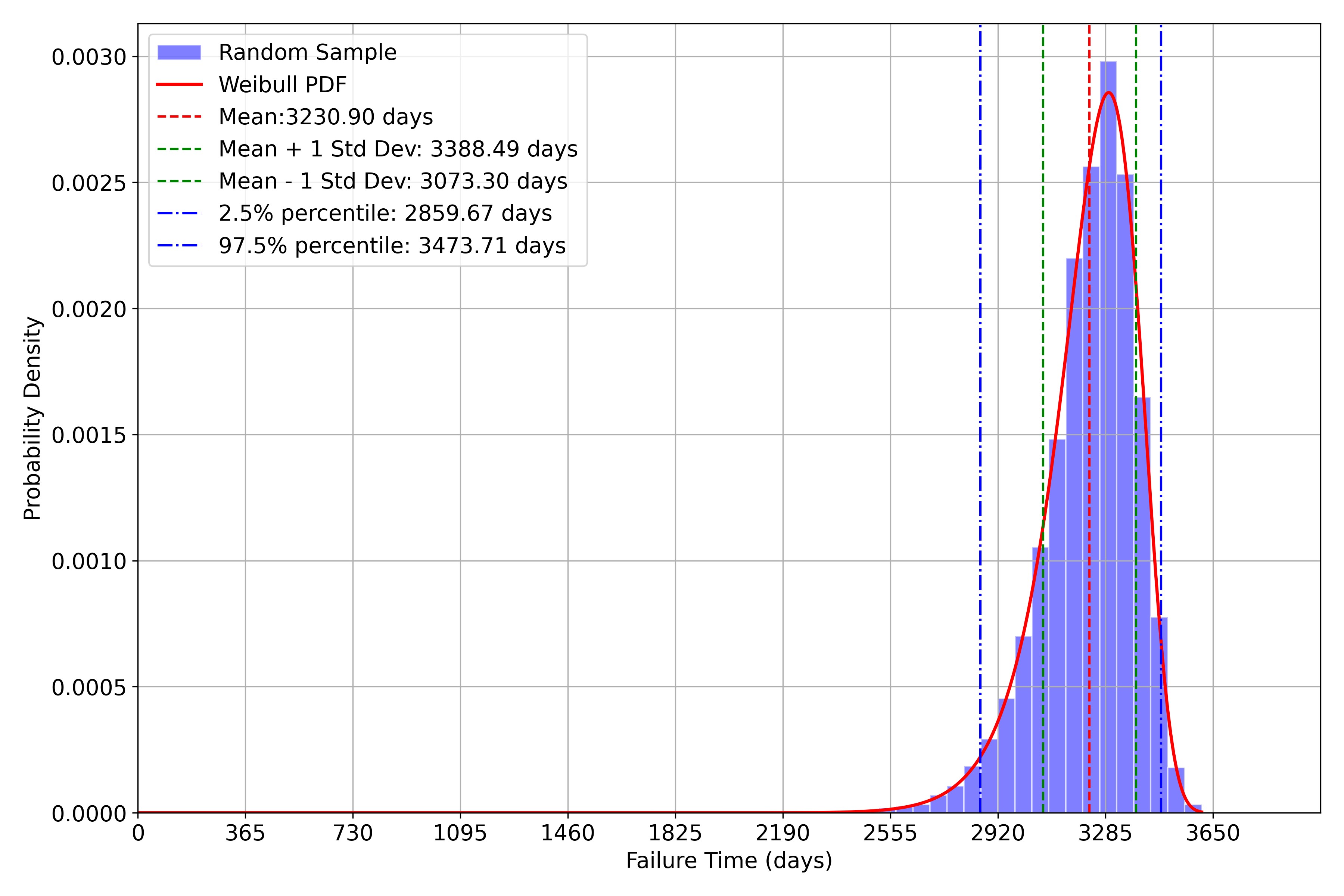} 
	\caption{S2 Weibull lifetime distribution}
	\label{fig:S2WLD} 
    \end{subfigure}
    \hfill
    \begin{subfigure}[t]{0.45\textwidth}
	\centering 
	\includegraphics[width=\textwidth]{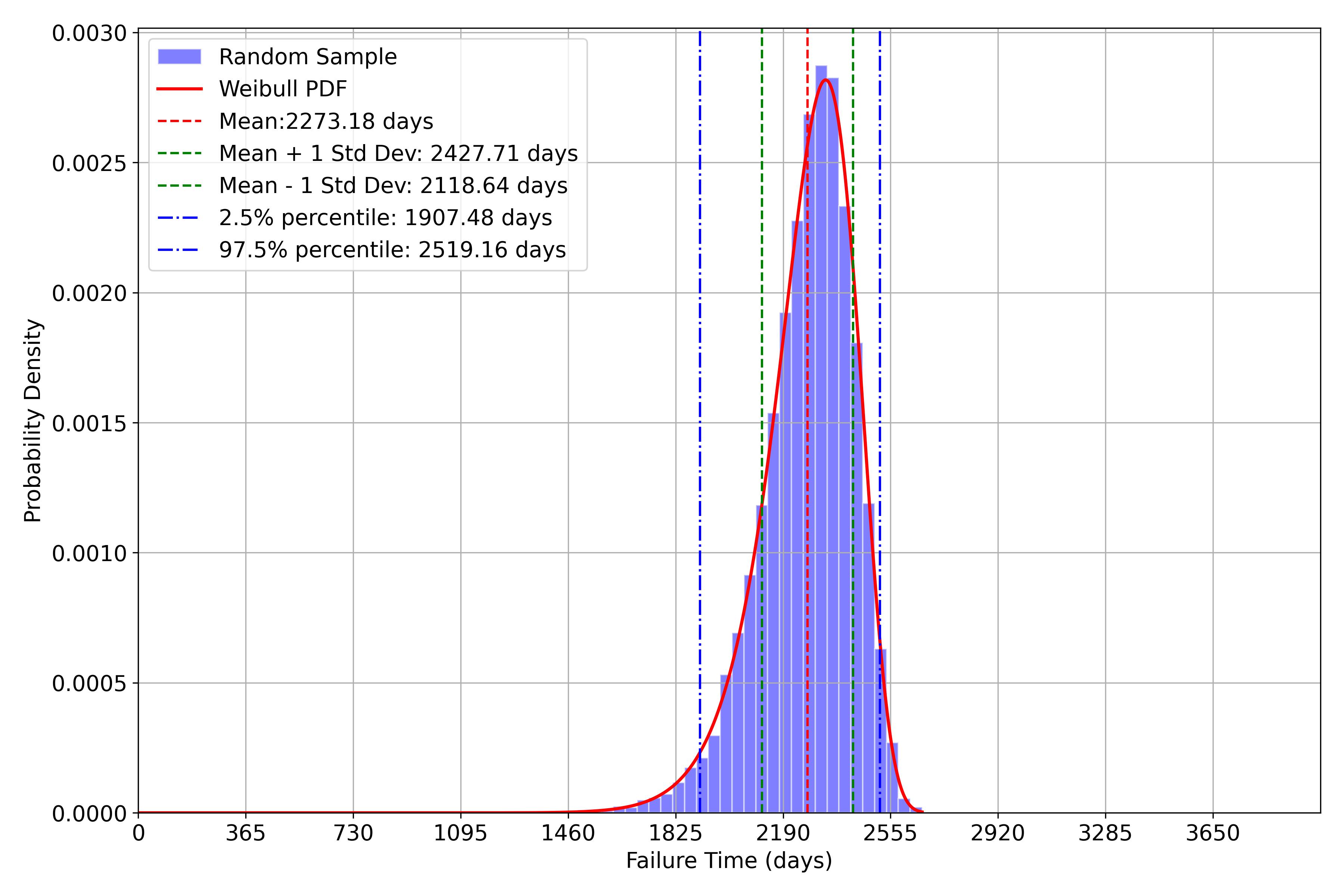} 
	\caption{S3 Weibull lifetime distribution}
	\label{fig:S3WLD} 
    \end{subfigure}
    \caption{Settings used in the sensitivity analysis}
    \label{fig:STSSA}
\end{figure}

To investigate the impact of variations in LED driver failure timing, a sensitivity analysis is conducted using two additional settings: Setting~S2 with parameters $(25.61, 3301.09)$ and Setting~S3 with parameters $(17.91, 2342.09)$. In S2, LED drivers fail later, with the mean driver failure time set three standard deviations later than the LED package MTTF, reflecting a scenario in which driver reliability is substantially better than package reliability, as shown in \autoref{fig:S2WLD}. In S3, LED drivers fail earlier, with the mean driver failure time set three standard deviations earlier than the package MTTF, representing comparatively poorer driver reliability, as shown in \autoref{fig:S3WLD}.

The Pareto front under S2 (\autoref{fig:PFS2}) exhibits trends broadly consistent with S1. Because drivers fail later, system renewal is primarily governed by package degradation, and many policies tolerate substantial degradation before replacement. At the upper-left end of the front, the solution with the lowest mean deficiency ratio ($\rev{R_{\text{SDR}}}$) represents a PM-driven maintenance strategy. For these Group 1 solutions, increasing the OM age $T_{\text{OM}}$ reduces OM replacements; however, this reduction is partially offset by a rise in CM replacements, resulting in a slight net decrease in $\rev{N_{\text{STR}}}$. As $T_{\text{OM}}$ approaches and then surpasses the package MTTF, the policy becomes more conservative (Group 2), resulting in fewer OM replacements but more site visits ($\rev{N_{\text{STV}}}$). Additionally, replacements triggered by CM visits partially offset the reduction in OM replacements, causing only a minor decrease in $\rev{N_{\text{STR}}}$. Compared with S1, S2 typically exhibits poorer system performance (higher $\rev{R_{\text{SDR}}}$) because maintenance becomes increasingly package-driven, allowing LEDs to reach more substantial degradation before replacement.

For setting S3 (\autoref{fig:PFS3}), the upper-left solutions likewise represent PM-driven maintenance, similar to S2. In Group 1, the aggressive OM policy results in slightly increased site visits ($\rev{N_{\text{STV}}}$). As $T_{\text{OM}}$ increases, the mean deficiency ratio ($\rev{R_{\text{SDR}}}$) worsens. When the OM policy shifts to more conservative strategies, OM replacements decrease significantly, but CM-driven replacements increase rapidly, indicating faster system deterioration. Notably, S3 reveals a pronounced upward shift in the mean total number of replacements ($\rev{N_{\text{STR}}}$). Given the earlier driver failures, the minimum number of replacements in S3 nearly equals the maximum number observed in S1 and S2, underscoring the critical impact of premature driver failure on maintenance planning. These shifts highlight how component reliability reshapes the Pareto trade-offs, and they underscore how the model can guide policy selection under differing component lifetime distributions.

Taken together, S2--S3 confirm that the proposed framework is structurally robust (the same trade-offs persist), while also demonstrating that relative luminaire reliability strongly influences where the Pareto front lies.

To examine the potential impact of thermal coupling between driver degradation and LED package aging, a sensitivity analysis is conducted by re-evaluating the two representative OM policies under an elevated service temperature. This setting is intended to represent a plausible coupled scenario in which a degrading driver induces thermal instability and thereby accelerates package degradation. Policies~3 and 4 are therefore re-evaluated at $55^\circ\mathrm{C}$, while keeping their decision variables unchanged, and the resulting policies are denoted as Policies~5 and 6, respectively. The comparison is summarized in \autoref{tab:temp_compare}.

\begin{table}[htbp]

\caption{Sensitivity of representative OM policies under a thermally coupled driver--package degradation scenario}
\label{tab:temp_compare}
    \begin{threeparttable}
    \resizebox{\linewidth}{!}{%
    \begin{tabular}{c c c c c c c c c c c c}
    \hline
    \makecell{Service-temperature\\scenario} & Policy & \makecell{$\rev{R_{\text{SDR}}}$ (Mean \\ deficiency \\ ratio (\%))} & \makecell{$\rev{N_{\text{STV}}}$ (No. of \\ visits)} & \makecell{$\rev{N_{\text{STR}}}$ (No. of \\replacements)} & \makecell{OM \\ threshold (fraction)} & \makecell{PM interval \\ (days)} & OM age (days) & \makecell{Mean PM \\ replacements} & \makecell{Mean CM \\ replacements} & OM (PM) & OM (CM) \\
    \hline
    $45^\circ\mathrm{C}$ & 3 & 0.1283 & 8.0917  & 608.0000 & 0.2 & 2190  & 1752 & 457.2635 & 2.0040  & 0.9075 & 147.8250 \\
    $45^\circ\mathrm{C}$ & 4 & 0.8719 & 12.0519 & 532.0176 & 0.8 & 11315 & 2263 & 0   & 12.2372 & 0 & 519.7804 \\
    $55^\circ\mathrm{C}$ & 5 & 0.1516 & 8.0667  & 608.0076 & 0.2 & 2190  & 1752 & 441.1591 & 2.2113  & 0.9900 & 163.6472 \\
    $55^\circ\mathrm{C}$ & 6 & 0.9302 & 12.3969 & 532.0347 & 0.8 & 11315 & 2263 & 0   & 13.1580 & 0 & 518.8767 \\
    \hline
    \end{tabular}%
    }
    \end{threeparttable}
\end{table}

As shown in \autoref{tab:temp_compare}, organizing the results by service-temperature scenario \rev{makes it clear that} introducing the thermally coupled degradation scenario leads to only modest changes in the maintenance outcomes of the representative OM policies. Comparing Policy~3 with Policy~5 and Policy~4 with Policy~6 shows that the mean deficiency ratio $\rev{R_{\text{SDR}}}$ increases slightly under $55^\circ\mathrm{C}$, indicating that the coupled scenario primarily manifests itself through mildly worse system performance. By contrast, the mean total number of site visits $\rev{N_{\text{STV}}}$ and the mean total number of replacements $\rev{N_{\text{STR}}}$ show only minor changes. These results suggest that, for the representative policies examined here, neglecting package--driver dependence does not materially overstate the effective service life of the maintained system or distort the maintenance-effort metrics. Instead, the main effect of the potential thermal coupling is a limited degradation in system performance, rather than a substantial shift in visit or replacement requirements.

\section{Conclusion}
This paper proposed an integrated, performance-driven, simulation-in-the-loop framework for long-term maintenance optimization of large-scale LED lighting systems. For each luminaire, gradual degradation of LED packages was modeled using a non-homogeneous Gamma process whose mean evolution follows an exponential lumen maintenance trend, while abrupt driver failures were represented by a Weibull lifetime model. The two mechanisms were combined through a competing-failure formulation to define luminaire states. Model parameters were calibrated from LM-80 data via Bayesian inference, enabling both parameter uncertainty and stochastic process variability to be propagated to service conditions for downstream Monte Carlo evaluation. Sensitivity studies further examined how thermally coupled degradation scenarios and driver reliability assumptions affect maintenance outcomes.

At the system level, luminaire states were mapped to the WP illuminance field through the lighting simulation model. Two standard commissioning indices, average illuminance and uniformity, were adopted to assess compliance at each time point. A deficiency ratio (DR) was then defined to evaluate the long-term performance of the lighting system over the entire operating horizon. This metric quantifies the fraction of operating time during which average illuminance or uniformity requirements are violated, thereby characterizing long-term system performance. To make large-scale policy evaluation computationally feasible, a surrogate model was developed to replace repeated Radiance evaluations. This surrogate yielded orders-of-magnitude acceleration while preserving high predictive fidelity.

Building on these elements, a performance-driven opportunistic maintenance policy was evaluated and optimized under three competing objectives: minimizing the mean DR, the mean total number of site visits, and the mean total number of luminaire replacements. Preventive maintenance intervals and opportunistic thresholds were used as decision variables, and policies were assessed through a discrete-event Monte Carlo simulator coupled with the performance evaluation module. A real office-zone case study demonstrated that the proposed framework could reveal clear Pareto trade-offs and provide decision support for selecting maintenance policies aligned with different performance and resource priorities. \rev{In practical terms, the DR enables operators to evaluate maintenance plans by the duration of illuminance non-compliance rather than by component age or failure counts alone. The Pareto front also supports policy selection. Performance-critical spaces can choose policies with lower DR values, while resource-constrained facilities can choose policies with fewer replacements or site visits. The operational interpretation of representative policies further shows how selected policies can inform spare parts provisioning and workforce allocation.}

Several limitations remain and suggest directions for future work. First, the degradation of the LED package and the failure of the LED driver were assumed to be independent. Potential interactions between these two mechanisms were not explicitly modeled, due to the lack of established quantitative relationships for their joint evolution. The thermal-coupling sensitivity analysis presented in this study suggests that a plausible coupled scenario mainly worsens system-performance indicators while having little influence on maintenance-effort metrics for the representative policies examined. Even so, future work may consider more detailed joint models of driver-induced thermal instability and package degradation. Second, the driver failure modeling in the case study relied on assumed parameter settings due to the limited availability of long-term empirical data. While the adopted Weibull formulation was consistent with standard reliability modeling practice, the parameterization was intended to provide representative scenarios for analysis. Future work with richer empirical driver lifetime data would enable more data-driven calibration and further strengthen the applicability of the results. Third, practical resource constraints such as spare parts inventory, crew capacity, and access scheduling could be integrated to produce more implementable policies, and richer cost models and real maintenance records would allow the current resource proxies (visits and replacements) to be mapped into monetary objectives and further strengthen field applicability.

\section*{Acknowledgments}
The authors would like to express their sincere gratitude to the reviewers for their valuable feedback and constructive comments. This research was funded by the Australian Research Council through the ARC Research Hub for Resilient and Intelligent Infrastructure Systems (RIIS) (IH210100048). The authors also appreciate the collaboration and support from their industry partner, Fredon, and Queensland University of Technology (QUT). Computational (and/or data visualization) resources and services used in this work were provided by the eResearch Office, Queensland University of Technology, Brisbane, Australia. 

\appendix
\numberwithin{equation}{section}
\numberwithin{figure}{section}
\numberwithin{table}{section}

\section{Supplementary algorithmic details}\label{app:alg_details}
\subsection{Discrete-event maintenance simulation}
\begin{algorithm}[H]
\caption{Discrete-event maintenance simulation for a given policy}
\label{alg:Discrete_event}
\scriptsize
\setstretch{0.85}
\begin{algorithmic}[1]
\State \textbf{Input:}
\begin{itemize}
    \item Degradation/failure models for luminaires: Gamma process parameters (or posterior draws) and Weibull parameters;
    \item Policy variables: OM threshold $H_{\text{OM}}$ and PM interval $T_{\text{PM}}$;
    \item Time parameters: operational horizon $T_{\text{over}}$, maintenance service times (e.g., $d_{\text{CM-df}}$, $d_{\text{CM-pf}}$);
    \item Recording schedule for performance evaluation (e.g., next record time rule for $t^{\text{record}}$);
    \item Simulation size: number of Monte Carlo replications $S$.
\end{itemize}
\State \textbf{Output:} For each run $s$: event-time vector $\mathbf{T}^{\text{ts-evt}}_{s}$ (\autoref{eq:GEVT}), state trajectory $\mathbf{L}_{s}$ (\autoref{eq:GLS}), and counts $(N_{\text{tv},s}, N_{\text{tr},s})$ (\rev{Equations~(\ref{eq:TV_sim})--(\ref{eq:TR_sim})}); and averages $(N_{\text{STV}},N_{\text{STR}})$ (\rev{Equations~(\ref{eq:TTV})--(\ref{eq:TTR})}).

\State Initialize accumulators for $N_{\text{STV}}$ and $N_{\text{STR}}$.
\For{$s=1$ \textbf{to} $S$}
    \State Initialize time $t \gets 0$, event index $k \gets 1$.
    \State Initialize luminaire states $\mathbf{L}(0)$ and schedule initial event vectors $\mathbf{t}_{s,1,j}$ for all $j$ using \autoref{eq:EV}.
    \State Initialize $N_{\text{cm,vis},s}\gets 0$, $N_{\text{pm,vis},s}\gets 0$, $N_{\text{cm},s}\gets 0$, $N_{\text{pm},s}\gets 0$, $N_{\text{om},s}\gets 0$.

    \While{$t < T_{\text{over}}$}
        \State Form the system event matrix $\mathbf{T}^{\text{evt}}_{s,k}$ using \autoref{eq:EMM}.
        \State Identify the next event time and its source:
        $ (t_{s,k}^{\text{ts-evt}},\, j^*,\, \text{type}) \gets \arg\min \mathbf{T}^{\text{evt}}_{s,k} $
        
        \State Set $t \gets t_{s,k}^{\text{ts-evt}}$ and append $t$ to $\mathbf{T}^{\text{ts-evt}}_{s}$.

        \If{$\text{type}=\text{PM}$ (i.e., $t=t_{s,k,j^*}^{\text{pm}}$)}
            \State $N_{\text{pm,vis},s}\gets N_{\text{pm,vis},s}+1$; replace luminaire $j^*$ ($N_{\text{pm},s}\gets N_{\text{pm},s}+1$).
            \State Set maintenance completion time $t_{s,k,j^*}^{\text{event-end}}\gets T_{\text{over}}$.
            \State \textbf{OM selection:} For each luminaire $\ell\neq j^*$, compute $H_{\text{mow},\ell}$ via \autoref{eq:hmov}.
            \If{$H_{\text{mow},\ell}\le H_{\text{OM}}$}
                \State Replace luminaire $\ell$ opportunistically ($N_{\text{om},s}\gets N_{\text{om},s}+1$).
                \State Update/reset its future event times in $\mathbf{t}_{s,k,\ell}$ (e.g., next PM and predicted CM times).
            \EndIf
            \State Update/reset future event times for luminaire $j^*$ (e.g., resample failure times after replacement, schedule next PM).
        
        \ElsIf{$\text{type}=\text{CM-df}$ (i.e., $t=t_{s,k,j^*}^{\text{cm,df}}$)}
            \State $N_{\text{cm,vis},s}\gets N_{\text{cm,vis},s}+1$; replace luminaire $j^*$ ($N_{\text{cm},s}\gets N_{\text{cm},s}+1$).
            \State Set $t_{s,k,j^*}^{\text{event-end}}\gets t + d_{\text{CM-df}}$.
            \State \textbf{OM selection:} For each luminaire $\ell\neq j^*$, compute $H_{\text{mow},\ell}$ via \autoref{eq:hmov}; if $H_{\text{mow},\ell}\le H_{\text{OM}}$, apply OM and update/reset its event times.
            \State Update/reset future event times for luminaire $j^*$ after replacement.

        \ElsIf{$\text{type}=\text{CM-pf}$ (i.e., $t=t_{s,k,j^*}^{\text{cm,pf}}$)}
            \State $N_{\text{cm,vis},s}\gets N_{\text{cm,vis},s}+1$; replace luminaire $j^*$ ($N_{\text{cm},s}\gets N_{\text{cm},s}+1$).
            \State Set $t_{s,k,j^*}^{\text{event-end}}\gets t + d_{\text{CM-pf}}$.
            \State \textbf{OM selection:} For each luminaire $\ell\neq j^*$, compute $H_{\text{mow},\ell}$ via \autoref{eq:hmov}; if $H_{\text{mow},\ell}\le H_{\text{OM}}$, apply OM and update/reset its event times.
            \State Update/reset future event times for luminaire $j^*$ after replacement.

        \ElsIf{$\text{type}=\text{END}$ (i.e., $t=t_{s,k,j^*}^{\text{event-end}}$)}
            \State Mark maintenance complete for luminaire $j^*$ and set $t_{s,k,j^*}^{\text{event-end}}\gets T_{\text{over}}$.
            \State Update $\mathbf{t}_{s,k,j^*}$ (e.g., enable future PM/CM events as applicable).

        \ElsIf{$\text{type}=\text{RECORD}$ (i.e., $t=t_{s,k,j^*}^{\text{record}}$)}
            \State No maintenance action; update the next record time(s) in the event vectors (e.g., advance $t^{\text{record}}$).
        \EndIf

        \State Record system state $\mathbf{L}(t)$ at the current event index and append to the trajectory $\mathbf{L}_{s}$.
        \State Increment event index $k \gets k + 1$.
    \EndWhile

    \State Compute totals for run $s$: $N_{\text{tv},s}$ and $N_{\text{tr},s}$ using \rev{Equations~(\ref{eq:TV_sim})--(\ref{eq:TR_sim})}.
\EndFor

    \State Average over runs: $N_{\text{STV}}$ and $N_{\text{STR}}$ using \rev{Equations~(\ref{eq:TTV})--(\ref{eq:TTR})}.
\end{algorithmic}
\end{algorithm}

\subsection{Performance evaluation module}
{\tiny
\begin{algorithm}[H]
\caption{Performance evaluation module (PEM) for one simulation run}
\label{alg:PEM}
\scriptsize
\setstretch{0.85}
\begin{algorithmic}[1]
\State \textbf{Input:}
\begin{itemize}
    \item Performance standards: $S_{E}$ and $S_{U}$;
    \item Event-time vector $\mathbf{T}^{\text{ts-evt}}_{s}=[t_{s,1},\ldots,t_{s,K}]$ (\autoref{eq:GEVT});
    
    \item Performance-recording interval $\Delta t_{\text{record}}=50$ days, which bounds the maximum spacing between successive performance evaluations to ensure interpolation accuracy;
    \item Luminaire state trajectory $\mathbf{L}_{s}$ (\autoref{eq:GLS});
    \item Performance mapping model (Radiance or surrogate; Section~\ref{sec:SAIE}).
\end{itemize}
\State \textbf{Output:} Deficiency ratio $R_{\text{DR},s}$ for run $s$ (\autoref{eq:NODR}) (and optionally illuminance sequence $\mathbf{E}_{s}$).
\Statex
\Comment{\textit{Step 1: map states to static performance indices at event times}}
\For{$k=1$ \textbf{to} $K$}
    \State Retrieve luminaire state vector $\mathbf{L}(t_{s,k})$ from $\mathbf{L}_{s}$.
    \State Evaluate the WP illuminance vector $\mathbf{E}(t_{s,k})$ using the performance mapping model.
    \State Compute $E_{\text{avg}}(t_{s,k})$ via \autoref{eq:avg} and $U(t_{s,k})$ via \autoref{eq:NUI}.
\EndFor
\Statex
\Comment{\textit{Step 2: compute interval-wise deficiency durations and PDD via linear interpolation}}
\State Initialize $\sum T_{\text{defi},s} \gets 0$.
\For{$k=2$ \textbf{to} $K$}
    \State Retrieve $E_{\text{avg}}$ and $U$ at both endpoints $t_{s,k-1}$ and $t_{s,k}$, and set $\Delta t_k = t_{s,k} - t_{s,k-1}$.
    
    \State \textbf{For each performance index} ($E_{\text{avg}}$ vs.\ $S_E$, and $U$ vs.\ $S_U$):
    \Statex \hspace{1.5em} (i) If both endpoint values lie below the threshold, the entire interval is counted as deficient: $T_{\bullet\_\text{defi},s}(k) \gets \Delta t_k$.
    \Statex \hspace{1.5em} (ii) If both endpoint values lie above the threshold, the interval is counted as compliant: $T_{\bullet\_\text{defi},s}(k) \gets 0$.
    \Statex \hspace{1.5em} (iii) If the two endpoint values lie on opposite sides of the threshold, the threshold-crossing time is estimated by linear interpolation between the endpoint values, and only the sub-interval that violates the requirement is counted toward $T_{\bullet\_\text{defi},s}(k)$.
    
    \State Compute the interval-level PDD using \autoref{eq:TDEFI}:
    \[
        T_{\text{defi},s}(k) \gets \max\!\left(T_{\text{E\_defi},s}(k),\, T_{\text{U\_defi},s}(k)\right).
    \]
    \State Accumulate $\sum T_{\text{defi},s} \gets \sum T_{\text{defi},s} + T_{\text{defi},s}(k)$.
\EndFor
\Statex
\Comment{\textit{Step 3: compute deficiency ratio for the run}}
\State Compute $R_{\text{DR},s} = \dfrac{\sum T_{\text{defi},s}}{T_{\text{over}}}$ using \autoref{eq:NODR}.
\end{algorithmic}
\end{algorithm}
}

\subsection{Sequential statistical screening of candidate policies}
{\tiny
\begin{algorithm}[H]
\caption{Sequential statistical screening of candidate policies}
\label{alg:stat_screen}
\scriptsize
\setstretch{0.85}
\begin{algorithmic}[1]
\State \textbf{Input:}
\begin{itemize}
    \item Objective means $\boldsymbol{\mu}_i=(\mu_{i,1},\mu_{i,2},\mu_{i,3})$ for candidate policies $i=1,\ldots,N$;
    \item Objective standard deviations $\boldsymbol{\sigma}_i=(\sigma_{i,1},\sigma_{i,2},\sigma_{i,3})$;
    \item Number of Monte Carlo runs $n_{\mathrm{simu}}$;
    \item Significance level $\alpha_{\mathrm{sv}}$.
\end{itemize}
\State \textbf{Output:} Retained policy set $\mathcal{S}_{\mathrm{ret}}$.
\Statex
\Comment{\textit{Step 1: initialize retention mask}}
\For{$i=1$ \textbf{to} $N$}
    \State Set $\texttt{retain}[i]\gets \texttt{True}$.
\EndFor
\Statex
\Comment{\textit{Step 2: sequential statistical screening}}
\For{$i=1$ \textbf{to} $N$}
    \If{$\texttt{retain}[i]=\texttt{False}$}
        \State \textbf{continue}
    \EndIf
    \For{$j=1$ \textbf{to} $N$}
        \If{$j=i$ \textbf{or} $\texttt{retain}[j]=\texttt{False}$}
            \State \textbf{continue}
        \EndIf
        \State Set $\texttt{remove}\gets \texttt{True}$.
        \For{$k=1$ \textbf{to} $3$}
            \State Perform a one-sided Welch's $t$-test for objective $k$:
            \[
            H_0:\mu_{j,k}\le \mu_{i,k}
            \qquad \text{vs.} \qquad
            H_1:\mu_{j,k}>\mu_{i,k},
            \]
            where smaller objective values are preferred.
            \State Let $p_{ijk}$ be the resulting $p$-value.
            \If{$p_{ijk}<\alpha_{\mathrm{sv}}$}
                \State Set $\texttt{remove}\gets \texttt{False}$.
                \State \textbf{break}
            \EndIf
        \EndFor
        \If{$\texttt{remove}=\texttt{True}$}
            \State Set $\texttt{retain}[i]\gets \texttt{False}$.
        \EndIf
    \EndFor
\EndFor
\Statex
\Comment{\textit{Step 3: collect retained policies}}
\State Return $\mathcal{S}_{\mathrm{ret}}=\{i:\texttt{retain}[i]=\texttt{True}\}$.
\end{algorithmic}
\end{algorithm}
}

\section{Impact of surrogate prediction error on performance indices} \label{sec:surrogate_error}

This appendix further examines how surrogate prediction error propagates to the performance indices used in maintenance evaluation. In the proposed framework, the surrogate model predicts the working plane illuminance vector from the luminaire-state vector $\mathbf{L}(t_k)$, after which the performance indices $E_{\mathrm{avg}}(t_k)$ and $U(t_k)$ are computed.

For the average illuminance metric, let the surrogate prediction error at the $i$-th working plane point be denoted by
\begin{equation}
    \hat{E}_i(t_k)=E_i(t_k)+\varepsilon_i(t_k),
\end{equation}
where the errors $\varepsilon_i(t_k)$ are assumed to be independent and approximately unbiased, with
\begin{equation}
    \mathbb{E}[\varepsilon_i(t_k)]=0, \qquad \mathrm{Var}(\varepsilon_i(t_k))=\sigma_\varepsilon^2.
\end{equation}

By the central limit theorem (CLT), for sufficiently large $N$, the average error is approximately normally distributed with variance $\frac{\sigma_{\varepsilon}^2}{N}$. 

In the present case study, a conservative error scale of $\sigma_\varepsilon \approx 3\times \mathrm{RMSE} \approx 3$ lux is adopted, with $N=265$. This yields a standard deviation of approximately
\begin{equation}
    \frac{\sigma_\varepsilon}{\sqrt{N}} \approx 0.18\ \text{lux}.
\end{equation}
This value is negligible relative to the maintained illuminance requirement of 500 lux.

For the uniformity metric, a first-order approximation can be developed near the compliance boundary. In the present case study, the threshold values are $E_{\mathrm{avg}}=500$ lux and $U=0.6$, corresponding to $E_{\min}=300$ lux. Using the CLT-based approximation for the average illuminance,
\begin{equation}
    \hat{E}_{\mathrm{avg}} \approx 500+\bar{\varepsilon},
\end{equation}
where
\begin{equation}
    \bar{\varepsilon} \sim \mathcal{N}\!\left(0,\frac{\sigma_\varepsilon^2}{N}\right),
\end{equation}
and representing the surrogate error at the minimum-illuminance point by
\begin{equation}
    \hat{E}_{\min}=300+\varepsilon_{\min},
\end{equation}
with
\begin{equation}
    \varepsilon_{\min}\sim\mathcal{N}(0,\sigma_\varepsilon^2),
\end{equation}
the predicted uniformity becomes
\begin{equation}
    \hat{U}=\frac{300+\varepsilon_{\min}}{500+\bar{\varepsilon}}.
\end{equation}

A first-order Taylor expansion of $U=E_{\min}/E_{\mathrm{avg}}$ about the compliance boundary gives
\begin{equation}
    \hat{U}
    \approx
    0.6+\frac{\varepsilon_{\min}}{500}
    -\frac{300}{500^2}\bar{\varepsilon}.
\end{equation}
This expression shows that the surrogate-induced perturbation in uniformity is governed primarily by the local prediction error at the minimum-illuminance point, whereas the contribution from the average-illuminance error is further attenuated through spatial averaging over the $N=265$ working plane points.

Using the conservative error scale $\sigma_\varepsilon \approx 3$ lux, the first term is on the order of $3/500 \approx 0.006$, whereas the second term is further reduced by spatial averaging and is on the order of $10^{-4}$. Accordingly, the overall perturbation in uniformity is expected to be around $10^{-2}$, which remains small relative to the compliance threshold.

Therefore, under the configuration considered in this study, the surrogate approximation error is expected to have a negligible impact on the performance indices and, consequently, on practical maintenance decision-making.

\section{Bayesian calibration convergence diagnostics}
To further assess the convergence and sampling efficiency of the Bayesian calibration, additional MCMC diagnostic plots are provided in this appendix. The diagnostics include trace plots, posterior density plots, and autocorrelation plots for the calibrated parameters $\ln A$, $b$, $\ln C$, and $E_a$.

\begin{figure}
    \centering
    \includegraphics[width=0.8\textwidth]{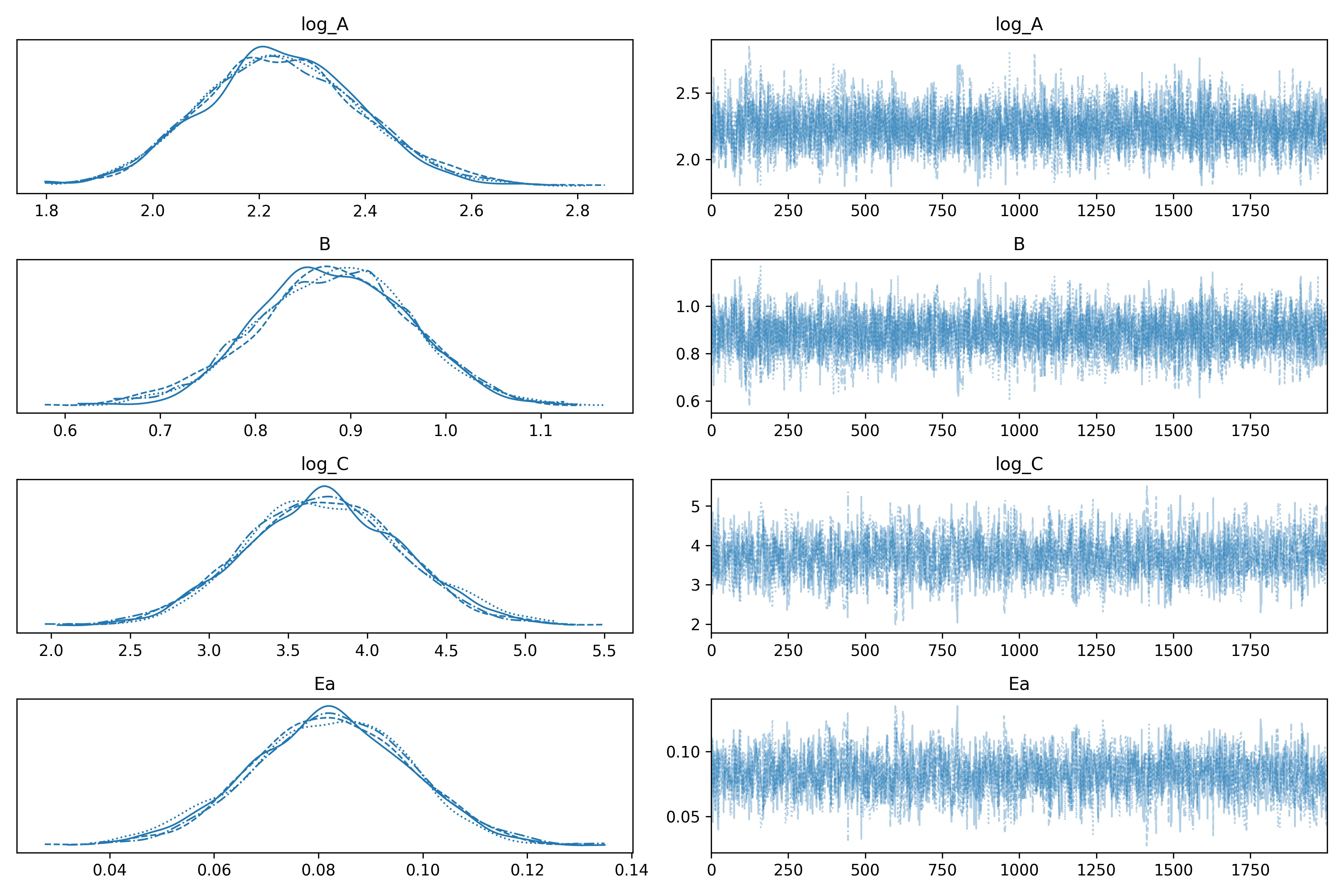}
    \caption{Trace plots of the posterior samples for the calibrated parameters across four MCMC chains.}
    \label{fig:mcmc_trace}
\end{figure}

\begin{figure}
    \centering
    \includegraphics[width=0.95\textwidth]{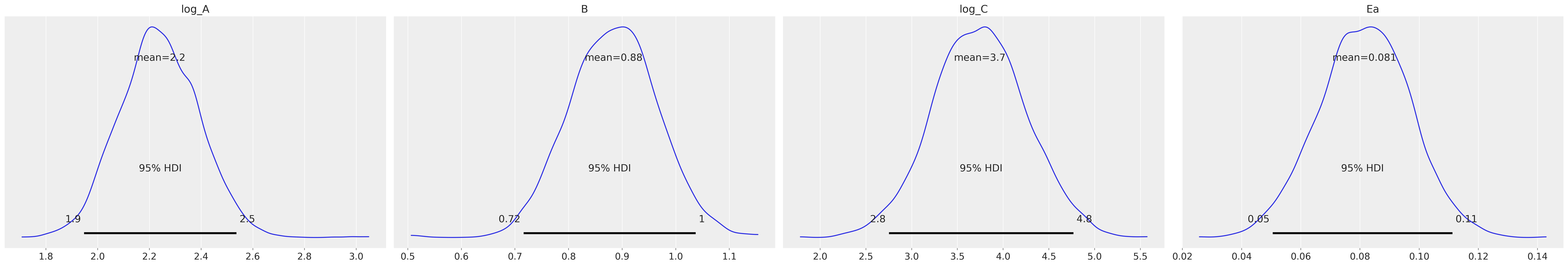}
    \caption{Posterior density plots of the calibrated parameters.}
    \label{fig:posterior_density}
\end{figure}

\begin{figure}
    \centering
    \includegraphics[width=0.8\textwidth]{F18c_autocorrelation.jpeg}
    \caption{Autocorrelation plots of the posterior samples for the calibrated parameters across four MCMC chains.}
    \label{fig:autocorrelation}
    {\footnotesize Note: Rapid decay with increasing lag indicates acceptable sampling efficiency.}
\end{figure}

\begin{table}[!htbp]
    \centering
    \caption{Posterior correlation matrix of the calibrated parameters.}
    \label{tab:posterior_corr}
    \begin{tabular}{lrrrr}
        \toprule
        & $\ln A$ & $b$ & $\ln C$ & $E_a$ \\
        \midrule
        $\ln A$ & 1.0000 & -0.9459 & 0.0038 & 0.0132 \\
        $b$     & -0.9459 & 1.0000 & 0.0259 & -0.0111 \\
        $\ln C$ & 0.0038 & 0.0259 & 1.0000 & -0.9934 \\
        $E_a$   & 0.0132 & -0.0111 & -0.9934 & 1.0000 \\
        \bottomrule
    \end{tabular}
\end{table}

\printcredits

\bibliographystyle{cas-model2-names}

\bibliography{references}


\end{document}